\documentclass[preprint,11pt,nopreprintline,sort&compress]{elsarticle}

\usepackage[left=1in,right=1in,top=1in,bottom=1in,a4paper]{geometry}

\usepackage{setspace,amsmath,latexsym,amssymb,epsfig,amsfonts,subfig}
\usepackage{graphicx}
\usepackage{dcolumn}
\usepackage{nicefrac}
\usepackage{siunitx}
\usepackage{tikz}
\usepackage{comment}
\usepackage{nth}
\usepackage{lipsum}
\usepackage{float}
\usepackage{hyperref} % keep hyperref LAST
\journal{J. Colloid Interface Sci.}

\begin{document}

\begin{frontmatter}

\title{Numerical simulations of simultaneous pair-drop impacts and their energetics}

% Oxford
\affiliation[1]{organization={Department of Engineering Science, University of Oxford},
            city={Oxford},
            postcode={OX1 3PJ}, 
            country={United Kingdom}}

% Authors

\author[1]{Ziyao Zhang\corref{cor1}}
    \ead{ziyao.zhang@eng.ox.ac.uk}
    \cortext[cor1]{Corresponding author}

\author[1]{Alfonso~A.~Castrej\'{o}n-Pita}

\author[1]{Wouter Mostert}

\begin{abstract}
  % We present three-dimensional (3-D) high-resolution direct numerical simulations of simultaneous pair-drop impacts. We investigate the dynamics and kinematics of the central rising sheet, formed by the interaction of two drops after impacts on the substrate. The simulation results align well with previously published experiments, including central sheet width, height, and the overall semilunar morphology, which has a semilunar character. We then further study the kinematics and energetics of the central sheet, where kinetic energy and surface tension energy dominate, with viscous dissipation playing an important role via the Reynolds number, while gravitational potential effects are of secondary importance. Next, a novel energetic model is proposed to predict the height of the central sheet, which agrees well with experiments and numerical data. Additionally, in the asymptotic regimes at high $We$ and $Re$, the spreading behaviours of central sheets resemble single drop spreading behaviors on the substrate. Finally a scaling law for the central sheet height is proposed and supported by numerical data. These insights promise a more complete understanding of the dynamics for the central rising sheet and lay the groundwork for further exploration of its fragmentation. 
  
We present three-dimensional direct numerical simulations of the simultaneous impact of two identical drops on an hydrophobic substrate, varying the relative strength of capillary and viscous effects respectively through Weber and Reynolds numbers of impact. The interaction between the two drops is characterized by the appearance of a lamella arising from the collision of the two droplets' spreading rims. We examine the width, the height, and the general morphological evolution of the central sheet; the numerical data is validated against prior experiments and used to guide the development of an energetic model for the maximum elevation of the central sheet. In particular, the rise  of the central sheet resembles the spreading behaviour single-drop impacts, especially at high Weber and Reynolds numbers. This fact can be used to estimate scalings in the capillary- and viscous-dominated regimes, which can be used to collapse the trajectories. These insights provide a route for a more complete understanding of the dynamics for the central rising sheet, and anticipate the detailed study of its fragmentation characteristics. 
\end{abstract}    
    
\begin{keyword}
    pair droplets \sep simultaneous impact \sep central rising sheet
\end{keyword}   

\end{frontmatter}

% Main text
\section{Introduction}\label{sec:Intro}
Drop impact is a widespread and ubiquitous phenomenon in nature and industry. Ranging from rain in nature, which brings the smell of the earth \citep{Joung2015NC}, and furthermore causes the erosion of soil \citep{Yarin2006ARFM}, drop impact is also important to applications in industry, such as inkjet printing, spray cooling, spray coating \citep{Josserand2016ARFM}, continuous and pulsant impacts on turbine blade \citep{Li2008IJMS}, and blood-splatter analysis \citep{Hulse-Smith2005JFS}. Understanding this process and its underlying mechanisms advances our knowledge of natural physics, and facilitates its controlled application in industrial settings.

While it is necessary first to understand the impact of single drops, it is not unusual that in many industrial settings multiple drop impacts may occur in close proximity, such as in spray coating, inkjet printing, etc. \citep{Josserand2016ARFM, Fest-Santini2021CF}. Such impacts should not be considered as a simple superposition of individual drop impacts because they may interact with one another. The inter-drop spacing $\Delta x$ and time lag $\Delta t$ between the two drops complicate this process, and generally simultaneous impacts, non-simultaneous impacts, or successive impacts can occur. For example, experimental and numerical work has been devoted to investigating successive impact and coalescence of drops while varying some of these parameters \citep{Guggulla2020ETFS, Chen2020POF, Wibowo2021IJTS, Luo2021IJHMT}.

When two drops are separated by a certain $\Delta x$, falling at low speed, they interact with each other upon impact, coalesce, and then form a large film\citep{Aarts2005PRL, Castrejon-Pita2013PRE}. When the drops fall at high kinetic energy, their spreading lamellae collide at the midpoint between the two drops and form an additional, vertical sheet of liquid. \citet{Barnes1999ICLASS} investigated the simultaneous and non-simultaneous impact on a small target through a monodisperse apparatus. It was found that the drops ejected from the central sheet were 50\% larger than those from single drop fragmentation. They also found that this 'splashing' behavior could occur at lower values of the splashing parameter K, than for single-drop impact, where K is defined as $K=We^{1/2}Re^{1/4}$, where $We$ is Weber number and $Re$ is Reynolds number. \citet{Roisman2002JCIS} also studied pair-drop impacts on a dry substrate, and showed that the maximum height of the central sheet increased with decreasing $\Delta x$. They also defined a dynamic model for the maximum height of the central sheet, with an explicit solution given for simultaneous impact. Afterward, \citet{Liang2020ActaM} and \citet{Gultekin2021ETFS}, in an experimental study, showed agreement with previous results and demonstrated that the maximum rising sheet height would increase with $We$. \citet{Gultekin2021ETFS} compared their results with the theoretical solution given by \citet{Roisman2002JCIS} but found disagreement in the case where the substrate was raised to an increased temperature. \citet{Ersoy2020POF} investigated the geometry and irregular splashing phenomenon of the central sheet, but faced challenges in their experiments in accounting for non-simultaneous drop impact. As in their experiments, the two drops impacted on the substrate within an inevitable time lag of $1~\text{ms}$. Most recently, \citet{Goswami2023JFM} studied simultaneous pair-drop impact on a dry substrate. They experimentally showed the height evolution of the central sheet with respect to different impact velocities and inter-drop spacing, and empirically found that the maximum height scaled as $H_{s,max}\sim We_{L,imp}^{0.48}$, where $We_{L,imp}$ is the Weber number associated with the impact between the two spreading lamellae.
%%%%%%%%%%%%%%%%%%%%%%%%

While previous studies have shed light on the interaction between two pair drops, the majority have used experiment, in which it is difficult, for example, to achieve precisely simultaneous impact of the pair drops. Numerical analysis can support an analysis of the problem not only by introducing precise control over impact parameters (see e.g. \cite{Raman2017JCIS,Sohag2023Splashing}), but also complete information on the flow fields in the liquid and gas phases with high spatiotemporal phases. The main purpose of this study is twofold: Firstly, to reproduce the problem of simultaneous pair-drop impact in high-resolution three-dimensional (3D) numerical simulations, validated against available experimental data; and secondly, to use the resulting numerical data propose an energetic model for the maximum height of the central uprising sheet. The study proceeds as follows. In \S$\,$\ref{sec:math_physical}, the physical and mathematical models are described first, and the numerical methods are explained and validated against experimental data. In \S$\,$\ref{sub:results}, we present our results: we first describe the whole process of pair-drop impacts; we present the effects of $We,Re$ on the issue, including maximum spreading radius, central sheet width, its height, and overall morphology. In \S$\,$\ref{sec:ki_en} , we propose a novel energetic model to explain the scaling law of the height as well as certain asymptotic behaviours of the height of the central sheet. Based on the analysis of scalings from energetic models, we obtain scaling laws for the maximum height of the central sheet and also the evolution time. We conclude in \S\ref{sec:conclusion} with remarks on future work.

\section{Formulation and methodology}\label{sec:math_physical}
\subsection{Mathematical and physical description}
We consider the simultaneous pair-drop impact problem. The problem formulation is shown in Fig. \ref{fig:problem_sketch}, with dimensional variables indicated by carets. Two drops of diameter $\hat D_0$ with the same density $\hat\rho_l$, viscosity $\hat\mu_l$, and surface tension $\hat\sigma$ to ambient air are considered to impact on a substrate with velocity $\hat{U}_0$ and lateral separation $\Delta\hat{x}$; they are contained in a gas of density and viscosity $\hat{\rho}_g, \hat{\mu}_g$. Gravitational acceleration $\hat{g}$ is included in these simulations but its effect on the dynamics of the impact and droplet interaction is negligible. In simulations, a domain width of $16\hat D_0$ is chosen to minimize the effects of the domain boundary \citep{Sanjay2023JFM1, Sanjay2023JFM2}, and the two drops are separated by $\Delta \hat x=1.8\hat D_0$, which is mostly fixed in this study. We consider a simultaneous impact problem, where the two drops are initialized at the same vertical position to avoid any deformation of the drops due to air drag prior to impact \citep{Goswami2023JFM}. Lateral and upper computational boundaries are set with outflow boundary conditions. The bottom computational boundary follows a non-slip velocity boundary condition, and we consider a hydrophobic surface here, namely the static contact angle $\theta_s = 90^\circ$.

The continuity and Navier-Stokes equation with surface tension for two-phase, viscous, immiscible flows can be written as \citep{Riviere2021JFM, Mostert2022JFM, Tang2023JFM, Wang2023JFM}:
\begin{equation}\label{eqn:interface-continuity}
    \frac{\partial \hat\rho}{\partial \hat t} + \nabla\cdot(\hat \rho\boldsymbol{\hat U})=0
\end{equation}
\begin{equation}\label{eqn:momentum}
    \hat\rho\left(\frac{\partial\boldsymbol{\hat U}}{\partial \hat t}+\boldsymbol{\hat U}\cdot\nabla\boldsymbol{\hat U}\right) =-\nabla \hat p+\nabla\cdot\left[\hat \mu\left(\nabla\boldsymbol{\hat U}+{\nabla\boldsymbol{\hat U}}^T\right)\right]+\hat\sigma\hat\kappa\delta_s\boldsymbol{n} + \hat\rho\boldsymbol{\hat g}
\end{equation}
\begin{equation}\label{eqn:continuity}
    \nabla\cdot \boldsymbol{\hat U}=0 
\end{equation}
which describe the conservation of mass and momentum. Here $\boldsymbol{\hat U}$ is the velocity vector, $\hat p$ is the fluid pressure, $\hat\kappa$ is the local curvature of the interface, $\boldsymbol{n}$ is the normal vector at the interface, and $\delta_s$ is Dirac delta function which is nonzero only at the interface of two different fluids \citep{Popinet2018ARFM}. One can use the characteristic length $\hat D_0$, velocity $\hat U_0$, time $\hat D_0/\hat U_0$ to get the dimensionless system (the dimensional variables are with carets, dimensionless are not), and thus the parameter space can be described in terms of four non-dimensional numbers \citep{Yarin2006ARFM, Wang2023JFM}:
\begin{equation}\label{eqns:non-dimensional number}
We = \frac{\hat\rho_l \hat D_0 \hat U_0^2}{\hat\sigma},\quad Re = \frac{\hat\rho_l \hat D_0 \hat U_0}{\hat\mu_l}, \quad \rho_r=\frac{\hat\rho_l}{\hat\rho_g},\quad \mu_r=\frac{\hat\mu_l}{\hat\mu_g}
\end{equation} 
which respectively describe the ratios of inertial and capillary effects, of inertial and viscous effects, and of density and viscosity between liquid and gas phases. For the purpose of comparison to the experimental setup of \citet{Goswami2023JFM} and \citet{GoswamiPhDthesis}, we will consider density and viscosity ratios relevant for water and 40\% glycerol-water mixtures, in air, namely $\rho_r=813,~\mu_r=56$ and $\rho_r=910,~\mu_r=269$, respectively. For the values of $We\in[15,200],~Re\in[938,8000]$ considered in this study, we have verified that gravity does not have an appreciable role. The remaining details of the parameter space are shown in the Supplementary.

\begin{figure}
\centering
   \includegraphics[width=0.47\columnwidth]{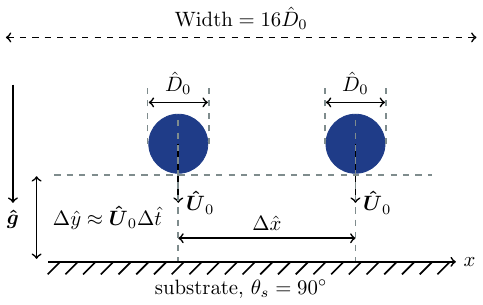}
\caption{Problem formulation illustrated using a two-dimensional schematic. Carets denote dimensional variables throughout the figure}
\label{fig:problem_sketch}
\end{figure}

\subsection{Numerical methods}
\begin{figure}
      \centering
    \includegraphics[width=0.7\textwidth]{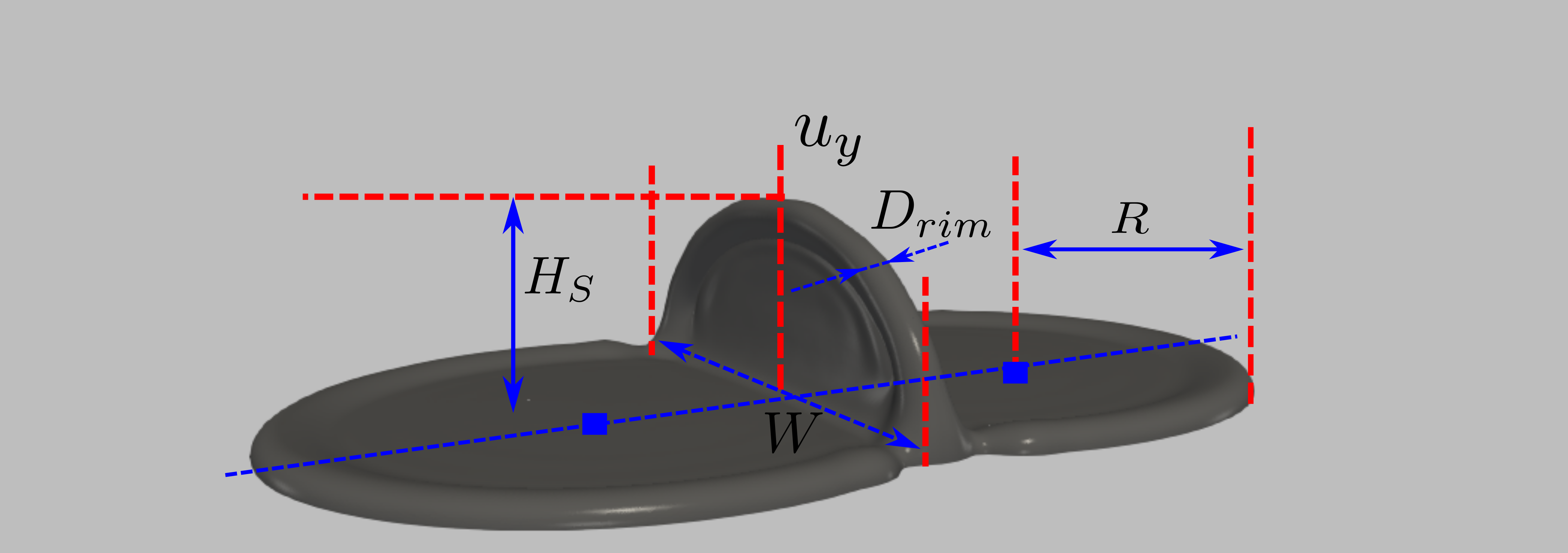}
      \caption{Definition of the measured variables of the central rising sheet formed by two drops after the second impact during the rising stage. The two blue squares indicate the assumed centers of the drops. $D_{rim}$, $W$, and $H_s$ are defined as the corresponding maximum values}
      \label{fig:Height_Hs}
\end{figure}

The open-source package \textit{Basilisk} is used to solve the nonlinear Navier-Stokes (NS) equations \eqref{eqn:interface-continuity}-\eqref{eqn:continuity} with an octree-based adaptive mesh refinement (AMR) method, which retains an accurate description of the flow while providing considerable computational savings over uniform-mesh approaches \citep{Popinet2020}. The geometric volume-of-fluid (VOF) method is used in a momentum-conserving formulation for the advection and numerical reconstruction of the liquid-air interface, which minimizes the parasitic currents induced by surface tension \citep{Popinet2018ARFM}. The VOF tracer is advected according to \citep{Popinet2009JCP}: 
\begin{equation}\label{eqn:volume_fraction}
    \frac{\partial f}{\partial \hat t}+\nabla\cdot(f\boldsymbol{\hat U})=0
\end{equation}
where $f$ refers to the volume fraction to differentiate liquid and gas, and it follows that,
\begin{equation}
    \begin{cases}
    f=1 &\text{in the liquid phase,}\\
    f=0 &\text{in the gas phase,} \\
    0<f<1 &\text{in the interface.}\\
    \end{cases}
\end{equation}
The density $\rho$ and viscosity $\mu$ can then be expressed as:
\begin{align}
    \hat\rho &= f\hat\rho_l + (1-f)\hat\rho_g,\\
    \hat\mu &= f\hat\mu_l + (1-f)\hat\mu_g.
\end{align}
The momentum equation is solved with the Bell-Colella-Glaz projection method \citep{Bell1989JCP}, and viscous terms are solved implicitly. The VOF advection scheme is mass- and momentum-conserving by construction, and surface tension is modelled with an implementation of the continuum surface force model \citep{Brackbill1992JCP,Popinet2009JCP}. Additional details can be found in \citet{Popinet2009JCP, Popinet2020}. The minimum grid size used in the simulations is specified by $\Delta = 16\hat D_0/2^L$, where $L$ is the maximum grid refinement level of AMR method. The results presented in this study are generally independent of mesh resolution at the chosen values of $L$; see the Supplementary Information for further details. We present the notations of the central sheet in Figure \ref{fig:Height_Hs} for following analyses.
% \begin{figure}
%       %\centering
%     \includegraphics[width=0.45\textwidth]{figures/front_Hs.png}
%     \hspace{0.05\textwidth}
%     \includegraphics[width=0.45\textwidth]{figures/side_Hs.png}
%       \caption{Notation of the maximum height of central rising sheet from front view and side views of two drops after the second impact at the rising stage.}
%       \label{fig:Height_Hs}
% \end{figure}

\section{Results}\label{sub:results}
\begin{figure}
\centering   
\begin{tabular}{c} 
 \includegraphics[width=0.50\columnwidth]{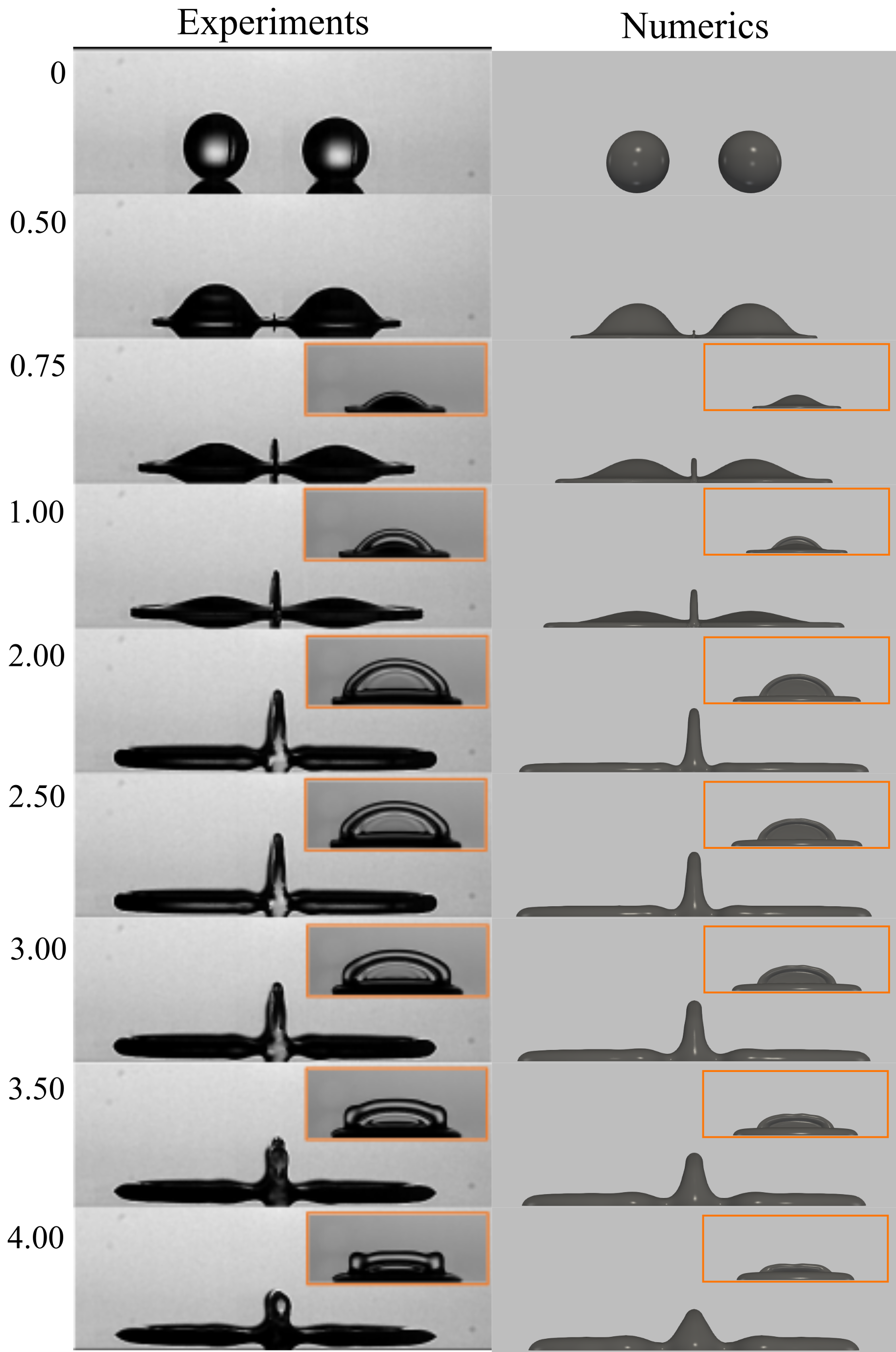}
\end{tabular}
\caption{Comparison of the evolution of the central sheet with non-dimensional time $\tau$ between numerical simulations and experiments for $We=130$ at $\phi=40\%$ solution. Note: the experimental images are adapted from \citet{GoswamiPhDthesis}, CC BY 4.0}
\label{fig:40_We130_f}
\end{figure}

%\subsection{Process}\label{subsec:overview}
A basic overview of the process is shown in Figure \ref{fig:40_We130_f}; the right column shows snapshots from the numerical simulations and the left column shows snapshots taken at the same times from the experiments of \cite{Goswami2023JFM} for comparison, which is found to be broadly favourable between simulations and experiment in the following description. (Detailed comparisons are given below and in the Supplementary Material.) We begin by giving a coherent description of two-drop post-impact interaction. For the parameters shown, and generally in this study, the droplets do not splash after impact, and spread radially along the substrate each forming a lamella bordered by a rim. The droplets' lamellae collide initially at the centrepoint $x=y=z=0$ at $0.4<\tau<0.5$ where $\tau \equiv \hat{t}\hat{U}_0/\hat{D}_0$, and the collision then proceeds symmetrically along the $y$ axis in a "tank-treading" fashion. The collision results in the vertically rising lamella, which has a semilunar shape in the $z-y$ plane during the interaction. The central sheet is restrained by surface tension and reaches a maximum height before beginning to fall back to the bulk. 

\begin{figure}
\centering
\begin{tabular}{cc}     
 \includegraphics[height=0.35\columnwidth]{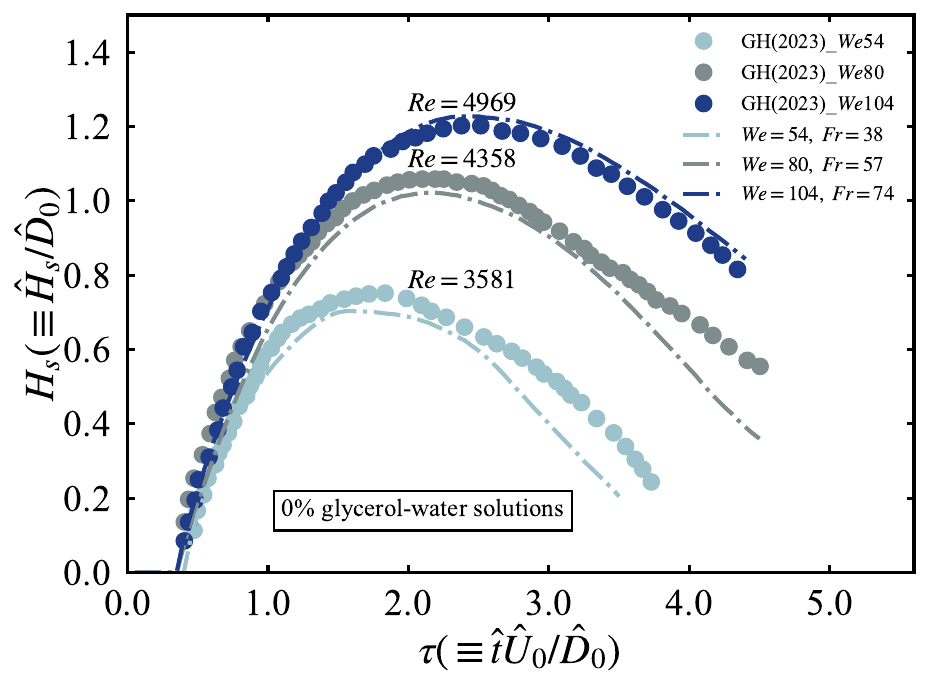} &
 \includegraphics[height=0.35\columnwidth]{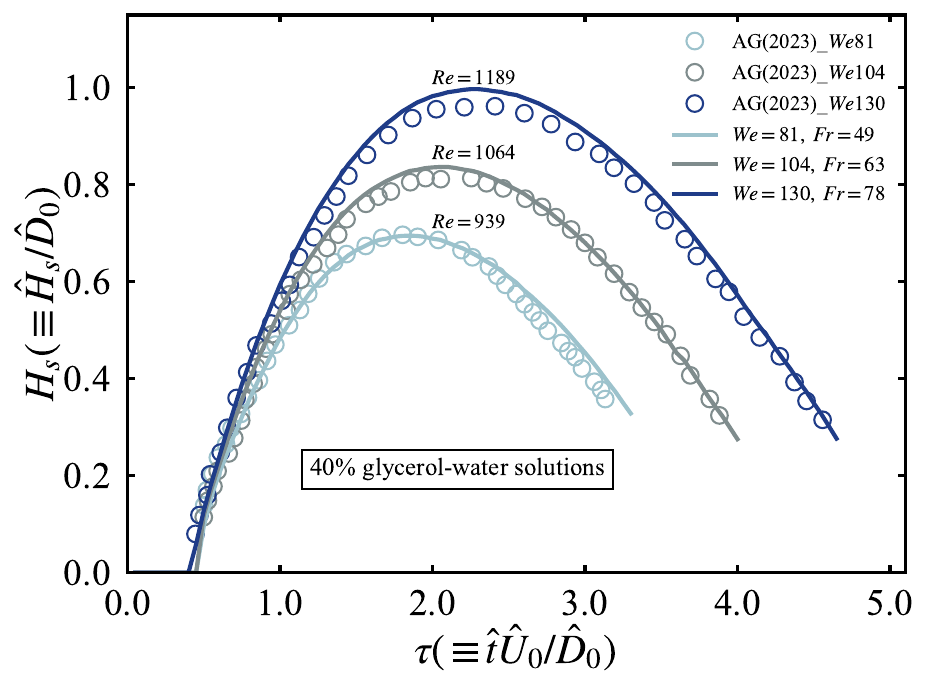} \\
 (a) & (b)
\end{tabular}
\caption{Comparison of the maximum heights between numerical results and experiments: (a) $\phi=0\%$ solutions and (b) $\phi=40\%$ solutions for three experimental cases with different $We$ (the corresponding $Re$ is obtained accordingly). Note: the experimental results are adapted from \citet{Goswami2023JFM}. For clarity, the definitions of variables are omitted from the labels in the following figures}
\label{fig:validation_quantitative}
\end{figure}

Figure \ref{fig:validation_quantitative} shows a quantitative comparison between experimental \citep{Goswami2023JFM} and numerical results in terms of the maximum height of the central sheet, which is the focus of the study. The height of the central sheet as a function of time is $\hat{H}_s(\hat{t})$, which can be non-dimensionalised by $H_s \equiv \hat{H}_s/\hat{D}_0$. The comparison between numerics and experiment is good up to the maximum value $H_{s,max}$ which is attained at $\tau=\tau_{H_s}$. After this time, the results deviate for the pure water case (Figure \ref{fig:validation_quantitative}a). In both experiments and in the present numerical simulations, these cases featured an unstable rim and occasional splashing; the disagreement between results may perhaps be attributed to differences in how the central sheet height is measured. But the agreement is significantly improved for the glycerol-water mixture ($\phi=40\%$), which tend not to splash, at both the rising and falling stages. Therefore, the numerical platform can accurately and precisely capture both the dynamic features of central rising sheet and its geometric height when the corrugation in the rim of the central sheet are not strong. The discrepancies that do exist may be caused by several effects. Firstly, the simulations use a static contact angle, which is an idealisation of the dynamic contact angle in the experiment. Secondly, because the two reference experimental studies \citep{Goswami2023JFM,GoswamiPhDthesis} report slightly different initial drop diameters, there may be a small error incurred in matching $\textrm{Re}$ within the numerical study. And thirdly, in experiment there are inevitable uncertainties in terms of impact time, inter-drop spacing, and drop morphology at impact moment.  Some of these imperfections cause corrugations at the rising or falling stage or lead to the maximum height differences directly at all stages. Nevertheless, in the face of these differences, the numerical simulations are in excellent agreement with the experimental results.

\subsection{Geometric considerations}\label{subsec:geometric}
We now consider the morphology of the pair-drop interaction in detail. Because the single-drop spreading on the substrates is well understood \citep{Yarin2006ARFM,Roisman2009POF,Eggers2010POF,Josserand2016ARFM}, we are here especially interested in the central sheet height $H_s$, which characterises well the pair-drop interaction problem. 

Now certain results from the literature on single-drop impacts are useful to provide a framework for understanding the pair-drop problem. In particular, the spreading behaviour of each drop on the side of the droplet opposite that of the interaction is relatively unchanged from the single-droplet case. Therefore we may consider the rim of the spreading lamella on the substrate to have radius $\hat{R}$ as measured on this far side. The spreading ratio is the (equivalent) dimensionless spreading diameter $\beta=\hat D/\hat D_0$, where $\hat D=2\hat R$. At the maximum extent of spreading, $\hat R=\hat R_{max}$ and $\beta_{max}=\hat D_{max}/\hat D_0$, and the central sheet reaches its maximum height at around the same time as maximum spreading (see below). Therefore $\beta_{max}$ plays an important role in the central sheet in two ways: (1) increased $\beta_{max}$ corresponds with a larger spreading time, which in turn allows for more of the total drop liquid to flow into the central sheet; and (2) it corresponds with a greater total width ($W$ as shown in Fig. \ref{fig:Height_Hs}) of the central rising sheet. 

It is desirable therefore to obtain an estimate of the maximum spreading ratio as a function of $\textrm{We}, \textrm{Re}$ from the literature on single-drop impacts, which has variously considered both energetic and dynamic models \citep{UK2005Lang, Wildeman2016JFM, Laan2014PRA, Gordillo2019JFM, Du2021Lang} - see the Supplementary Information for a comparison. In short, our numerical data are in closest agreement with \citet{Wildeman2016JFM}, whose theory will be used in the following development.

\begin{figure}[H]
\centering
  \includegraphics[width=0.5\columnwidth]{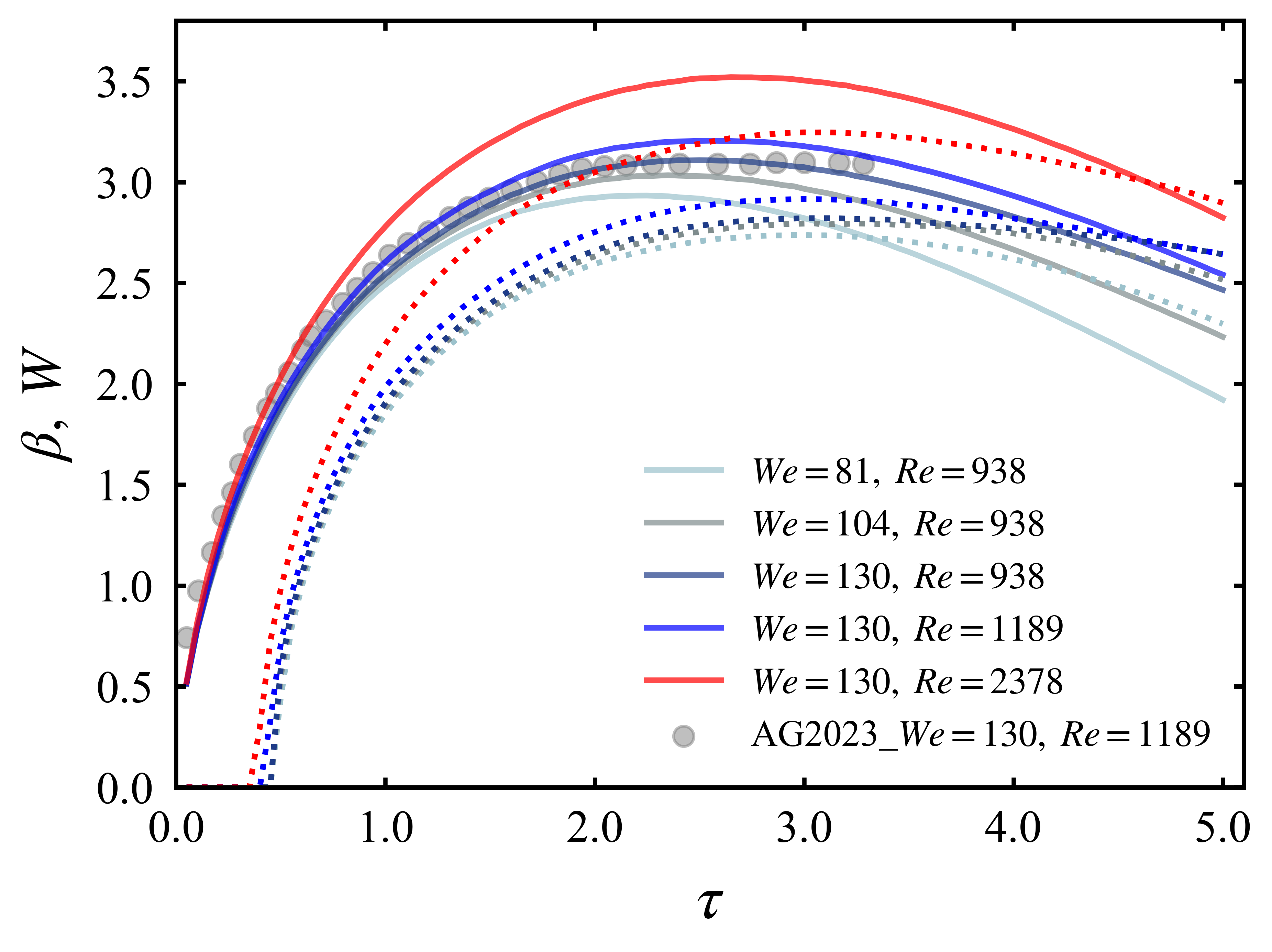}
\caption{Evolution of the maximum spreading ratio $\beta \equiv D/D_0$, shown by solid lines, and the dimensionless width of the central interaction surface $W$, shown by dotted lines of the same color, for various $\mathrm{We}$ and $\mathrm{Re}$. Reference experimental data adapted from \citet{Goswami2023JFM}, are plotted as filled circles and show good agreement with the corresponding numerical results.}
\label{fig:central_width}
\end{figure}

The evolution of $\beta$ is compared with experiment, and as central sheets are also affected by the width $W$, the evolution of the width of central sheet under different $We$ and $Re$ are shown in Fig. \ref{fig:central_width}. For $\beta$ evolution, grey circles are data obtained from experiment \citep{GoswamiPhDthesis}, with which simulation results are again in good agreement. It is shown that $\beta$ increases at first and then decreases after it reaches its maximum because the drops begin to retract. Because the width is confined by $\beta_{max}$ and $\Delta x$ geometrically, the trends of $W$ roughly match those of  $\beta_{max}$. 
% The peak values of both $\beta$ (i.e. $\beta_{max}$) and W are more sensitive to Re than to We, with larger values of Re yielding larger peak values; see Figure \ref{supp-fig:Rmax_comparison}, which shows that $\beta_{max}$ increases distinctly with $Re$ but remains the same for lowest e.g. $Re$ at different $We$.
%\subsubsection{Geometric description of central rising sheet}

\section{Energetics of the central rising sheet}\label{sec:ki_en}
\begin{figure}
\centering  
 \includegraphics[height=0.5\columnwidth]{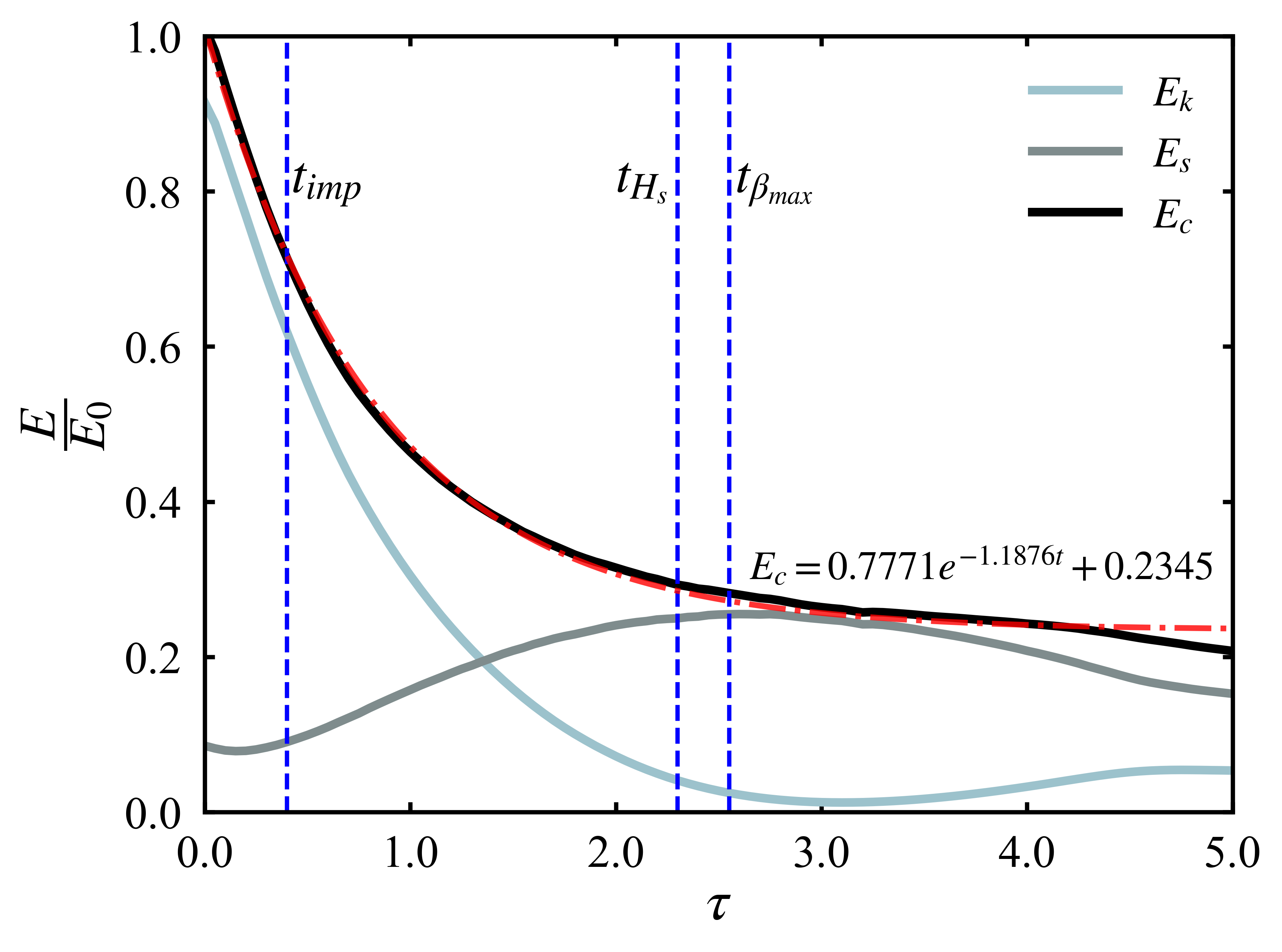}
\caption{Evolution of the energy normalized by the initial total energy during the impact process for $We=130$ and $Re=1189$. Here, $t_{imp}$ denotes the impact time, $t_{H_s}$ the time of maximum sheet height, and $t_{\beta_{max}}$ the time at which the remaining drops on the substrate reach the maximum spreading distance. The conservative energy $E_c = E_s + E_k$ exhibits an exponential decay over time}
\label{fig:Energy_instant}
\end{figure}

% $u_y$ is similar to the velocity profile of single-drop spreading on the substrate (shown in \citet{Riboux2017PRE}). However, no negative velocity regime appears.
%\subsection{Energetics over the impact process}\label{subsec:energy_evolution}
Many studies focus on the energetics of single drop impacts \citep{UK2005Lang, Attane2007POF, Wildeman2016JFM, Du2021Lang}, but these are not fully understood for simultaneous pair-drop impacts. In this subsection, we investigate kinetic energy $\hat E_k$ (KE), surface energy $\hat E_s$ (SE), local viscous dissipation $\hat \varepsilon_d$, and total dissipation $\hat E_d$, which are expressed as \citep{Wildeman2016JFM,Mostert2022JFM,Ray2024JFM}:
\begin{equation}
    \hat E_k = \int_\Omega \dfrac{1}{2}\hat \rho_l \left(\boldsymbol{\hat u}\cdot\boldsymbol{\hat u}\right)\:\mathrm{d}\Omega
\end{equation}
\begin{equation}
    \hat E_s = \int_{\hat A_{LV}} \hat \sigma\:\mathrm{d}\hat A_{LV} - \int_{\hat A_{SL}} \hat \sigma \cos{\theta_{eq}}\:\mathrm{d}\hat A_{SL}
\end{equation}
\begin{align}\label{eqn:ed}
    \hat \varepsilon_d = & 2\hat\mu\left[\left(\frac{\partial \hat u_x}{\partial \hat x}\right)^2+\left(\frac{\partial \hat u_y}{\partial \hat x}\right)^2+\left(\frac{\partial \hat u_z}{\partial \hat x}\right)^2\right] \nonumber \\
    & + \hat \mu\left[\left(\frac{\partial \hat u_x}{\partial \hat y}+\frac{\partial \hat u_y}{\partial \hat x}\right)^2 + \left(\frac{\partial \hat u_y}{\partial \hat z}+\frac{\partial \hat u_z}{\partial \hat y}\right)^2 + \left(\frac{\partial \hat u_z}{\partial \hat x}+\frac{\partial \hat u_x}{\partial \hat z}\right)^2 \right]
\end{align}
\begin{equation}\label{eqn:Ed}
    \hat E_d = \int_0^{\hat t} \left(\int_\Omega \hat \varepsilon_d\:\mathrm{d}\Omega\right)\:\mathrm{d}\hat t
\end{equation}
where $\hat \Omega$ is fluid volume, $\hat A_{LV}$ is liquid-air interface area, and $\hat A_{SL}$ is solid-liquid interface area. We used $\hat \rho_l\hat U_0^2D_0^3$ to obtain dimensionless form of energies. In Fig. \ref{fig:Energy_instant}, we plot the evolution of the energy (where $E_g$ is gravitational energy and negligible, $E_c$ is the conservative energy and is thus defined with $E_c=E_k+E_s$) over time for $We=130,~Re=1189$. It shows that kinetic energy is partly converted to surface tension energy, which reaches a peak at $\tau \simeq 2.5$, but the kinetic energy is otherwise dissipated directly. The highest dissipation rates are achieved early in the entire process, having reached close to zero by the time of the maximum in surface energy.
% the local viscous dissipation rate at $t_{H_s}$ and Fig. \ref{fig:Energy_instant}(a) shows that Region $\mathrm{i}$ contains a region of viscous dissipation, with only a small amount of dissipation at the junction between the central lamella and the rim (Region $\mathrm{iv}$. Indeed most of the viscous dissipation is confined to the spreading lamellae on the substrate. This appears consistent with the observation by \cite{Goswami2023JFM} that the maximum sheet height may be related to only the Weber number of the colliding rims on the substrate, through an inviscid scaling.

%Kinetic energy mainly converts to surface tension energy and is dissipated by the viscous dissipation.

For single drop impacts, the viscous dissipation is always difficult to incorporate into a theoretical model \citep{UK2005Lang,Wildeman2016JFM}, and there are similar challenges in pair-drop impacts. However, we see here that for pair-drop impacts, $E_c$ follows an exponential decay, which for the case $We=130,~Re=1189$ presented in Fig. \ref{fig:Energy_instant}, is found through fitting to be $\Delta E_c = 0.7771e^{-1.1876t}+0.2345$. Thus we estimate the dissipated energy instead by the deviation of the conservative energy from its initial value, yielding $\Delta E_d = -\Delta E_c$. This estimate is preferable to the direct evaluation \eqref{eqn:ed}-\eqref{eqn:Ed} because it is simplest to establish convergence with respect to numerical resolution, for the resolutions presented here - see the Supplementary Information.  We also the important times $t_{imp}$, $t_{H_s}$, and the moment at which the maximum spreading radius is reached on the substrate, $t_{\beta_{max}}$. There is no energy discontinuity at $t_{imp}$, but $E_s$ noticeably increases after $t_{imp}$, and reaches its maximum at around $t_{\beta_{max}}$. $E_k$ reaches its minimum at a short time after $t_{\beta_{max}}$, suggesting that some kinetic energy is left in the drops at $t_{H_s}$. However, the minimum is quite broad and there is little variation in $E_k$ after $t_{H_s}$. Thus for the purposes of constructing a predictive energetic model, we will in the following approximate $t_{H_s}\simeq t_{\beta_max}$ and that $E_k$ reaches its minimum at $t_{H_s}$.

%\section{Energetic model prediction of the central sheet height}\label{sec:energy_model}
\subsection{Cylindrical disk and lollipop model for the central sheet}
\begin{figure}
\centering
\begin{tabular}{c}
  \includegraphics[width=0.5\columnwidth]{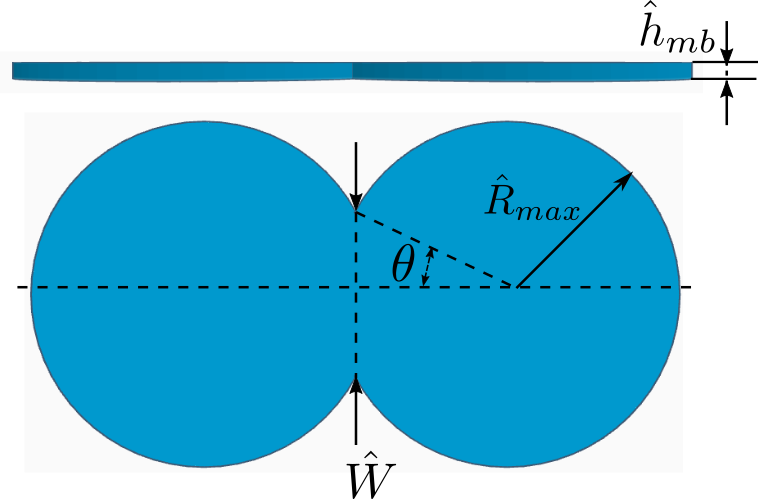}\\
    (a)
\end{tabular}
\begin{tabular}{cc}     
 \includegraphics[width=0.4\columnwidth]{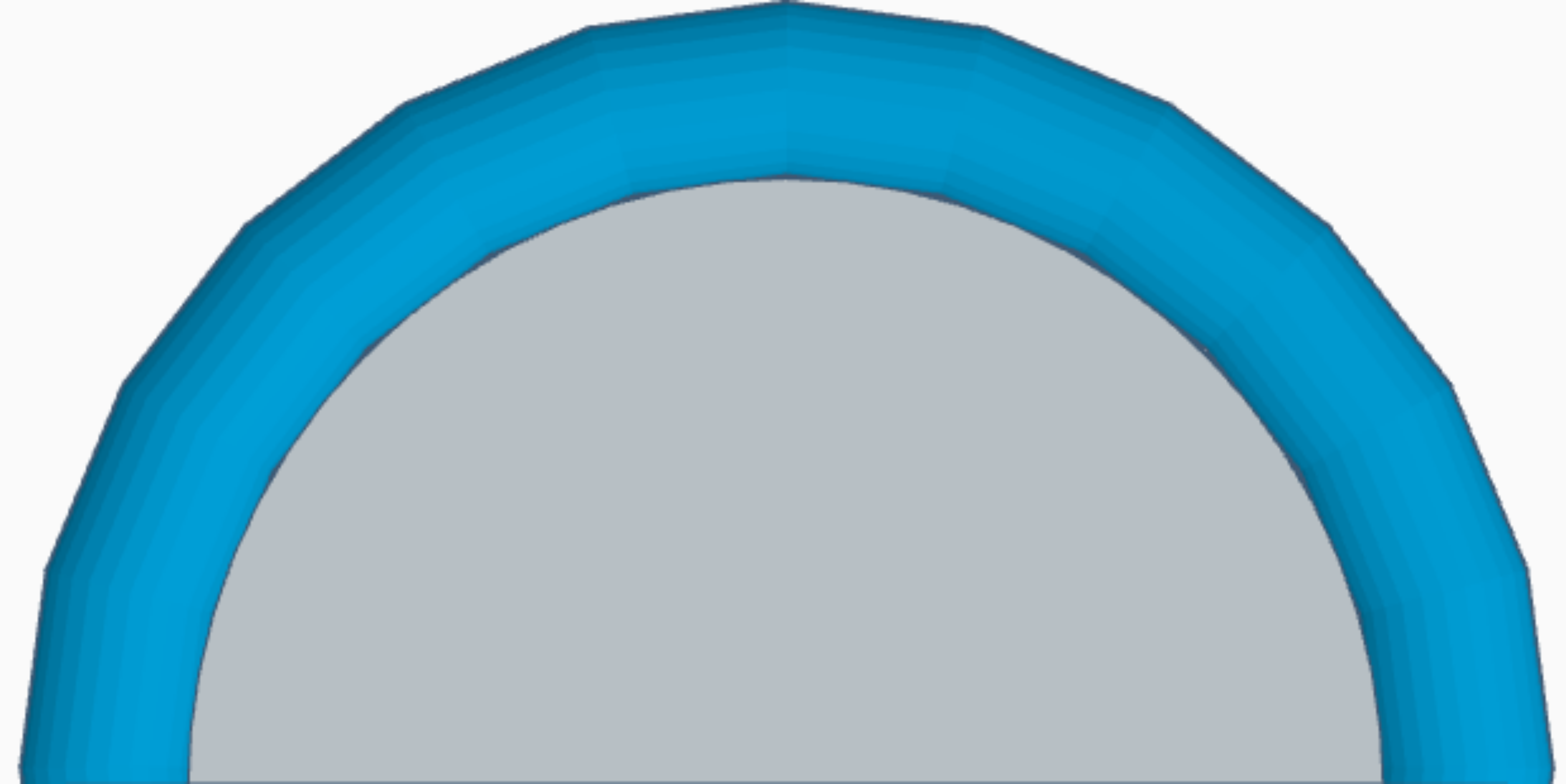} &
 \includegraphics[width=0.25\columnwidth]{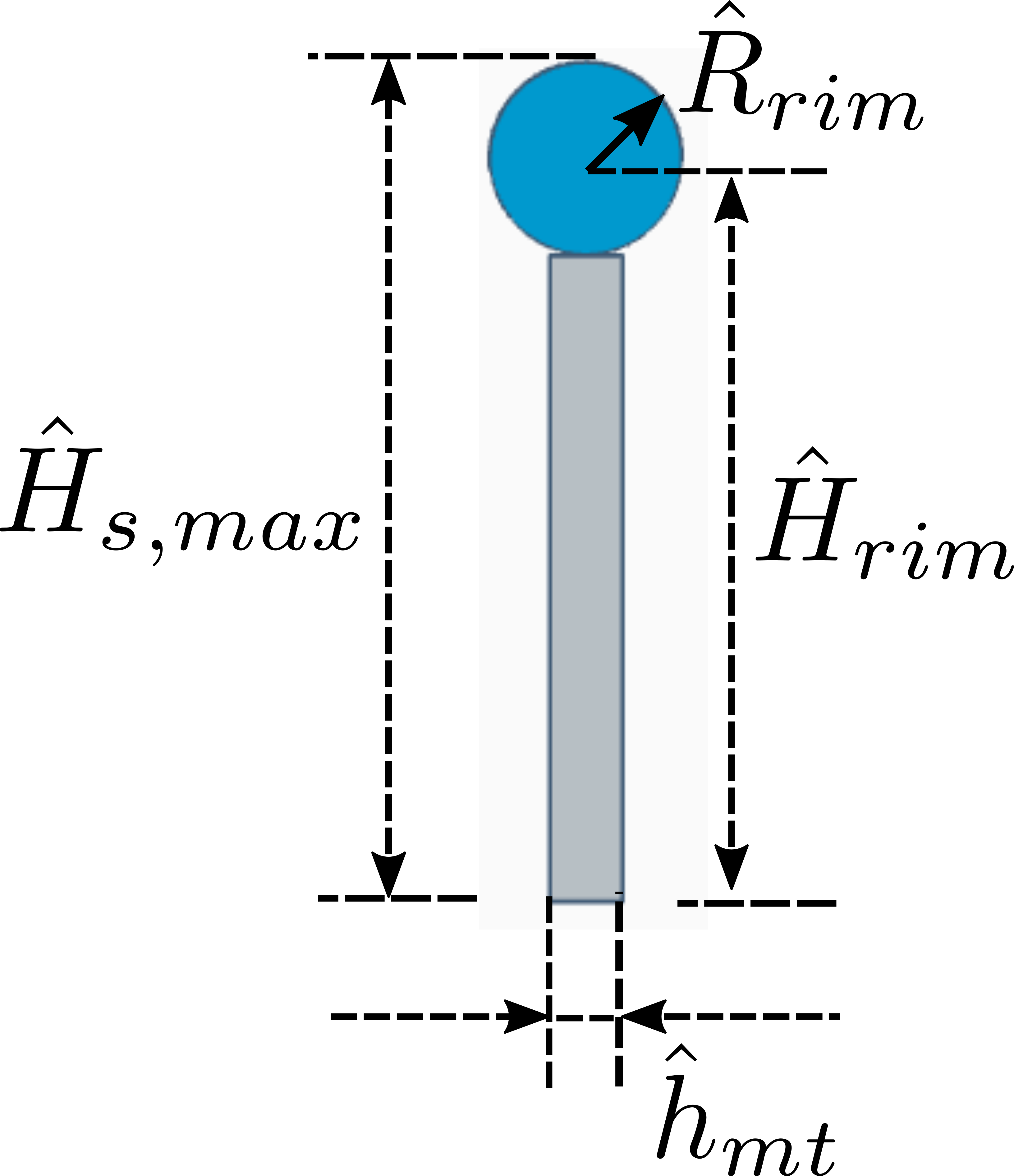} \\
 (b) & (c)
\end{tabular}
\caption{Schematics of the energetic model prediction, illustrating two geometric assumptions: the cylindrical disk and the lollipop models, with the lollipop model shown. For the cylindrical disk assumption, $\hat{R}_{rim}$ is zero}
\label{fig:Lollipop_model}
\end{figure}

In this section, we advance an energetic theory to predict the height of the central sheet at $\hat t_{H_s}=t_{H_s}\hat D_0/\hat U_0$. In the following we will use dimensionless (uncareted) variables. A schematic of the energetic model is shown in Fig. \ref{fig:Lollipop_model}. We model the drops on the substrate with a thin cylindrical disk in Fig. \ref{fig:Lollipop_model}(a), with the maximum spreading radius $\hat R_{max}$, uniform thickness $\hat h_{mb}$, half central angle $\theta$, and width $\hat W$ for the central sheet. While \cite{Goswami2023JFM} model the central sheet as a circular segment, we model the lamella itself as an half-ellipse, bordered by a rim with a circular cross-section. In this way, the transverse cross-section in the $x-z$ plane resembles a lollipop. 

With this geometry, the variables relevant to the central sheet are the maximum height $\hat H_{s,max}$, the elevation of the rim at the lamella periphery $\hat H_{rim}$, the radius of the rim $\hat R_{rim}$, and the thickness of the central lamella $\hat h_{mt}$. When the rim dominates the central sheet dynamics, the lamellar part is thin and thus we assume $\hat h_{mt}\simeq0$; whereas if $\hat R_{rim}\simeq0$, then $\hat H_{s,max}=\hat H_{rim}$, and $\hat h_{mt}\neq0$, the model resembles a elliptical disk. We consider both of these cases in the following.

Based on the observation from Fig. \ref{fig:Energy_instant} that there is little variation among all the energetic contributions between $\hat t_{H_s}$ and $\hat t_{\beta_{max}}$, we assume the central sheet reaches its maximum height at the same time that the drops on the substrate reach their maximum spreading radii, i.e. $\hat t_{H_s} \approx \hat t_{\beta_{max}}$. We also approximate the remaining kinetic energy in the system to be zero at this time, and use the same equilibrium contact angle $\theta_{eq}=90^\circ$ as in the simulations. Given the morphology assumptions in Fig. \ref{fig:Lollipop_model}, the properties of volume conservation, energy conservation, and geometric confinement can be expressed as \citep{UK2005Lang,Wildeman2016JFM}:
\begin{equation}\label{eqn:energy_conservation}
    \hat E_{s0} + \hat E_{k0} = \hat E_s + \hat E_d 
\end{equation}
\begin{align}\label{eqn:volumn_conservation}
    % 2\cdot\frac{4}{3}\pi\left(\frac{D_0}{2}\right)^3 = &\left(\frac{2\pi-2\theta}{2\pi}\pi R_{max}^2  +\frac{1}{2}R_{max}^2\sin 2\theta\right)\cdot h_{mb} \cdot2 \nonumber \\
    % & + \frac{1}{2}\pi \left(H_{rim}-R_{rim}\right)\frac{W}{2}h_{mt} \nonumber \\
    % & + \frac{\pi R_{rim}^2}{2}\bigg[ 3\pi\left(H_{rim}+\frac{W}{2}\right) \nonumber \\
    % &\quad - \pi\sqrt{\left(3H_{rim}+\frac{W}{2}\right)\left(\frac{3W}{2}+H_{rim}\right)}\bigg]
    2\hat V_{drop} = \hat V_{rs} + \hat V_{cs}
\end{align}
\begin{equation}\label{eqn:geometic_conservation}
    \left(\frac{\hat W}{2}\right)^2+\left(\frac{\Delta \hat x}{2}\right)^2=\hat R_{max}^2
\end{equation}
Where $\hat E_{k0}$ and $\hat E_{s0}$ are the initial kinetic energy and surface tension energy respectively, $\hat V_{drop}$ is the initial volume of a (single) drop, $\hat V_{rs}$ is the volume of the drop remaining on the substrate (i.e. excluding the central sheet), $\hat V_{cs}$ is the drop volume of the central sheet, and $\hat H_{s,max}=\hat H_{rim}+\hat R_{rim}$. Therefore, in dimensionless form:
\begin{align}
    2+\frac{1}{6}We = & 2\frac{\pi-\theta}{\pi}\beta_{max}h_{mb} + \beta_{max}^2\left(\frac{\pi-\theta}{2\pi}+\frac{\sin{2\theta}}{4\pi}\right) \nonumber \\
    &+\frac{1}{2}\beta_{max}(H_{rim}-R_{rim})\sin{\theta} \nonumber \\
    & + \pi R_{rim}\bigg[\left(3H_{rim}+\frac{3}{2}\beta_{max}\sin{\theta}\right) \nonumber \\
    &\quad - \sqrt{\left(3H_{rim}+\frac{1}{2}\beta_{max}\sin{\theta}\right)\left(\frac{3}{2}\beta_{max}\sin{\theta}+H_{rim}\right)}\bigg] \nonumber \\
    & + \frac{\hat E_d}{\pi\sigma D_0^2} \label{energycons}
\end{align}
\begin{align}
   \frac{1}{3} = &2h_{mb}\left(\frac{\pi-\theta}{4\pi} \beta_{max}^2  +\frac{1}{8}\beta_{max}^2\sin{2\theta}\right) + \frac{1}{4}\left(H_{rim} - R_{rim}\right)h_{mt}\beta_{max}\sin{\theta} \nonumber \\
    & + \frac{\pi R_{rim}^2}{2}\bigg[ 3\left(H_{rim}+\frac{\beta_{max}\sin{\theta}}{2}\right) \nonumber \\ 
    &\quad -\sqrt{\left(3H_{rim}+\frac{\beta_{max}\sin{\theta}}{2}\right)\left(\frac{3\beta_{max}\sin{\theta}}{2}+H_{rim}\right)}\bigg]\label{masscons}
\end{align}

\begin{equation}\label{eqn:geometric_dimensionless}
    \theta = \arccos{\frac{\Delta x}{\beta_{max}}}
\end{equation}

\begin{equation}\label{eqn:Hmax}
    H_{s,max}=H_{rim}+R_{rim}
\end{equation}
In this system of four nonlinear algebraic equations, there are seven unknown variables: $h_{mb},~\beta_{max}$, $~W,~h_{mt},~H_{s,max},~H_{rim},~R_{rim}$. By using the lollipop or cylindrical disk model this is reduced to six as $h_{mt}\simeq0$ or $R_{rim}\simeq0$ respectively. Next, based on the maximum spreading ratio comparison in Figure \ref{fig:Rmax_comparison} and as reported in \citet{Goswami2023JFM}, the central rim collision has little effect on the maximum spreading radius on the parts of the substrate away from the central sheet. Therefore, to close the system, we use the theory of single drop impacts on the substrate for $h_{mb},~\beta_{max}$. In \citet{Wildeman2016JFM}, $h_{mb}=2/(3\beta_{max}^2)$, and $\beta_{max}$ can be obtained from:
\begin{equation}
    \frac{3(1-\cos{\theta})}{We}\beta_{max}^2 + \frac{0.7}{\sqrt{Re}}\beta_{max}\sqrt{\beta_{max}-1} = \frac{12}{We} + \frac{1}{2} \label{betamax_eq}
\end{equation}
It remains to estimate $\hat{E}_d$ in \eqref{energycons}. We calculate the viscous dissipation in eqn. (\ref{energycons}) using an exponential model by assuming $\hat E_d=(C_0-C_0e^{-Bt})\hat E_0$, and obtain $C_0\approx0.87(\frac{We^{1.2}}{Re})^{0.09}$, $B\approx1.75(\frac{We^{1.2}}{Re})^{0.3}$ by fitting according to the dissipation data obtained in the numerical results, so that the dissipation term in eqn. \eqref{energycons} becomes $(2+We/6)(C_0-C_0e^{-Bt})$. Finally, for $t_{H_s}$ in the dissipation model, we ch the scaling  $t_{H_s}=0.08(We^{1/2}Re^{1/10})+t_{imp}$ by fitting (see  \S\ref{sec:asymptotic} below) and by subtracting the average $\overline{t}_{imp}\approx0.425$ for all cases. By solving the energetic system model in \eqref{energycons}-\eqref{betamax_eq} with the \texttt{SymPy} library using the nonlinear equation solver \texttt{nsolve} in Python, we can then obtain the height of the central sheet for different cases.

\begin{figure}
\centering
  \includegraphics[width=0.7\columnwidth]{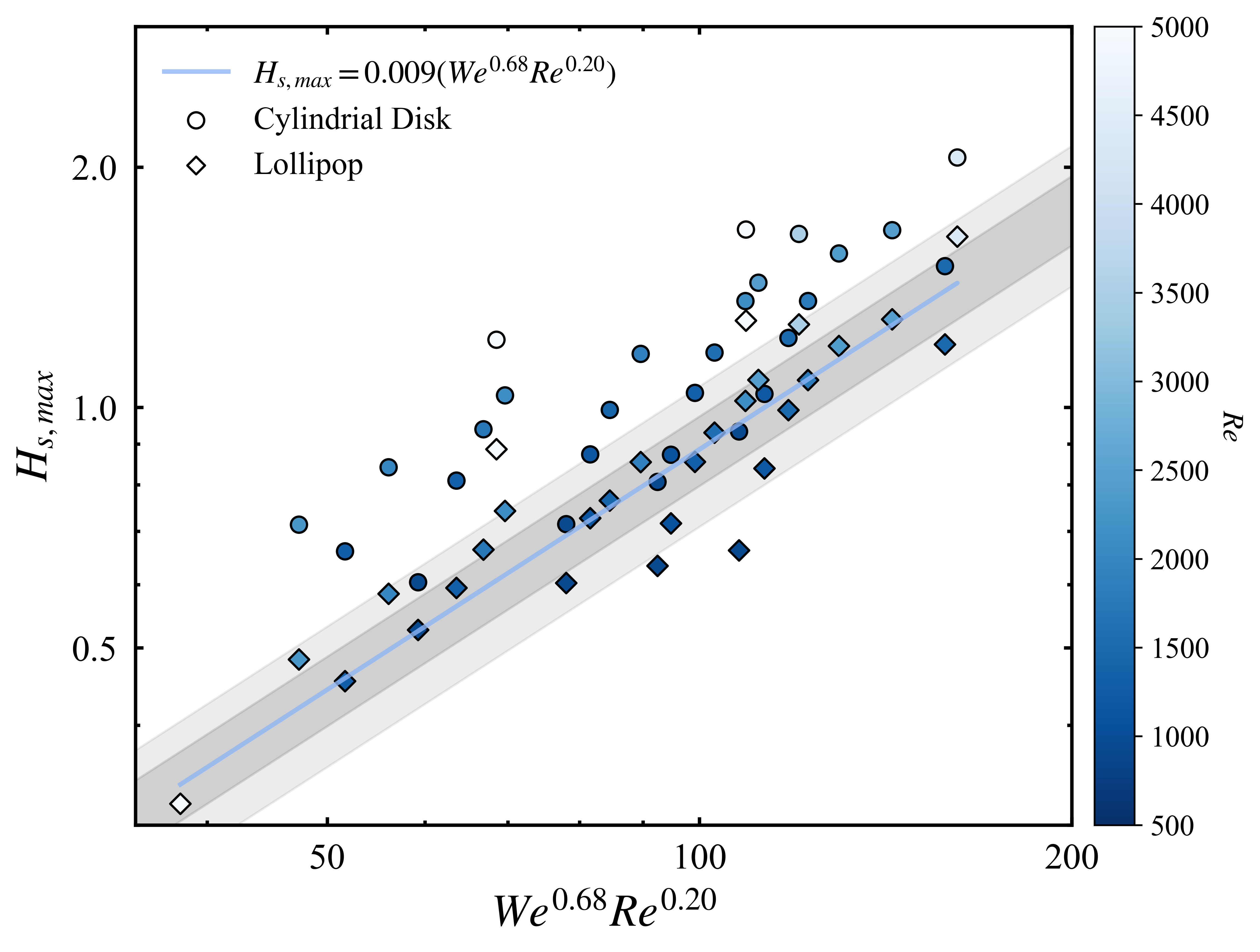}
\caption{Comparison of the maximum height between the numerical results, shown by the fitted solid line, and the energetic model predictions under the two geometric assumptions}
\label{fig:Energy_model}
\end{figure}

We now test this energetic model against the numerical data. Fig. \ref{fig:Energy_model}, shows $H_{s,max}$ plotted against a scaling of $We^{0.68}Re^{0.2}$ (see \S \ref{sec:asymptotic}) on the abscissa, including the numerical results and those from the energetic model in both the lollipop and cylindrical disk expressions. The data are coloured by the value of $Re$. In comparison with the numerical data, the cylindrical disk model overestimates the height for the most cases, but the lollipop model performs well overall within a 10\% error region. For the cylindrical disk model, the model prediction attains 20\% error against the numerical result for $Re \gtrsim 1500$, but for $Re\lesssim 1000$  the model nearly overlaps with the numerical results, showing better prediction than the lollipop model. But the lollipop model performs better overall, especially for $1000 \lesssim Re \lesssim 4000$. For $Re\gtrsim 4000$, both models overestimate the maximum height. Overall, the energetic model, especially in the lollipop expression, shows a good ability to predict the height of the central sheet. Discrepancies in the model may be explained by intrinsic errors in the energy model for single drop impacts, nonsimultaneity of $t_{H_s}, t_{\beta_{max}}$ and the kinetic energy minimum, the actual value of the kinetic energy at $t_{H_s}$, the choice of shape for the central sheet and rim, and the nonuniform thickness of drops on the substrate and of the central sheet.
%%%%%%%%%%%%%%%%%change the figure

\subsection{Asymptotic height prediction at different $\Delta x$}\label{sec:asymptotic}
\begin{figure}
\centering
  \includegraphics[width=0.5\columnwidth]{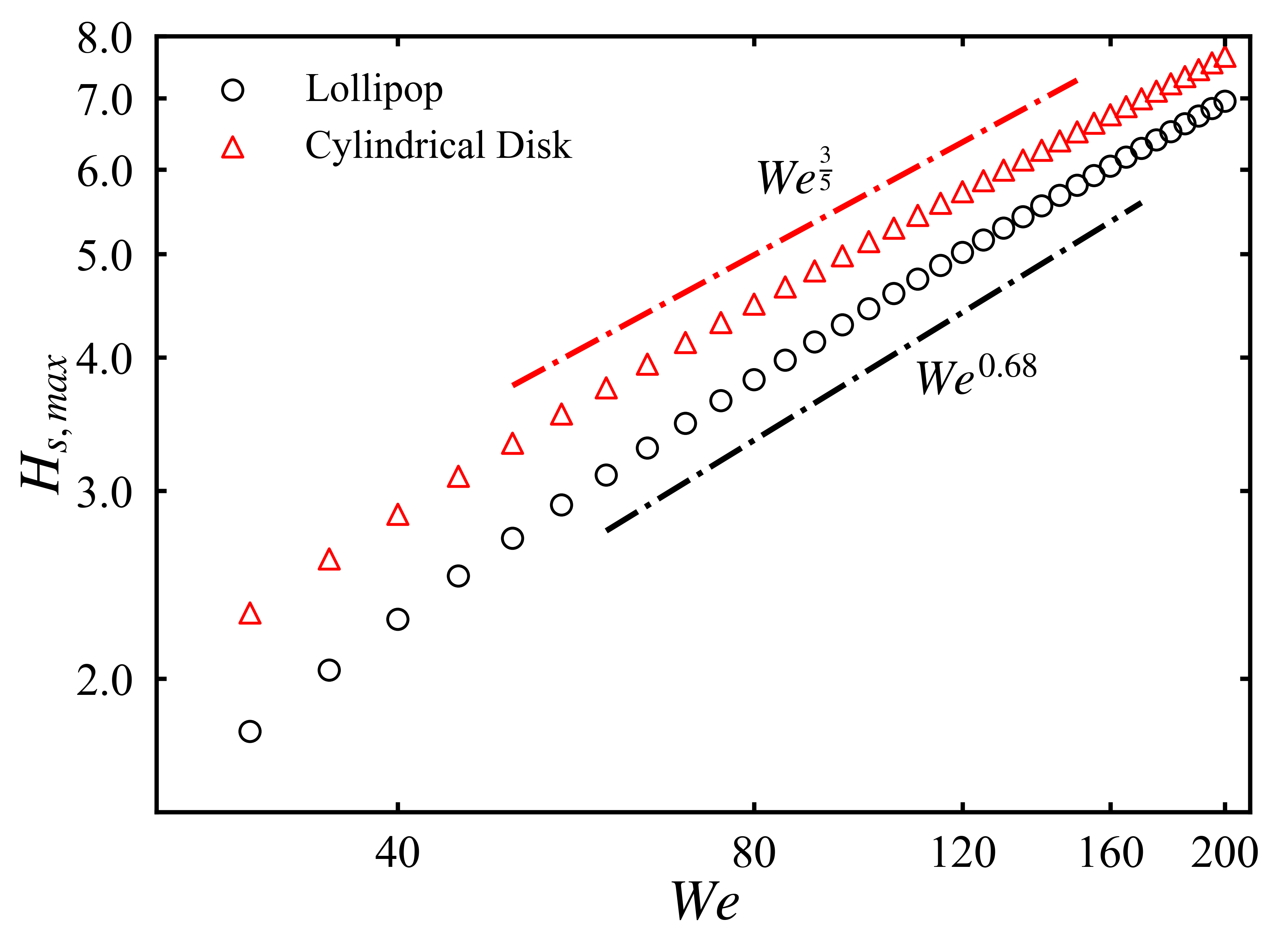}
\caption{Scaling of $H_{s,\max}$ with $We$ as predicted by the energetic model under the lollipop and cylindrical disk geometric assumptions}
\label{fig:scaling_law}
\end{figure}

We now consider asymptotic behaviour at large $Re$ and $We$. First, consider large $We$ in the distinguished limit $Oh \to 0$ where $Oh = \sqrt{We}/Re$; for comparison, the $Oh$ considered in the numerical simulations does not exceed $0.012$. In this limit, surface tension forces dominate, resulting in an asymptotic scaling $\beta_{max}\sim We^{\frac{1}{2}}$ \citep{Eggers2010POF,Wildeman2016JFM} for single drop impacts on the substrate. The leading order result is that, $H_{s,max} \to \sqrt{We}$ (see below), but this also requires $\beta_{max} \gg \Delta x$, which is not plausible for the problem formulation. This may in principle be corrected by expanding \eqref{energycons}-\eqref{betamax_eq} in an asymptotic series in powers of $We^{-1/2}$, but instead we proceed by simply solving the equations numerically in the limit $Re \to \infty$, and neglecting $\hat{E}_d$, for a range of values $We$. For $\Delta x=1.8$, the results for $H_{s,max}$ are plotted in Figure \ref{fig:scaling_law} for the lollipop and cylindrical disk models respectively. In each case, while the results do not follow a true power law, they can be approximated by one, using an exponent of $0.68, 0.6$ respectively.

%Essentially, the scaling law obtained from these two models should be the same, and here we discuss the reasons for the discrepancies in detail. Even though the thickness of the lamella part of the central sheet is thin, the volume of this part does not remain negligible as the central sheet continues to rise, and during which time the rim radius decreases. These subtleties in the development of the lamellar part of the central sheet account for the difference between two models. However, the difference is small considering the effect that no fitting parameters are used in the model aside from the construction of the dissipation model.

\begin{figure}
\centering
\begin{tabular}{c}
  \includegraphics[width=0.5\columnwidth]{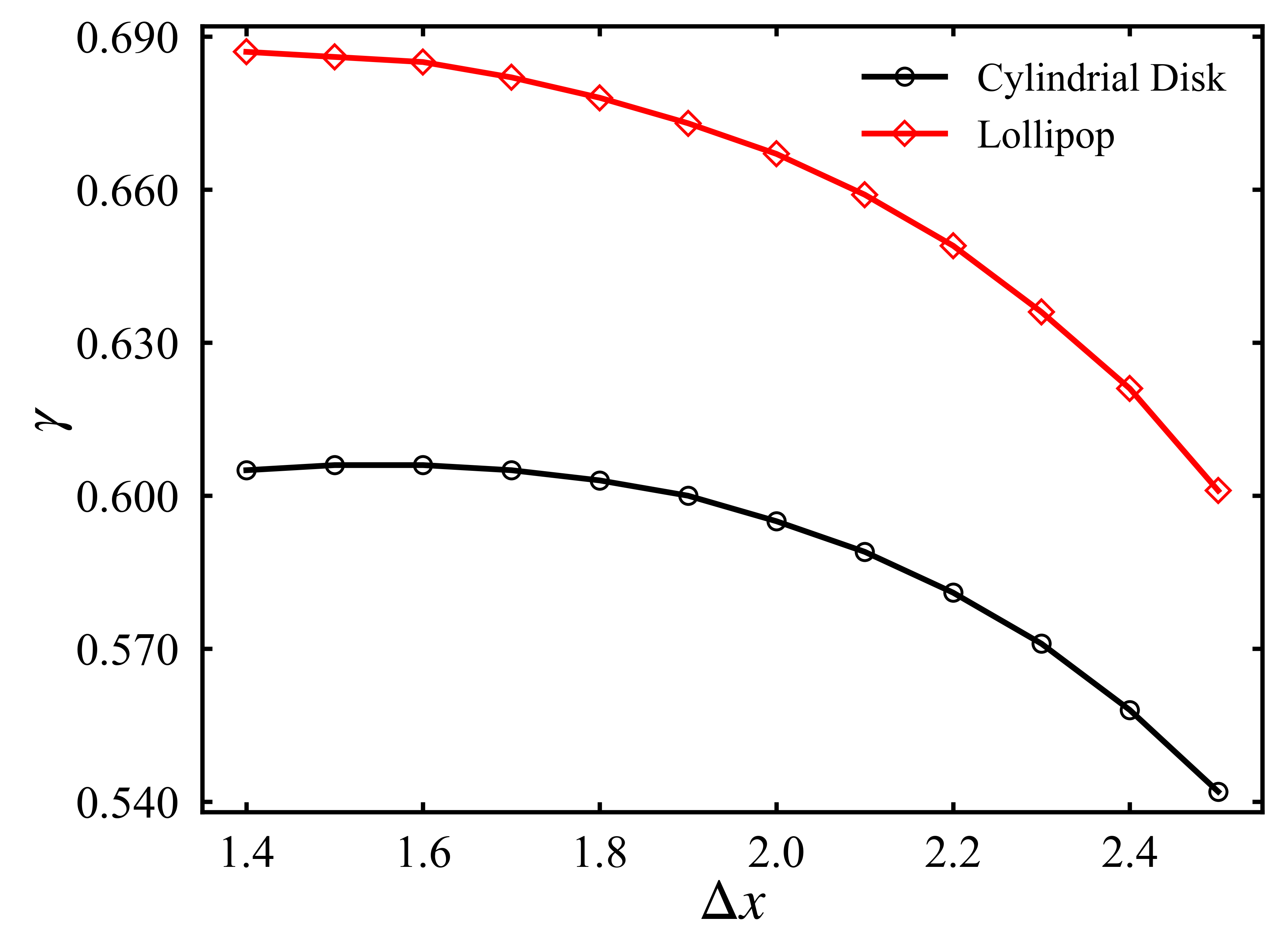}
\end{tabular}
\caption{Scaling of $H_{s,\max}$ with $We^{\gamma}$ at different $\Delta x$ as predicted by the energetic model under the lollipop and cylindrical disk geometric assumptions}
\label{fig:scalinglaw_Deltax}
\end{figure}

The closeness of these values to the fractions $2/3, 3/5$ is tantalizing but fortuitous, given that the fitting parameter depends on the range of $We$ being considered, as well as on the value of $\Delta x$. Figure \ref{fig:scalinglaw_Deltax} shows different approximate scaling laws for $H_{s,max}$ at different $\Delta x$. It shows the exponent of $We$, namely $H_{s,max}\sim We^{\gamma}$, changing with $\Delta x$, diamond line points for the lollipop model, and circle line points for the cylindrical disk model. As \citet{Goswami2023JFM} pointed out, the morphology of the central sheet is flat-topped for $\Delta x=1.3$, so we do not consider smaller inter-drop distance. As $\Delta x$ increases, $\gamma$ decreases for both models because more drop bulk is left on the substrate instead of flowing into the central sheet. In addition, the distance has a weaker effect on $\gamma$ at $\Delta x\lesssim2$. As $\Delta x$ increases more, the distance effect plays a more important role on the height, which is more obvious for the cylindrical disk model. We provide further evidence supporting this scaling of $H_{s,max}$ at different $\Delta x$ in the supplementary section \ref{appC}.

 Next we seek an heuristic way to estimate the scaling in $Re$. The above process is difficult to repeat in the distinguished limit $Oh\to \infty$, namely by taking the limit $We \to \infty$ and then solving \eqref{energycons}-\eqref{betamax_eq} numerically for a range of finite $Re$, because the dissipation term in \eqref{energycons} becomes difficult to handle. In this case, we revert to the leading-order result, namely that the scaling law for $H_{s,max}$ is the same as $\beta_{max}$. In the viscous regime $Oh \gg 1$, we have $H_{s,max} \propto \beta_{max}\sim Re^{\frac{1}{5}}$ \citep{Eggers2010POF,Wildeman2016JFM,Gordillo2019JFM}. The reason for rejecting the leading-order scaling in $We$ in the capillary limit above, but accepting the leading-order scaling for $Re$ in the viscous limit, is that the error incurred by the latter is expected to be somewhat less given that $H_{s,max}$ is less sensitive to $Re$ than $We$.

\begin{figure}
\centering
\begin{tabular}{c}
  \includegraphics[width=0.7\columnwidth]{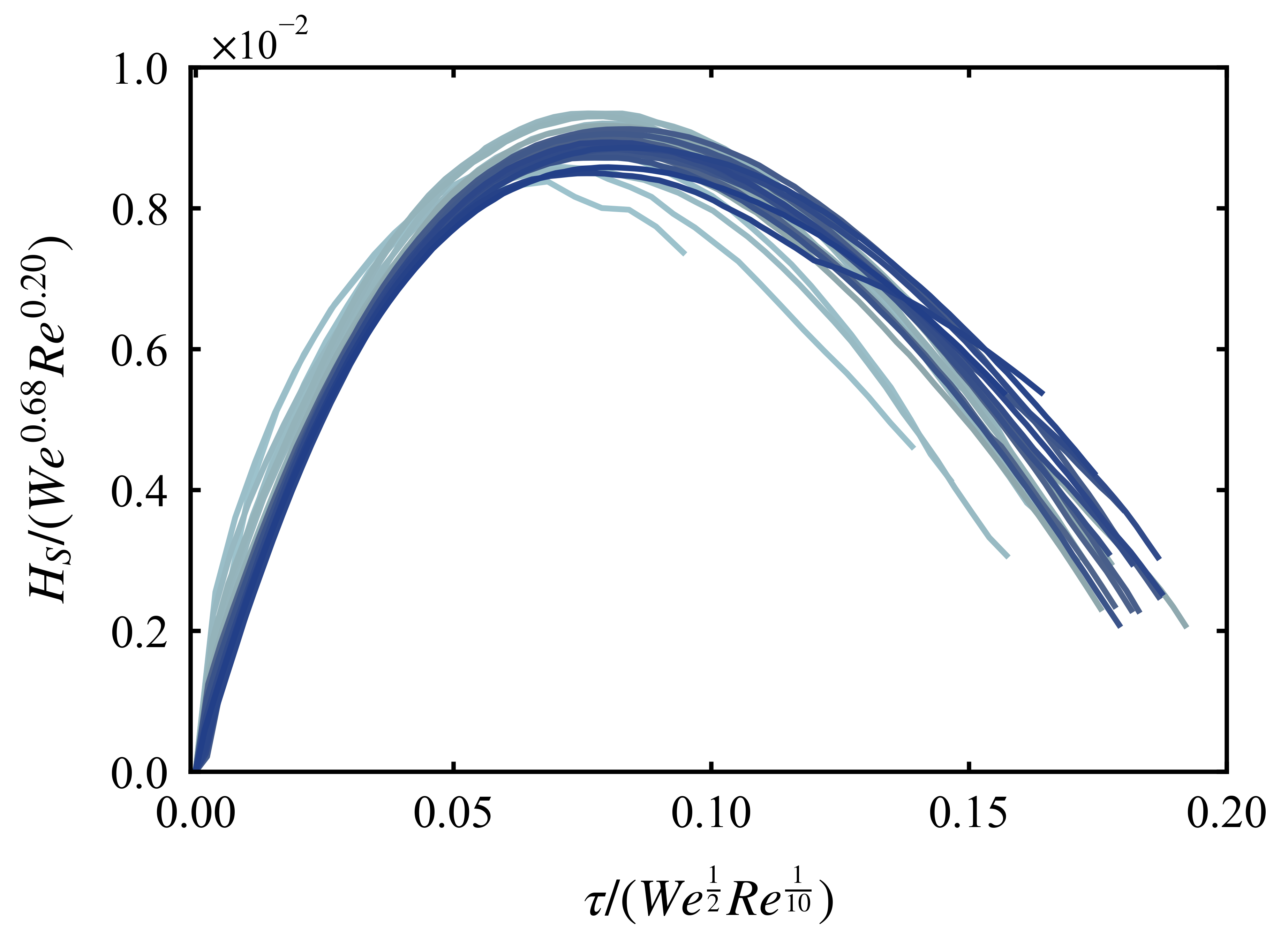}
\end{tabular}
\caption{Collapse of the central rising sheet evolution using the scaling law derived from the energetic model in terms of $We$ and $Re$. The time is collapsed using the capillary time scale ($We^{1/2}$) with a $Re$ correction}
\label{fig:rescaling}
\end{figure}

The resulting approximate exponents on $We$ (of either $0.68$ or $0.6$, for $\Delta x = 1.8$) and $Re$ (of $0.2$ to leading order, irrespective of $\Delta x$) have been obtained in different limits of the parameter $Oh$ and therefore it is not technically appropriate to combine them in a single scaling. We attempt it nevertheless in Fig. \ref{fig:rescaling} where the vertical coordinate is rescaled height ($H_{S}/(We^{0.68}Re^{0.2})$). The horizontal coordinate has been scaled by ($\tau/(We^{\frac{1}{2}}Re^{\frac{1}{10}})$), partly informed by the inviscid asymptotic scaling of $t_{\beta_{max}}\sim\textrm{We}^{1/2}$ \citep{Amirfazli2024PRSA}, although the $\textrm{Re}$-scaling is \emph{ad-hoc}. All the evolution of heights nearly collapse, which suggests that $H_{s,max}$ and $t_{H_s}$ follows these scalings. This is the reason for the abscissa scaling in Fig. \ref{fig:Energy_model}. Note also that, because $\tau_{cap}\equiv\sqrt{\rho_lD_0^3/\sigma}$, the implication is $t_{H_s}\sim Re^{\frac{1}{10}}\tau_{cap}$ .

Finally we remark on the differences in the scaling presented here compared to those of \citet{Goswami2023JFM, GoswamiPhDthesis}. They show a scaling law on the maximum height given impact Weber number $H_{s,max}\sim 0.16We_{L,imp}^{0.53}$, where $We_{L,imp}\equiv2\hat\rho_l \hat T_L\hat V_{L,imp}^2/\hat\sigma$, and $\hat T_L$ and $\hat V_{L,imp}$ are the average thickness and velocity of the lamella edges. As we discussed above in \S\,\ref{subsec:geometric}, the dynamics of central rising sheet consists of its lamella and rim dynamics. The thickness of the central lamella is weakly dependent of $We$ and $Re$, but $We$ dominates the evolution of rims, which means the maximum height is mainly determined by $We_{imp}$ as reported in \citet{Goswami2023JFM, GoswamiPhDthesis}. The advantage of the present scaling is that it depends directly on the $We$, $Re$ corresponding to the initial impact, and it is therefore simpler to characterize. Furthermore, the scaling $H_{s,max}\sim We^{0.68}$ obtained from asymptotic behaviour collapses very well the numerical results, capturing the effect of a near-complete conversion of initial kinetic to surface energy, with dissipation playing a secondary role. The role of $Re$ comes in through boundary effects; the exponents on $Re$ in the above scalings appear to work well in explaining the data, but given that this scaling is not technically compatible with that of $We$, we do not yet have a complete model to explain these, and there may be other choices that also work well. 
%%%%% will they ask us to extract We_imp Re_imp

\section{Conclusion}\label{sec:conclusion}
We present high-resolution three-dimensional simulations on the simultaneous pair-drop impacts by solving two-phase Navier-Stokes equations using the \textit{Basilisk} numerical library. A large range of $We$ and $Re$ parameter space is considered in the paper, $15\leq We \leq200$ and $938\leq We \leq8000$. We mainly investigate the dynamics and energetics of the central sheet, and, in formulating the numerical model, we assumed the dynamics contact angle in the experiment to be approximated by a static contact angle in the numerical simulations. Despite this assumption the simulations were shown to agree well with experiment \citep{Goswami2023JFM,GoswamiPhDthesis}, aside from minor discrepancies. The excellent comparison with experiment highlights the utility of the numerical approach for this problem, and opens up possibilities for many different kinds of multiple-drop impact problems.

%The evolution of the central sheet is mainly controlled by surface tension, and to a lesser extent by viscosity. The velocity profile $u_y$ of the central sheet may be divided into five regimes distributed spatially throughout the sheet, but curiously the velocity profile is not well-described by the self-similar inertial solution $u_y \propto y/t$ observed in many studies \citep{Yarin1995JFM,Roisman2009POF,Eggers2010POF,Wang2017JFM,Gordillo2019JFM,Tang2024JFM}. We avoid the complications of assembling a dynamic model, and instead consider the energetics of the problem. 

A novel energetic model based on simplified "cylindrical disk" and "lollipop" morphologies is proposed, which well predicts $H_{s,max}$ within the parameters in the paper, with marginal error. The viscous dissipation follows an exponential decay that can be easily captured by a fit. The cylindrical disk model overestimates $H_{s,max}$ mostly, especially for $Re\gtrsim 1500$ where the model prediction attains 20\% error against the numerical results, but works well for $Re\lesssim 1000$. The lollipop model performs better overall with most of the results falling within 10\% error against the numerical results, but shows an overestimation for $Re\gtrsim 4000$. Furthermore, numerically estimated asymptotic solutions are considered under both morphological models. We find approximate scalings of $H_{s,max}\propto We^{0.6}$ for the cylindrical disk model and $H_s\propto We^{0.68}$ for the lollipop model, wherein all the initial energy converts to surface energy at $t_{H_s}$ The scaling of $We$ for $H_{s,max}$ is weakly dependent on the drop spacing for $\Delta x \lesssim 2$, but decreases beyond this inter-drop distance. Similarly a scaling proportional to $Re^{0.2}$ is suggested in a different limit of the equations. Following these results, an heuristic scaling law $H_{s,max}\sim We^{0.68}Re^{0.2}$ is proposed in terms of the scaling laws obtained from the energetic model. A complete explanation of the $Re$-scaling is left for future work.
% Here we provide a straightforward scaling law for $H_{s,max}$ by origin $We$ and $Re$, which is different from \citet{GoswamiPhDthesis}.

This work contributes to the mechanisms underlying not only pair-drop impacts but also multiple-drop impacts, and further fragmentation investigation of the central sheet. It also opens up many possibilities for other problems including impacts on the thin films or deep pools, and takes the necessary fundamental steps to consider the effects of viscoelasticity or surfactant effects in such impacts. 

\section*{Acknowledgements}
Z. Zhang acknowledges financial support of the Clarendon Fund Scholarships and Magdalen Graduate Scholarships at the Magdalen College of the University of Oxford. Use of the University of Oxford Advanced Research Computing (ARC) facility is also acknowledged. WM and AACP acknowledge the Natural Environment Research Council grant UKRI1271. Moreover, AAC-P acknowledges funds from the UK Engineering and Physical Sciences Research Council via grant no. EP/W016036/1).

\section*{Declaration of interests}
The authors report no conflict of interest.

\section*{Author ORCIDs} 
Z. Zhang https://orcid.org/0000-0003-3795-7044; A.A. CAstrej\'on-Pita  https://orcid.org/0000-0003-4995-2582; W. Mostert https://orcid.org/0000-0001-6251-4136

%\clearpage
%\newpage

% References using BibTeX
\bibliographystyle{elsarticle-num-names}
\bibliography{main}

\begin{thebibliography}{47}
\expandafter\ifx\csname natexlab\endcsname\relax\def\natexlab#1{#1}\fi
\providecommand{\url}[1]{\texttt{#1}}
\providecommand{\href}[2]{#2}
\providecommand{\path}[1]{#1}
\providecommand{\DOIprefix}{doi:}
\providecommand{\ArXivprefix}{arXiv:}
\providecommand{\URLprefix}{URL: }
\providecommand{\Pubmedprefix}{pmid:}
\providecommand{\doi}[1]{\href{http://dx.doi.org/#1}{\path{#1}}}
\providecommand{\Pubmed}[1]{\href{pmid:#1}{\path{#1}}}
\providecommand{\bibinfo}[2]{#2}
\ifx\xfnm\relax \def\xfnm[#1]{\unskip,\space#1}\fi
%Type = Article
\bibitem[{Joung and Buie(2015)}]{Joung2015NC}
\bibinfo{author}{Y.~S. Joung}, \bibinfo{author}{C.~R. Buie},
\newblock \bibinfo{title}{Aerosol generation by raindrop impact on soil},
\newblock \bibinfo{journal}{Nature communications} \bibinfo{volume}{6}
  (\bibinfo{year}{2015}) \bibinfo{pages}{6083}.
%Type = Article
\bibitem[{Yarin(2006)}]{Yarin2006ARFM}
\bibinfo{author}{A.~L. Yarin},
\newblock \bibinfo{title}{Drop impact dynamics: splashing, spreading, receding,
  bouncing…},
\newblock \bibinfo{journal}{Annu. Rev. Fluid Mech.} \bibinfo{volume}{38}
  (\bibinfo{year}{2006}) \bibinfo{pages}{159--192}.
%Type = Article
\bibitem[{Josserand and Thoroddsen(2016)}]{Josserand2016ARFM}
\bibinfo{author}{C.~Josserand}, \bibinfo{author}{S.~T. Thoroddsen},
\newblock \bibinfo{title}{Drop impact on a solid surface},
\newblock \bibinfo{journal}{Annual review of fluid mechanics}
  \bibinfo{volume}{48} (\bibinfo{year}{2016}) \bibinfo{pages}{365--391}.
%Type = Article
\bibitem[{Li et~al.(2008)Li, Zhou, Chen, Xu, Hui, and Zhang}]{Li2008IJMS}
\bibinfo{author}{N.~Li}, \bibinfo{author}{Q.~Zhou}, \bibinfo{author}{X.~Chen},
  \bibinfo{author}{T.~Xu}, \bibinfo{author}{S.~Hui},
  \bibinfo{author}{D.~Zhang},
\newblock \bibinfo{title}{Liquid drop impact on solid surface with application
  to water drop erosion on turbine blades, part i: Nonlinear wave model and
  solution of one-dimensional impact},
\newblock \bibinfo{journal}{International Journal of Mechanical Sciences}
  \bibinfo{volume}{50} (\bibinfo{year}{2008}) \bibinfo{pages}{1526--1542}.
%Type = Article
\bibitem[{Hulse-Smith et~al.(2005)Hulse-Smith, Mehdizadeh, and
  Chandra}]{Hulse-Smith2005JFS}
\bibinfo{author}{L.~Hulse-Smith}, \bibinfo{author}{N.~Z. Mehdizadeh},
  \bibinfo{author}{S.~Chandra},
\newblock \bibinfo{title}{Deducing drop size and impact velocity from circular
  bloodstains},
\newblock \bibinfo{journal}{Journal of Forensic Sciences} \bibinfo{volume}{50}
  (\bibinfo{year}{2005}) \bibinfo{pages}{JFS2003224--10}.
%Type = Article
\bibitem[{Fest-Santini et~al.(2021)Fest-Santini, Steigerwald, Santini, Cossali,
  and Weigand}]{Fest-Santini2021CF}
\bibinfo{author}{S.~Fest-Santini}, \bibinfo{author}{J.~Steigerwald},
  \bibinfo{author}{M.~Santini}, \bibinfo{author}{G.~E. Cossali},
  \bibinfo{author}{B.~Weigand},
\newblock \bibinfo{title}{Multiple drops impact onto a liquid film: Direct
  numerical simulation and experimental validation},
\newblock \bibinfo{journal}{Computers \& Fluids} \bibinfo{volume}{214}
  (\bibinfo{year}{2021}) \bibinfo{pages}{104761}.
%Type = Article
\bibitem[{Guggilla et~al.(2020)Guggilla, Narayanaswamy, and
  Pattamatta}]{Guggulla2020ETFS}
\bibinfo{author}{G.~Guggilla}, \bibinfo{author}{R.~Narayanaswamy},
  \bibinfo{author}{A.~Pattamatta},
\newblock \bibinfo{title}{An experimental investigation into the spread and
  heat transfer dynamics of a train of two concentric impinging droplets over a
  heated surface},
\newblock \bibinfo{journal}{Experimental Thermal and Fluid Science}
  \bibinfo{volume}{110} (\bibinfo{year}{2020}) \bibinfo{pages}{109916}.
%Type = Article
\bibitem[{Chen et~al.(2020)Chen, Chen, Yan, Li, and Lin}]{Chen2020POF}
\bibinfo{author}{C.-K. Chen}, \bibinfo{author}{S.-Q. Chen},
  \bibinfo{author}{W.-M. Yan}, \bibinfo{author}{W.-K. Li},
  \bibinfo{author}{T.-H. Lin},
\newblock \bibinfo{title}{Experimental study on two water drops successively
  impinging on a solid surface},
\newblock \bibinfo{journal}{AIP Advances} \bibinfo{volume}{10}
  (\bibinfo{year}{2020}).
%Type = Article
\bibitem[{Wibowo et~al.(2021)Wibowo, Widyatama, Kamal et~al.}]{Wibowo2021IJTS}
\bibinfo{author}{T.~Wibowo}, \bibinfo{author}{A.~Widyatama},
  \bibinfo{author}{S.~Kamal}, et~al.,
\newblock \bibinfo{title}{The effect of ethylene glycol concentration on the
  interfacial dynamics of the successive droplets impacting onto a horizontal
  hot solid surface},
\newblock \bibinfo{journal}{International Journal of Thermal Sciences}
  \bibinfo{volume}{159} (\bibinfo{year}{2021}) \bibinfo{pages}{106594}.
%Type = Article
\bibitem[{Luo et~al.(2021)Luo, Wu, Xiao, and Chen}]{Luo2021IJHMT}
\bibinfo{author}{J.~Luo}, \bibinfo{author}{S.-Y. Wu},
  \bibinfo{author}{L.~Xiao}, \bibinfo{author}{Z.-L. Chen},
\newblock \bibinfo{title}{Hydrodynamics and heat transfer of multiple droplets
  successively impacting on cylindrical surface},
\newblock \bibinfo{journal}{International Journal of Heat and Mass Transfer}
  \bibinfo{volume}{180} (\bibinfo{year}{2021}) \bibinfo{pages}{121749}.
%Type = Article
\bibitem[{Aarts et~al.(2005)Aarts, Lekkerkerker, Guo, Wegdam, and
  Bonn}]{Aarts2005PRL}
\bibinfo{author}{D.~G. Aarts}, \bibinfo{author}{H.~N. Lekkerkerker},
  \bibinfo{author}{H.~Guo}, \bibinfo{author}{G.~H. Wegdam},
  \bibinfo{author}{D.~Bonn},
\newblock \bibinfo{title}{Hydrodynamics of droplet coalescence},
\newblock \bibinfo{journal}{Physical review letters} \bibinfo{volume}{95}
  (\bibinfo{year}{2005}) \bibinfo{pages}{164503}.
%Type = Article
\bibitem[{Castrej{\'o}n-Pita et~al.(2013)Castrej{\'o}n-Pita, Kubiak,
  Castrej{\'o}n-Pita, Wilson, and Hutchings}]{Castrejon-Pita2013PRE}
\bibinfo{author}{J.~Castrej{\'o}n-Pita}, \bibinfo{author}{K.~Kubiak},
  \bibinfo{author}{A.~Castrej{\'o}n-Pita}, \bibinfo{author}{M.~Wilson},
  \bibinfo{author}{I.~Hutchings},
\newblock \bibinfo{title}{Mixing and internal dynamics of droplets impacting
  and coalescing on a solid surface},
\newblock \bibinfo{journal}{Physical Review E—Statistical, Nonlinear, and
  Soft Matter Physics} \bibinfo{volume}{88} (\bibinfo{year}{2013})
  \bibinfo{pages}{023023}.
%Type = Inproceedings
\bibitem[{Barnes et~al.(1999)Barnes, Hardalupas, Taylor, Wilkins
  et~al.}]{Barnes1999ICLASS}
\bibinfo{author}{H.~Barnes}, \bibinfo{author}{Y.~Hardalupas},
  \bibinfo{author}{A.~Taylor}, \bibinfo{author}{J.~Wilkins}, et~al.,
\newblock \bibinfo{title}{An investigation of the interaction between two
  adjacent impinging droplets},
\newblock in: \bibinfo{booktitle}{Proceedings of the 15th International
  Conference on Liquid Atomisation and Spray Systems (ILASS), Toulouse},
  \bibinfo{year}{1999}, pp. \bibinfo{pages}{1--7}.
%Type = Article
\bibitem[{Roisman et~al.(2002)Roisman, Prunet-Foch, Tropea, and
  Vignes-Adler}]{Roisman2002JCIS}
\bibinfo{author}{I.~V. Roisman}, \bibinfo{author}{B.~Prunet-Foch},
  \bibinfo{author}{C.~Tropea}, \bibinfo{author}{M.~Vignes-Adler},
\newblock \bibinfo{title}{Multiple drop impact onto a dry solid substrate},
\newblock \bibinfo{journal}{Journal of colloid and interface science}
  \bibinfo{volume}{256} (\bibinfo{year}{2002}) \bibinfo{pages}{396--410}.
%Type = Article
\bibitem[{Liang et~al.(2020)Liang, Yu, Chen, and Shen}]{Liang2020ActaM}
\bibinfo{author}{G.~Liang}, \bibinfo{author}{H.~Yu}, \bibinfo{author}{L.~Chen},
  \bibinfo{author}{S.~Shen},
\newblock \bibinfo{title}{Interfacial phenomena in impact of droplet array on
  solid wall},
\newblock \bibinfo{journal}{Acta Mechanica} \bibinfo{volume}{231}
  (\bibinfo{year}{2020}) \bibinfo{pages}{305--319}.
%Type = Article
\bibitem[{Gultekin et~al.(2021)Gultekin, Erkan, Ozdemir, Colak, and
  Suzuki}]{Gultekin2021ETFS}
\bibinfo{author}{A.~Gultekin}, \bibinfo{author}{N.~Erkan},
  \bibinfo{author}{E.~Ozdemir}, \bibinfo{author}{U.~Colak},
  \bibinfo{author}{S.~Suzuki},
\newblock \bibinfo{title}{Simultaneous multiple droplet impact and their
  interactions on a heated surface},
\newblock \bibinfo{journal}{Experimental Thermal and Fluid Science}
  \bibinfo{volume}{120} (\bibinfo{year}{2021}) \bibinfo{pages}{110255}.
%Type = Article
\bibitem[{Ersoy and Eslamian(2020)}]{Ersoy2020POF}
\bibinfo{author}{N.~E. Ersoy}, \bibinfo{author}{M.~Eslamian},
\newblock \bibinfo{title}{Central uprising sheet in simultaneous and
  near-simultaneous impact of two high kinetic energy droplets onto dry surface
  and thin liquid film},
\newblock \bibinfo{journal}{Physics of Fluids} \bibinfo{volume}{32}
  (\bibinfo{year}{2020}).
%Type = Article
\bibitem[{Goswami and Hardalupas(2023)}]{Goswami2023JFM}
\bibinfo{author}{A.~Goswami}, \bibinfo{author}{Y.~Hardalupas},
\newblock \bibinfo{title}{Simultaneous impact of droplet pairs on solid
  surfaces},
\newblock \bibinfo{journal}{Journal of Fluid Mechanics} \bibinfo{volume}{961}
  (\bibinfo{year}{2023}) \bibinfo{pages}{A17}.
%Type = Article
\bibitem[{Raman(2018)}]{Raman2017JCIS}
\bibinfo{author}{K.~A. Raman},
\newblock \bibinfo{title}{Dynamics of simultaneously impinging drops on a dry
  surface: Role of inhomogeneous wettability and impact shape},
\newblock \bibinfo{journal}{Journal of colloid and interface science}
  \bibinfo{volume}{516} (\bibinfo{year}{2018}) \bibinfo{pages}{232--247}.
%Type = Article
\bibitem[{Sohag et~al.(2023)Sohag, Zhang, and Yang}]{Sohag2023Splashing}
\bibinfo{author}{M.~Sohag}, \bibinfo{author}{W.~Zhang},
  \bibinfo{author}{X.~Yang},
\newblock \bibinfo{title}{Three-dimensional numerical study of two drops
  impacting on a heated solid surface by smoothed particle hydrodynamics},
\newblock \bibinfo{journal}{Physics of Fluids} \bibinfo{volume}{35}
  (\bibinfo{year}{2023}).
%Type = Article
\bibitem[{Sanjay et~al.(2023{\natexlab{a}})Sanjay, Lakshman, Chantelot,
  Snoeijer, and Lohse}]{Sanjay2023JFM1}
\bibinfo{author}{V.~Sanjay}, \bibinfo{author}{S.~Lakshman},
  \bibinfo{author}{P.~Chantelot}, \bibinfo{author}{J.~H. Snoeijer},
  \bibinfo{author}{D.~Lohse},
\newblock \bibinfo{title}{Drop impact on viscous liquid films},
\newblock \bibinfo{journal}{Journal of Fluid Mechanics} \bibinfo{volume}{958}
  (\bibinfo{year}{2023}{\natexlab{a}}) \bibinfo{pages}{A25}.
%Type = Article
\bibitem[{Sanjay et~al.(2023{\natexlab{b}})Sanjay, Chantelot, and
  Lohse}]{Sanjay2023JFM2}
\bibinfo{author}{V.~Sanjay}, \bibinfo{author}{P.~Chantelot},
  \bibinfo{author}{D.~Lohse},
\newblock \bibinfo{title}{When does an impacting drop stop bouncing?},
\newblock \bibinfo{journal}{Journal of Fluid Mechanics} \bibinfo{volume}{958}
  (\bibinfo{year}{2023}{\natexlab{b}}) \bibinfo{pages}{A26}.
%Type = Article
\bibitem[{Riviere et~al.(2021)Riviere, Mostert, Perrard, and
  Deike}]{Riviere2021JFM}
\bibinfo{author}{A.~Riviere}, \bibinfo{author}{W.~Mostert},
  \bibinfo{author}{S.~Perrard}, \bibinfo{author}{L.~Deike},
\newblock \bibinfo{title}{Sub-hinze scale bubble production in turbulent bubble
  break-up},
\newblock \bibinfo{journal}{Journal of Fluid Mechanics} \bibinfo{volume}{917}
  (\bibinfo{year}{2021}) \bibinfo{pages}{A40}.
%Type = Article
\bibitem[{Mostert et~al.(2022)Mostert, Popinet, and Deike}]{Mostert2022JFM}
\bibinfo{author}{W.~Mostert}, \bibinfo{author}{S.~Popinet},
  \bibinfo{author}{L.~Deike},
\newblock \bibinfo{title}{High-resolution direct simulation of deep water
  breaking waves: transition to turbulence, bubbles and droplets production},
\newblock \bibinfo{journal}{Journal of Fluid Mechanics} \bibinfo{volume}{942}
  (\bibinfo{year}{2022}) \bibinfo{pages}{A27}.
%Type = Article
\bibitem[{Tang et~al.(2023)Tang, Adcock, and Mostert}]{Tang2023JFM}
\bibinfo{author}{K.~Tang}, \bibinfo{author}{T.~Adcock},
  \bibinfo{author}{W.~Mostert},
\newblock \bibinfo{title}{Bag film breakup of droplets in uniform airflows},
\newblock \bibinfo{journal}{Journal of Fluid Mechanics} \bibinfo{volume}{970}
  (\bibinfo{year}{2023}) \bibinfo{pages}{A9}.
%Type = Article
\bibitem[{Wang et~al.(2023)Wang, Liu, Bayeul-Lain{\'e}, Murphy, Katz, and
  Coutier-Delgosha}]{Wang2023JFM}
\bibinfo{author}{H.~Wang}, \bibinfo{author}{S.~Liu}, \bibinfo{author}{A.-C.
  Bayeul-Lain{\'e}}, \bibinfo{author}{D.~Murphy}, \bibinfo{author}{J.~Katz},
  \bibinfo{author}{O.~Coutier-Delgosha},
\newblock \bibinfo{title}{Analysis of high-speed drop impact onto deep liquid
  pool},
\newblock \bibinfo{journal}{Journal of Fluid Mechanics} \bibinfo{volume}{972}
  (\bibinfo{year}{2023}) \bibinfo{pages}{A31}.
%Type = Article
\bibitem[{Popinet(2018)}]{Popinet2018ARFM}
\bibinfo{author}{S.~Popinet},
\newblock \bibinfo{title}{Numerical models of surface tension},
\newblock \bibinfo{journal}{Annual Review of Fluid Mechanics}
  \bibinfo{volume}{50} (\bibinfo{year}{2018}) \bibinfo{pages}{49--75}.
%Type = Phdthesis
\bibitem[{Goswami(2023)}]{GoswamiPhDthesis}
\bibinfo{author}{A.~Goswami}, \bibinfo{title}{Impact of Multiple Droplets on
  Solid Surfaces}, \bibinfo{type}{Phd thesis}, Imperial College London,
  \bibinfo{year}{2023}.
%Type = Article
\bibitem[{Popinet(2020)}]{Popinet2020}
\bibinfo{author}{S.~Popinet},
\newblock \bibinfo{title}{Basilisk flow solver and pde library},
\newblock \bibinfo{journal}{available at available at http://basilisk. fr}
  (\bibinfo{year}{2020}).
%Type = Article
\bibitem[{Popinet(2009)}]{Popinet2009JCP}
\bibinfo{author}{S.~Popinet},
\newblock \bibinfo{title}{An accurate adaptive solver for
  surface-tension-driven interfacial flows},
\newblock \bibinfo{journal}{Journal of Computational Physics}
  \bibinfo{volume}{228} (\bibinfo{year}{2009}) \bibinfo{pages}{5838--5866}.
%Type = Article
\bibitem[{Bell et~al.(1989)Bell, Colella, and Glaz}]{Bell1989JCP}
\bibinfo{author}{J.~B. Bell}, \bibinfo{author}{P.~Colella},
  \bibinfo{author}{H.~M. Glaz},
\newblock \bibinfo{title}{A second-order projection method for the
  incompressible navier-stokes equations},
\newblock \bibinfo{journal}{Journal of computational physics}
  \bibinfo{volume}{85} (\bibinfo{year}{1989}) \bibinfo{pages}{257--283}.
%Type = Article
\bibitem[{Brackbill et~al.(1992)Brackbill, Kothe, and
  Zemach}]{Brackbill1992JCP}
\bibinfo{author}{J.~U. Brackbill}, \bibinfo{author}{D.~B. Kothe},
  \bibinfo{author}{C.~Zemach},
\newblock \bibinfo{title}{A continuum method for modeling surface tension},
\newblock \bibinfo{journal}{Journal of computational physics}
  \bibinfo{volume}{100} (\bibinfo{year}{1992}) \bibinfo{pages}{335--354}.
%Type = Article
\bibitem[{Roisman(2009)}]{Roisman2009POF}
\bibinfo{author}{I.~V. Roisman},
\newblock \bibinfo{title}{Inertia dominated drop collisions. ii. an analytical
  solution of the navier--stokes equations for a spreading viscous film},
\newblock \bibinfo{journal}{Physics of Fluids} \bibinfo{volume}{21}
  (\bibinfo{year}{2009}).
%Type = Article
\bibitem[{Eggers et~al.(2010)Eggers, Fontelos, Josserand, and
  Zaleski}]{Eggers2010POF}
\bibinfo{author}{J.~Eggers}, \bibinfo{author}{M.~A. Fontelos},
  \bibinfo{author}{C.~Josserand}, \bibinfo{author}{S.~Zaleski},
\newblock \bibinfo{title}{Drop dynamics after impact on a solid wall: theory
  and simulations},
\newblock \bibinfo{journal}{Physics of fluids} \bibinfo{volume}{22}
  (\bibinfo{year}{2010}).
%Type = Article
\bibitem[{Ukiwe and Kwok(2005)}]{UK2005Lang}
\bibinfo{author}{C.~Ukiwe}, \bibinfo{author}{D.~Y. Kwok},
\newblock \bibinfo{title}{On the maximum spreading diameter of impacting
  droplets on well-prepared solid surfaces},
\newblock \bibinfo{journal}{Langmuir} \bibinfo{volume}{21}
  (\bibinfo{year}{2005}) \bibinfo{pages}{666--673}.
%Type = Article
\bibitem[{Wildeman et~al.(2016)Wildeman, Visser, Sun, and
  Lohse}]{Wildeman2016JFM}
\bibinfo{author}{S.~Wildeman}, \bibinfo{author}{C.~W. Visser},
  \bibinfo{author}{C.~Sun}, \bibinfo{author}{D.~Lohse},
\newblock \bibinfo{title}{On the spreading of impacting drops},
\newblock \bibinfo{journal}{Journal of fluid mechanics} \bibinfo{volume}{805}
  (\bibinfo{year}{2016}) \bibinfo{pages}{636--655}.
%Type = Article
\bibitem[{Laan et~al.(2014)Laan, de~Bruin, Bartolo, Josserand, and
  Bonn}]{Laan2014PRA}
\bibinfo{author}{N.~Laan}, \bibinfo{author}{K.~G. de~Bruin},
  \bibinfo{author}{D.~Bartolo}, \bibinfo{author}{C.~Josserand},
  \bibinfo{author}{D.~Bonn},
\newblock \bibinfo{title}{Maximum diameter of impacting liquid droplets},
\newblock \bibinfo{journal}{Physical Review Applied} \bibinfo{volume}{2}
  (\bibinfo{year}{2014}) \bibinfo{pages}{044018}.
%Type = Article
\bibitem[{Gordillo et~al.(2019)Gordillo, Riboux, and
  Quintero}]{Gordillo2019JFM}
\bibinfo{author}{J.~M. Gordillo}, \bibinfo{author}{G.~Riboux},
  \bibinfo{author}{E.~S. Quintero},
\newblock \bibinfo{title}{A theory on the spreading of impacting droplets},
\newblock \bibinfo{journal}{Journal of Fluid Mechanics} \bibinfo{volume}{866}
  (\bibinfo{year}{2019}) \bibinfo{pages}{298--315}.
%Type = Article
\bibitem[{Du et~al.(2021)Du, Wang, Li, Min, and Wu}]{Du2021Lang}
\bibinfo{author}{J.~Du}, \bibinfo{author}{X.~Wang}, \bibinfo{author}{Y.~Li},
  \bibinfo{author}{Q.~Min}, \bibinfo{author}{X.~Wu},
\newblock \bibinfo{title}{Analytical consideration for the maximum spreading
  factor of liquid droplet impact on a smooth solid surface},
\newblock \bibinfo{journal}{Langmuir} \bibinfo{volume}{37}
  (\bibinfo{year}{2021}) \bibinfo{pages}{7582--7590}.
%Type = Article
\bibitem[{Attan{\'e} et~al.(2007)Attan{\'e}, Girard, and Morin}]{Attane2007POF}
\bibinfo{author}{P.~Attan{\'e}}, \bibinfo{author}{F.~Girard},
  \bibinfo{author}{V.~Morin},
\newblock \bibinfo{title}{An energy balance approach of the dynamics of drop
  impact on a solid surface},
\newblock \bibinfo{journal}{Physics of Fluids} \bibinfo{volume}{19}
  (\bibinfo{year}{2007}).
%Type = Article
\bibitem[{Ray et~al.(2024)Ray, Han, Yue, Guo, Chao, and Cheng}]{Ray2024JFM}
\bibinfo{author}{S.~Ray}, \bibinfo{author}{Y.~Han}, \bibinfo{author}{Z.~Yue},
  \bibinfo{author}{H.~Guo}, \bibinfo{author}{C.~Y.~H. Chao},
  \bibinfo{author}{S.~Cheng},
\newblock \bibinfo{title}{New insights into head-on bouncing of unequal-size
  droplets on a wetting surface},
\newblock \bibinfo{journal}{Journal of Fluid Mechanics} \bibinfo{volume}{983}
  (\bibinfo{year}{2024}) \bibinfo{pages}{A25}.
%Type = Article
\bibitem[{Amirfazli et~al.(2024)Amirfazli, Bustamante, Hu, and
  {\'O}~N{\'a}raigh}]{Amirfazli2024PRSA}
\bibinfo{author}{A.~Amirfazli}, \bibinfo{author}{M.~D. Bustamante},
  \bibinfo{author}{Y.~Hu}, \bibinfo{author}{L.~{\'O}~N{\'a}raigh},
\newblock \bibinfo{title}{Bounds on the spreading radius in droplet impact: the
  inviscid case},
\newblock \bibinfo{journal}{Proceedings of the Royal Society A}
  \bibinfo{volume}{480} (\bibinfo{year}{2024}) \bibinfo{pages}{20230760}.
%Type = Article
\bibitem[{Batzdorf et~al.(2017)Batzdorf, Breitenbach, Schlawitschek, Roisman,
  Tropea, Stephan, and Gambaryan-Roisman}]{Batzdorf2017IJHMT}
\bibinfo{author}{S.~Batzdorf}, \bibinfo{author}{J.~Breitenbach},
  \bibinfo{author}{C.~Schlawitschek}, \bibinfo{author}{I.~V. Roisman},
  \bibinfo{author}{C.~Tropea}, \bibinfo{author}{P.~Stephan},
  \bibinfo{author}{T.~Gambaryan-Roisman},
\newblock \bibinfo{title}{Heat transfer during simultaneous impact of two drops
  onto a hot solid substrate},
\newblock \bibinfo{journal}{International Journal of Heat and Mass Transfer}
  \bibinfo{volume}{113} (\bibinfo{year}{2017}) \bibinfo{pages}{898--907}.
%Type = Article
\bibitem[{Zhang et~al.(2022)Zhang, Sanjay, Shi, Zhao, Lv, Feng, and
  Lohse}]{Zhang2022PRL}
\bibinfo{author}{B.~Zhang}, \bibinfo{author}{V.~Sanjay},
  \bibinfo{author}{S.~Shi}, \bibinfo{author}{Y.~Zhao}, \bibinfo{author}{C.~Lv},
  \bibinfo{author}{X.-Q. Feng}, \bibinfo{author}{D.~Lohse},
\newblock \bibinfo{title}{Impact forces of water drops falling on
  superhydrophobic surfaces},
\newblock \bibinfo{journal}{Physical review letters} \bibinfo{volume}{129}
  (\bibinfo{year}{2022}) \bibinfo{pages}{104501}.
%Type = Article
\bibitem[{Wang and Bourouiba(2017)}]{Wang2017JFM}
\bibinfo{author}{Y.~Wang}, \bibinfo{author}{L.~Bourouiba},
\newblock \bibinfo{title}{Drop impact on small surfaces: thickness and velocity
  profiles of the expanding sheet in the air},
\newblock \bibinfo{journal}{Journal of Fluid Mechanics} \bibinfo{volume}{814}
  (\bibinfo{year}{2017}) \bibinfo{pages}{510--534}.
%Type = Article
\bibitem[{Tang et~al.(2024)Tang, Adcock, and Mostert}]{Tang2024JFM}
\bibinfo{author}{K.~Tang}, \bibinfo{author}{T.~Adcock},
  \bibinfo{author}{W.~Mostert},
\newblock \bibinfo{title}{Fragmentation of colliding liquid rims},
\newblock \bibinfo{journal}{Journal of Fluid Mechanics} \bibinfo{volume}{987}
  (\bibinfo{year}{2024}) \bibinfo{pages}{A18}.
%Type = Article
\bibitem[{Yarin and Weiss(1995)}]{Yarin1995JFM}
\bibinfo{author}{A.~L. Yarin}, \bibinfo{author}{D.~A. Weiss},
\newblock \bibinfo{title}{Impact of drops on solid surfaces: self-similar
  capillary waves, and splashing as a new type of kinematic discontinuity},
\newblock \bibinfo{journal}{Journal of fluid mechanics} \bibinfo{volume}{283}
  (\bibinfo{year}{1995}) \bibinfo{pages}{141--173}.

\end{thebibliography}

\clearpage
\appendix
\setcounter{figure}{0}
\renewcommand{\thefigure}{S\arabic{figure}}

\begin{frontmatter}

\title{The central rising sheet of simultaneous pair-drop impacts: Supplementary Material}

\end{frontmatter}

% Main text
\section{Numerical verification and validation}\label{sec:validation}
% \subsection{Validation and Verification} %%%%%
Here, we present extra information about verification and validation for the simulations. In the experiments \citep{Goswami2023JFM,GoswamiPhDthesis}, a smooth acrylic substrate with advancing angle $80\pm 2^\circ$ and receding angle $58\pm2^\circ$. For the numerical problem, superhydrophobic or fully wetting boundary conditions for the volume fraction $f$ are not suitable. For superhydrophobic substrates, splashing behavior appears easily but requires a relatively fine grid to resolve satisfactorily. On the other hand, when fully wetting boundary conditions are considered, it is thin film impact \citep{Yarin2006ARFM}. Therefore, in the numerical simulations a static contact angle $\theta_{s}=90^\circ$ is used on the impact substrate as an approximation. As shown in Fig. \ref{fig:problem_sketch}, the whole simulation domain is presented and the grid is refined around two drops at initialization.
\begin{figure}[h]
\centering
\begin{tabular}{c}     
  \includegraphics[width=0.47\columnwidth]{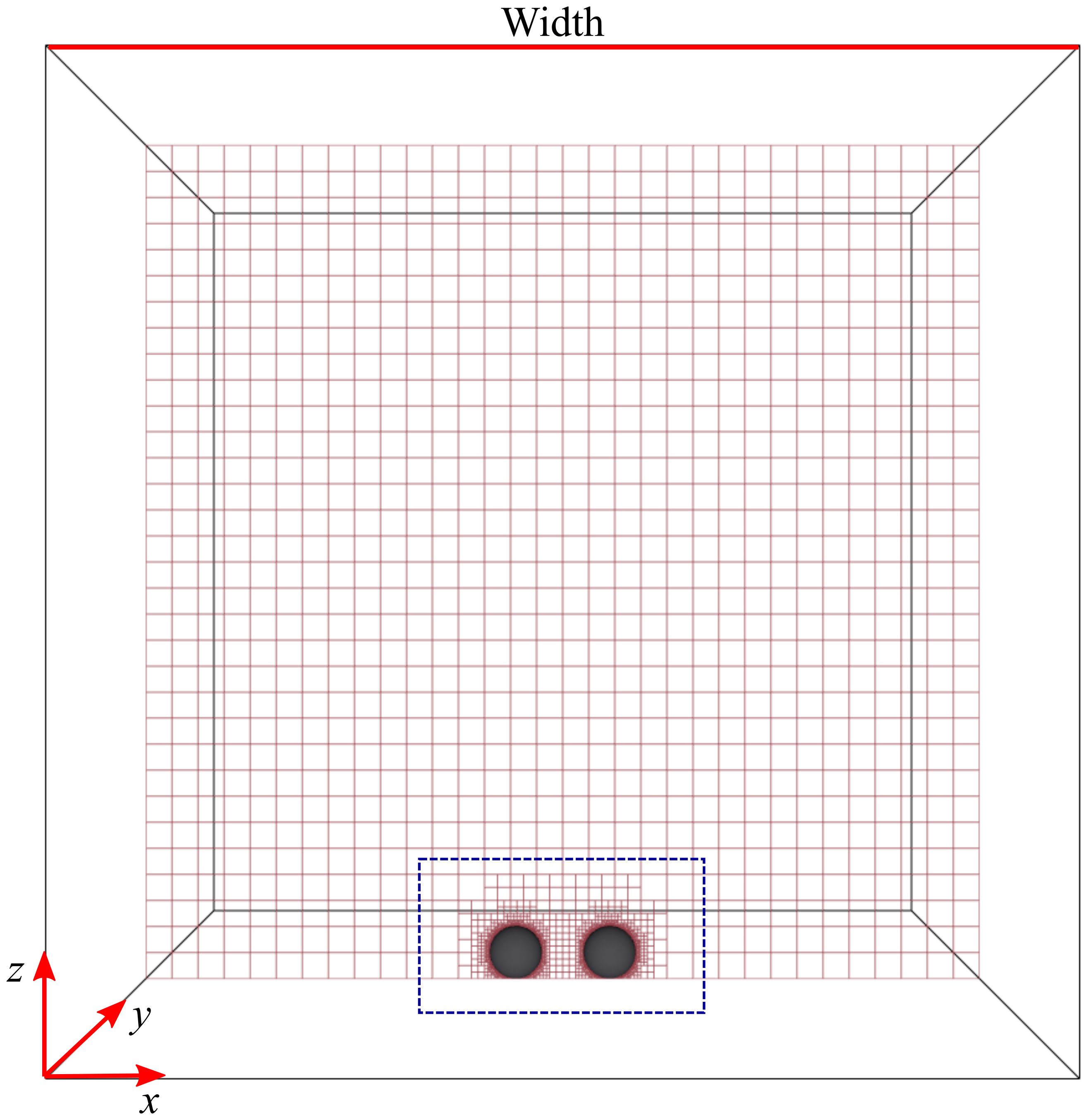}
\end{tabular}
\caption{Computational domain illustrating the adaptive grid refinement around the two drops}
\label{fig:problem_sketch}
\end{figure}

\subsection{Mesh convergence study}
\begin{figure}
      \centering
      \begin{tabular}{cc} 
      \includegraphics[width=0.465\columnwidth]{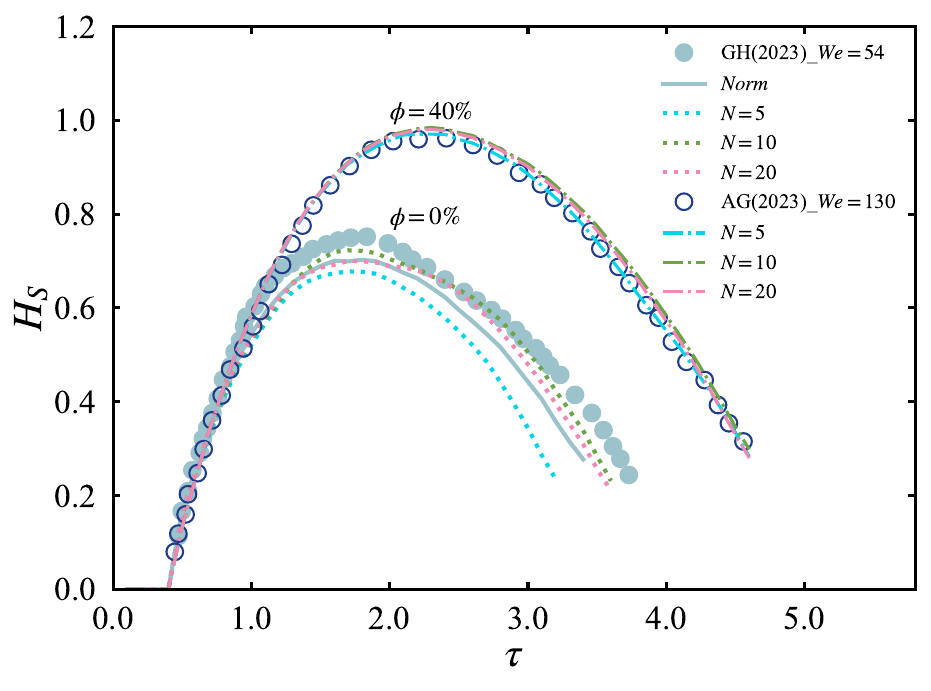} &
      \includegraphics[width=0.465\columnwidth]{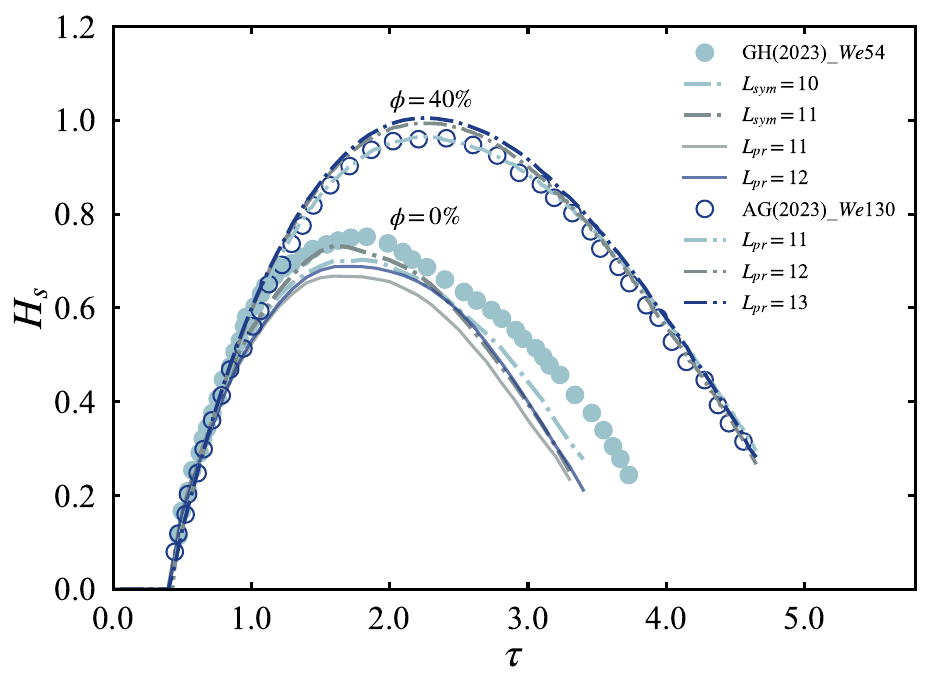} \\
      (a) & (b)
      \end{tabular}
      \caption{Verification of the numerical method through comparisons of the central rising sheet height: (a) comparison with experiments for different iteration numbers $N$, and (b) comparison between two numerical configurations and maximum resolution levels for two representative cases, namely $\phi=0\%$, $We=54$, $Re=3581$ and $\phi=40\%$, $We=130$, $Re=1189$. Here, $H_S=\hat H_s/\hat D_0$ and $\tau=\hat t \hat U_0/\hat D_0$. Note: the experimental data are adapted from \citet{Goswami2023JFM, GoswamiPhDthesis}}
      \label{fig:verification}
\end{figure}

Mesh convergence test is shown in this section. In Figure \ref{fig:verification} the nondimensional heights of rising sheet ($H_s=\hat H_s/\hat D_0$, we omit the star for clarity in the following) at central of two drops for $\phi=0\%,~We=54,~Re=3581,~\rho_r=813,~\mu_r=56$ and $\phi=40\%,~We=130,~Re=1189,~\rho_r=910,~\mu_r=269$ evolving with nondimensional time ($\tau=tU_0/D_0$) are compared with experimental data. The large density ratio leads to a stiff coefficient matrix and causes numerical problems in two-phase flow, which is a general problem for the simulations of 40\% glycerol-water solutions. Therefore, to ensure stability, the Poisson solver within the numerical code run for a fixed number of iterations $N$. We verify first that the numerical results are insensitive to the choice of $N$, and that they compare well with experiment.

For the case of distilled water drops, we compare numerical results obtained with or without the minimum iteration method with experimental results. In Figure \ref{fig:verification}(a), the evolution of $H_s$ is shown versus $\tau$. The points are experimental results in \citet{Goswami2023JFM, GoswamiPhDthesis}, and the lines represent numerical data with different choices of $N$. In terms of $\phi=0\%$, namely water drops, three different minimum iteration numbers $N=5,~10,~20$ are taken into consideration and presented as dotted lines, obtained with a default Poisson solver, which is considered the reference numerical case. It is shown that the data is insensitive to the choice of $N$. The relative errors for maximum height are 3.56\%, 2.88\%, 0.38\% for $N=5,~10,~20$ respectively concerning the reference numerical case. Moreover, they all collapse very well with experiment results. For $\phi=40\%$, hollow points are experimental results in \citet{GoswamiPhDthesis}, and dashed point lines are numerical results. The numerical and experimental results collapse extremely well, and relative errors at maximum height are 1.45\% and 0.43\% as $N$ increase from 5 to 10 and from 10 to 20 respectively. Therefore, $N=20$ can solve well the parameter space considered in this paper. 

The numerical results in Figure \ref{fig:verification}(a) are obtained through a symmetric configuration, in which means one drop is initialized at the beginning and a symmetric boundary condition is set at the middle interaction surface of two drops \citep{Batzdorf2017IJHMT}. This is illustrated and compared in detail in the following. We next determine the effect of the symmetry plane, compared with the approach where both drops are resolved within the domain. For this study, the grid convergence test is carried out using drops of a 40\% glycerol-water solution with the same parameters as used by \citet{GoswamiPhDthesis} at $We=130$. As presented in Figure \ref{fig:verification}(b), it indicates that results obtained from symmetric and pair impact configuration are fairly close and agree well with experimental results with $\phi=0\%$. The size of the computational domain for symmetric configuration is half that of the pair impact configuration, namely $8D_0$. Therefore, the maximum grid resolution is the same for $L_{sym}=10 \text{ and } L_{pr}=11$, and $L_{sym}=11 \text{ and } L_{pr}=12$, and thus the relative errors at the maximum height between symmetric and drop pair configuration at two resolutions are 4.94\% and 6.07\% respectively. Hence, the drop pair configuration is chosen for most numerical cases as it gives the better agreement with experiment. 

\begin{figure}
\centering
\begin{tabular}{c}
  \includegraphics[width=0.8\columnwidth]{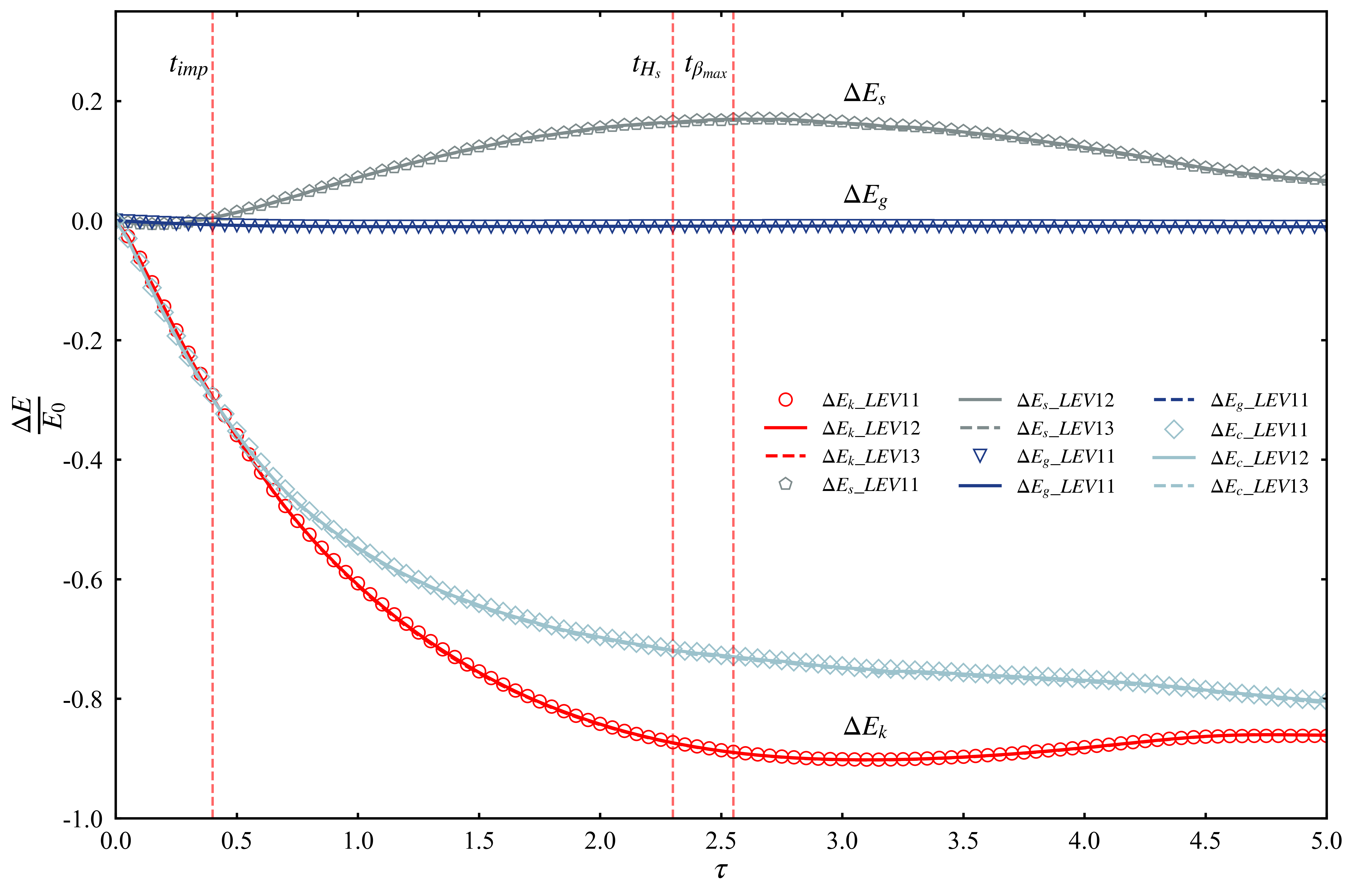}\\
\end{tabular}
\caption{Evolution of the normalized energy variations for different maximum grid resolutions at $We=130$ and $Re=1189$. Here, $t_{imp}$ denotes the impact time, $t_{H_s}$ the time of maximum sheet height, and $t_{\beta_{max}}$ the time at which the remaining drops on the substrate reach the maximum spreading distance. The change in conservative energy $\Delta E_c$, exhibits an exponential decay over time. $\Delta E_s$ denotes the change in surface energy, $\Delta E_g$ the change in gravitational energy, which remains negligible, and $\Delta E_k$ the change in kinetic energy}
\label{fig:Energy_change}
\end{figure}

In order to study the grid convergence of the energetics of the process, we plot the volume-integrated energy normalized by its initial value, comprising the initial kinetic (KE) and surface tension (SE) energies, in Fig. \ref{fig:Energy_change} for $We=130,~Re=1189$. We include different maximum grid levels to inspect sensitivity to mesh resolution. The energetics of the conservative quantities (namely KE, SE) are individually grid-converged, as is their sum. 
% While it is difficult to establish grid-convergence for the directly resolved total dissipation $E_d$, which is computed from the deformation tensor in eqns. (\ref{eqn:ed}, \ref{eqn:Ed}).

\begin{figure}
\centering
\begin{tabular}{c}     
 \includegraphics[height=0.5\columnwidth]{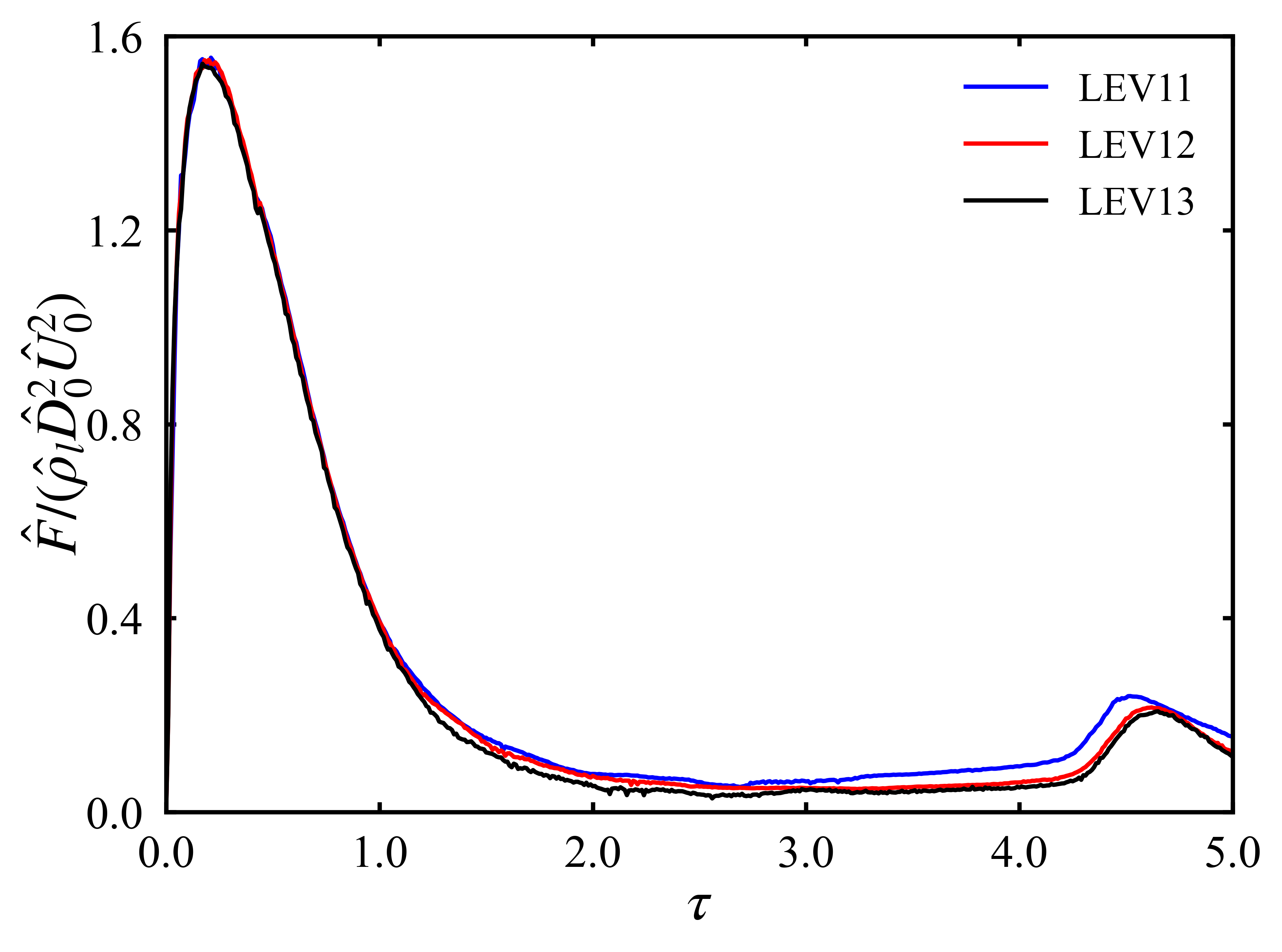}
\end{tabular}
\caption{Convergence test of the evolution of the normalized normal force on the substrate for different grid resolution levels at $\phi=40\%$, $We=130$, and $Re=1189$}
\label{fig:F_bottom}
\end{figure}

Furthermore, a grid resolution study is carried out through pair configuration at $\phi=40\%,~We=130$, compared with experimental results as presented in Figure \ref{fig:verification}(b). As $L$ increases, the errors at $t_{H_s}$ are 2.92\% and 1.11\% respectively, which means the results are grid converged. Moreover, we compared the normal force on the substrate for the same parameters and it can be expressed as equation \ref{eqn:force} \citep{Zhang2022PRL}:
\begin{equation}
    \hat F(\hat t) = (\int_{\hat A}(\hat p-\hat p_0)d\hat A)\boldsymbol{z}
\label{eqn:force}
\end{equation}
where $\hat p$ is the dynamic pressure on the substrate, $\hat p_0$ is environmental pressure, $\hat A$ is substrate's area, and $\boldsymbol{z}$ is the unit vector normal to the substrate. As shown in Figure \ref{fig:F_bottom}, the evolution of $\hat F$ has two peak values. The first peak is because of the impact of these two drops, and the second peak is due to the impact of the central sheet. The force at early time collapses fairly well with different $L$. We thus compared the relative errors of the second peak, which are 9.77\% and 3.72\% respectively. Therefore, we run the simulations with $L=12$ in this research.

\subsection{Experimental comparison}
Since we get converged results using $N=20,~L=12$ in the section above, we compare the results obtained from numerical simulations with experiments \citep{Goswami2023JFM, GoswamiPhDthesis} by using the settings in the following. Various $We$ are examined with corresponding $Re$ for the distilled water ($\phi=0\%$) and 40\% glycerol-water ($\phi=40\%$) mixtures. 

\begin{figure}
\centering
\begin{tabular}{cc}     
 \includegraphics[height=0.7\columnwidth]{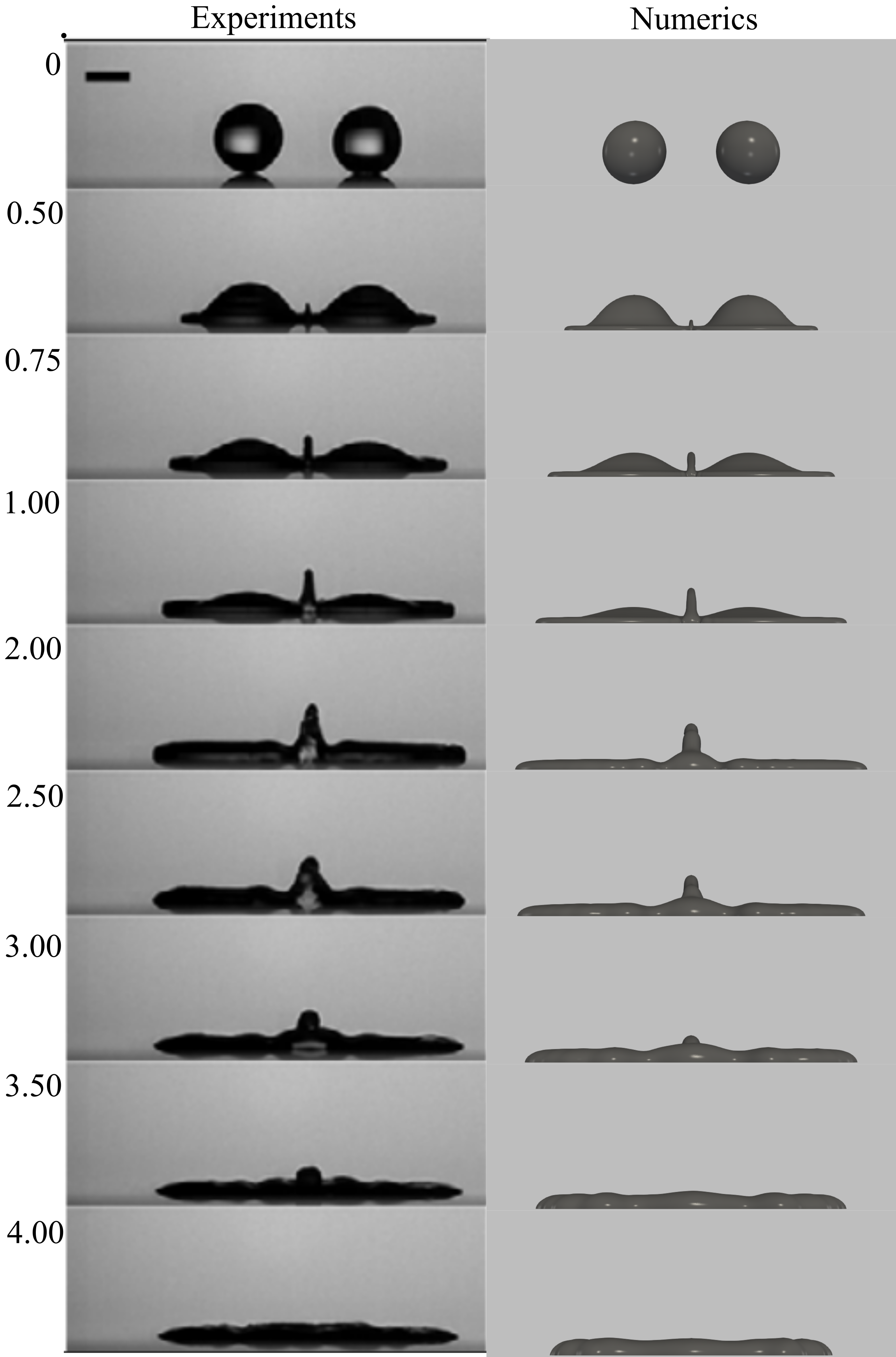} &
 \includegraphics[height=0.693\columnwidth]{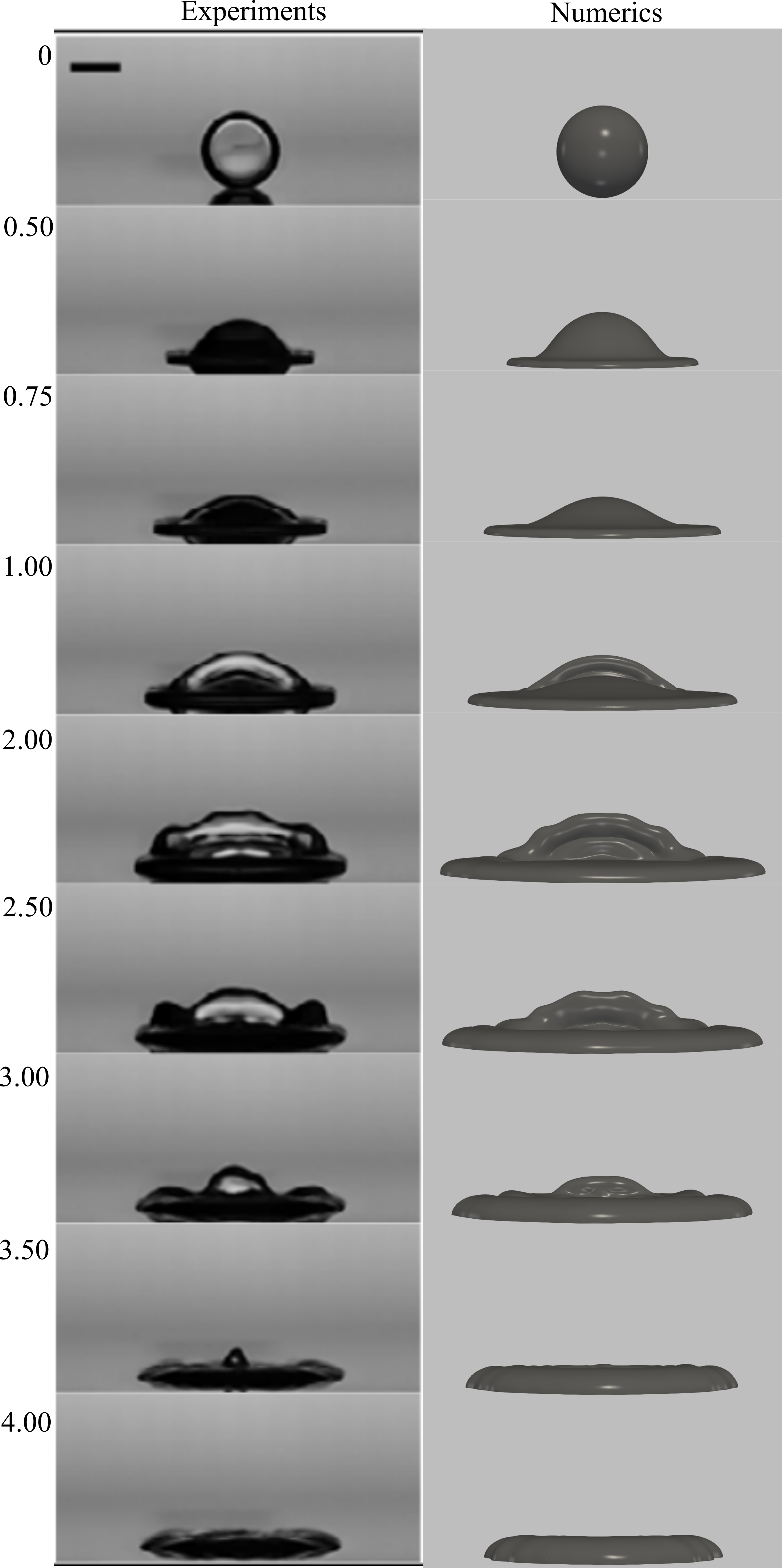} \\
 (a) & (b)
\end{tabular}
\caption{Comparison of the evolution of the central sheet with non-dimensional time between numerical simulations and experiments for distilled water drops at $We=54$ and $Re=3581$: (a) front view and (b) side view. Note: the experimental images are adapted from \citet{Goswami2023JFM}, CC BY 4.0}
\label{fig:0_We54_f}
\end{figure}

Fig. \ref{fig:0_We54_f} presents pair-drop impact at $We=54,~Re=3581$, which corresponds to a distilled-water problem. There is general agreement between numerical and experimental process. First (Fig. \ref{fig:0_We54_f}(a)), the sheet rises due to inertia and then falls because of surface tension \citep{ Goswami2023JFM, Roisman2002JCIS}. The typical semilunar shape is captured numerically \citep{Goswami2023JFM, Liang2020ActaM, Ersoy2020POF}, as shown in the side view in Fig. \ref{fig:0_We54_f}(b). 

\begin{figure}
\centering   
\begin{tabular}{c} 
 \includegraphics[width=0.705\columnwidth]{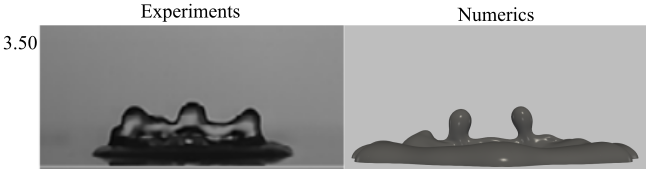} \\
 (a)\\
 \includegraphics[width=0.71\columnwidth]{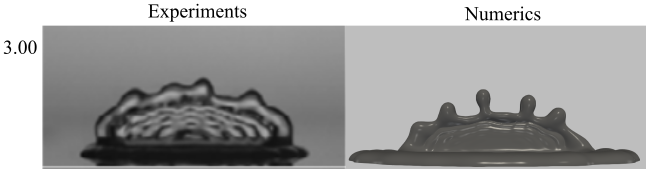} \\
 (b)\\
 \includegraphics[width=0.7\columnwidth]{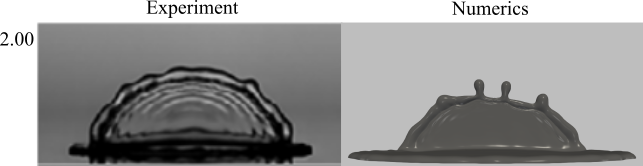}\\
 (c)
\end{tabular}
\caption{Snapshots of experimental and numerical results at the same non-dimensional time for different $We$ at $\phi=0\%$: (a) $We=80$, $Re=4358$; (b) $We=104$, $Re=4969$; and (c) $We=128$, $Re=5513$. Note: the experimental snapshots are adapted from \citet{Goswami2023JFM}, CC BY 4.0}
\label{fig:0_falling_comparison_f}
\end{figure}

The corresponding quantitative comparison of central sheet height evolution between numerical data and experimental results in \citet{Goswami2023JFM, GoswamiPhDthesis} is presented in the main text Fig. \ref{fig:validation_quantitative}(a) and Fig. \ref{fig:validation_quantitative}(b). There are small differences in nominal diameter between \citet{GoswamiPhDthesis} and \citet{Goswami2023JFM}; here we use the former study to determine Reynolds numbers for comparison.  As shown in the main text Fig. \ref{fig:validation_quantitative}(a), the height evolution matches well with experimental data at three different $We$. According to Fig. \ref{fig:0_We54_f}, the structure of the central sheet resembles a well-defined semilunar shapes. At larger times, when the central sheet starts to fall, there is a slight deviation between experimental and numerical curves. To explain this, in Fig. \ref{fig:0_falling_comparison_f}, the comparison of volume function $f$ with experiments at different $\tau$ with different $We$ is used to illustrate the differences. As shown in Fig. \ref{fig:0_falling_comparison_f}(a) and (b), the corrugations develop at the falling stage, and thus induce cusps. These cusps influence the manner in which the maximum height is measured: in the present study we use the geometrical maximum value of the height, which is defined in the main text, while \citet{Goswami2023JFM} determined the maximum height using a fitted circle. 

% \begin{figure}
% \centering
% \begin{tabular}{cc}     
%  \includegraphics[height=0.35\columnwidth]{figures/Hmax_0_cut.pdf} &
%  \includegraphics[height=0.35\columnwidth]{figures/Hmax_40.pdf} \\
%  (a) & (b)
% \end{tabular}
% \caption{Comparison of the maximum heights between numerical data and experiments: (a) $\phi=0\%$ solutions and (b) $\phi=40\%$ solutions with three different $We$ experimental cases ($Re$ is obtained correspondingly). Note: the experimental results are from \citet{Goswami2023JFM}}
% \label{fig:validation_quantitative}
% \end{figure}

\begin{figure}
\centering
\begin{tabular}{cc} 
    \includegraphics[width=0.4\columnwidth]{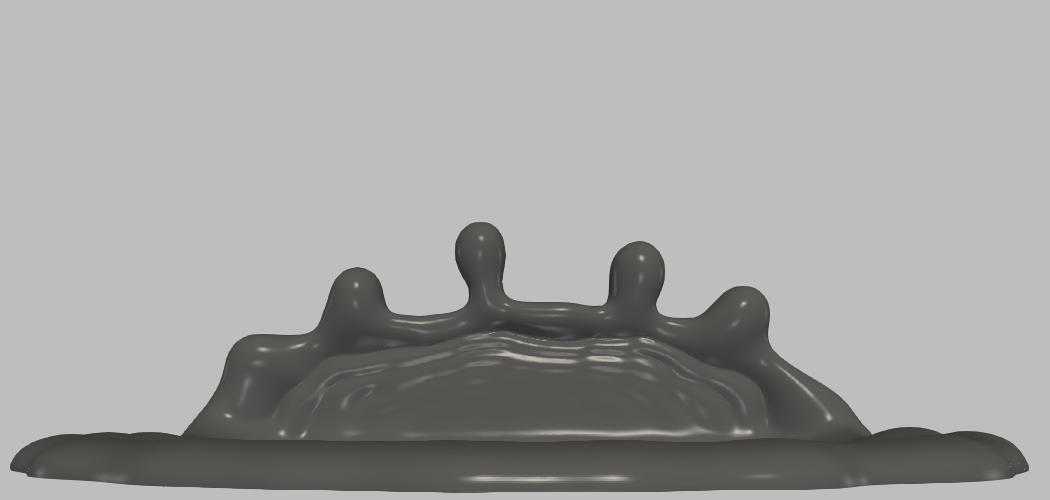} & 
    \includegraphics[width=0.4\columnwidth]{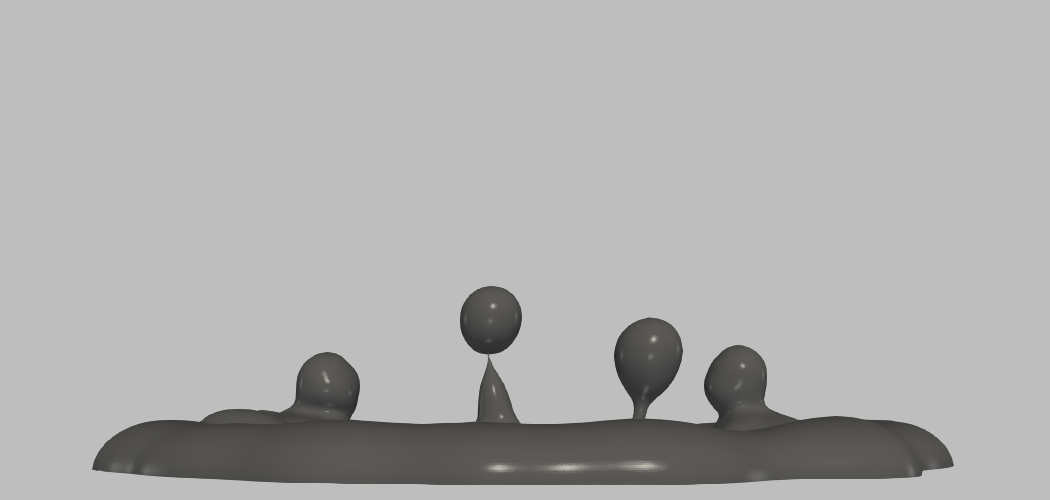} \\
    (a) & (b) \\
    \includegraphics[width=0.4\columnwidth]{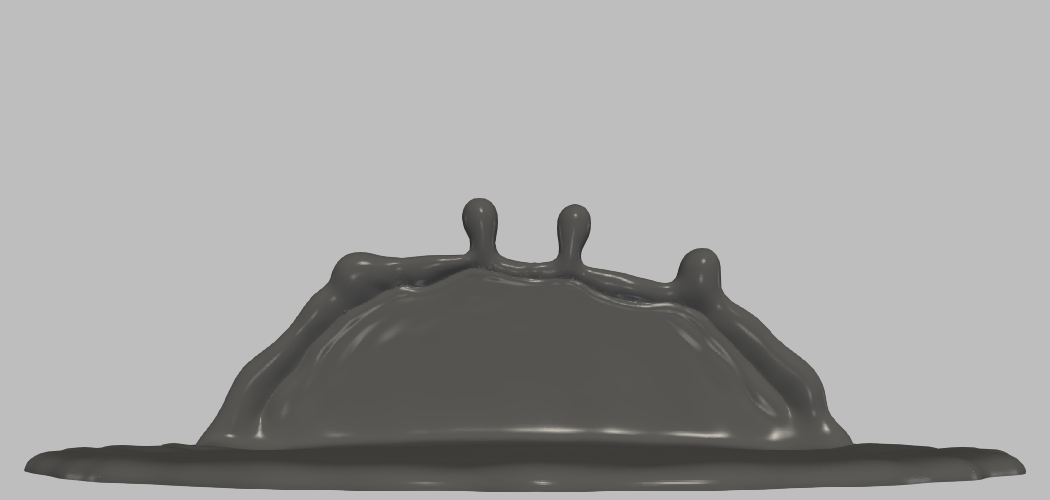} & 
    \includegraphics[width=0.4\columnwidth]{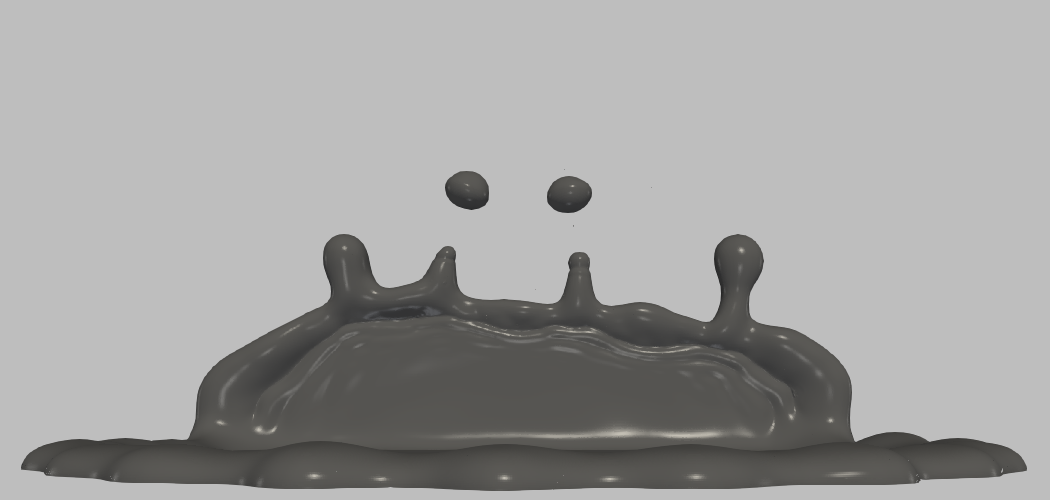} \\
    (c) & (d)
\end{tabular}
\caption{Snapshots of numerical simulation results $f$ for $\phi=0\%$ at $We=104$, $Re=4969$: (a) $\tau=3$ and (b) $\tau=4.4$, and at $We=128$, $Re=5513$: (c) $\tau=2$ and (d) $\tau=3$}
\label{fig:0_falling_explanation_f}
\end{figure}

%%Moreover, the difference of static contact angle and dynamic contact angle may matter more when corrugation appears, and it thus causes a mismatch here. 
The cusps seen in Fig. \ref{fig:0_falling_comparison_f} are associated with instability of the central sheet. At higher $We$ and $Re$, some cusps may pinch off, signalling splashing, as shown in Fig. \ref{fig:0_falling_explanation_f}. There are specific differences for cusp growth and splashing behaviours between numerical and experimental data. These differences may be attributed to many effects, such as the differences in contact angle, deviation from spherically symmetric drops, or from true simultaneous impact in the experiments, which cannot be quantitatively accounted for in this comparison. Nevertheless the overall agreement between simulation and experiment is excellent. And as we mentioned in the main text, the agreement between experimental and numerical results is further improved for the glycerol-water mixture ($\phi=40\%$).

% This agreement between experimental and numerical results is further improved for the glycerol-water mixture ($\phi=40\%$). As shown in the main text Fig. \ref{main-fig:validation_quantitative}(b) for $\phi=40\%$ solutions, the heights of central rising sheet at three different $We$ match well with experimental data at either the rising stage or the falling stage. Additionally, the corresponding structures of front view and side view (the inset) at each $\tau$ depicted in the main text Fig. \ref{main-fig:40_We130_f} are more regular. This regular morphology makes our measurement closer to experimental data.

\begin{figure}
\centering
\begin{tabular}{c} 
    \includegraphics[width=0.7\columnwidth]{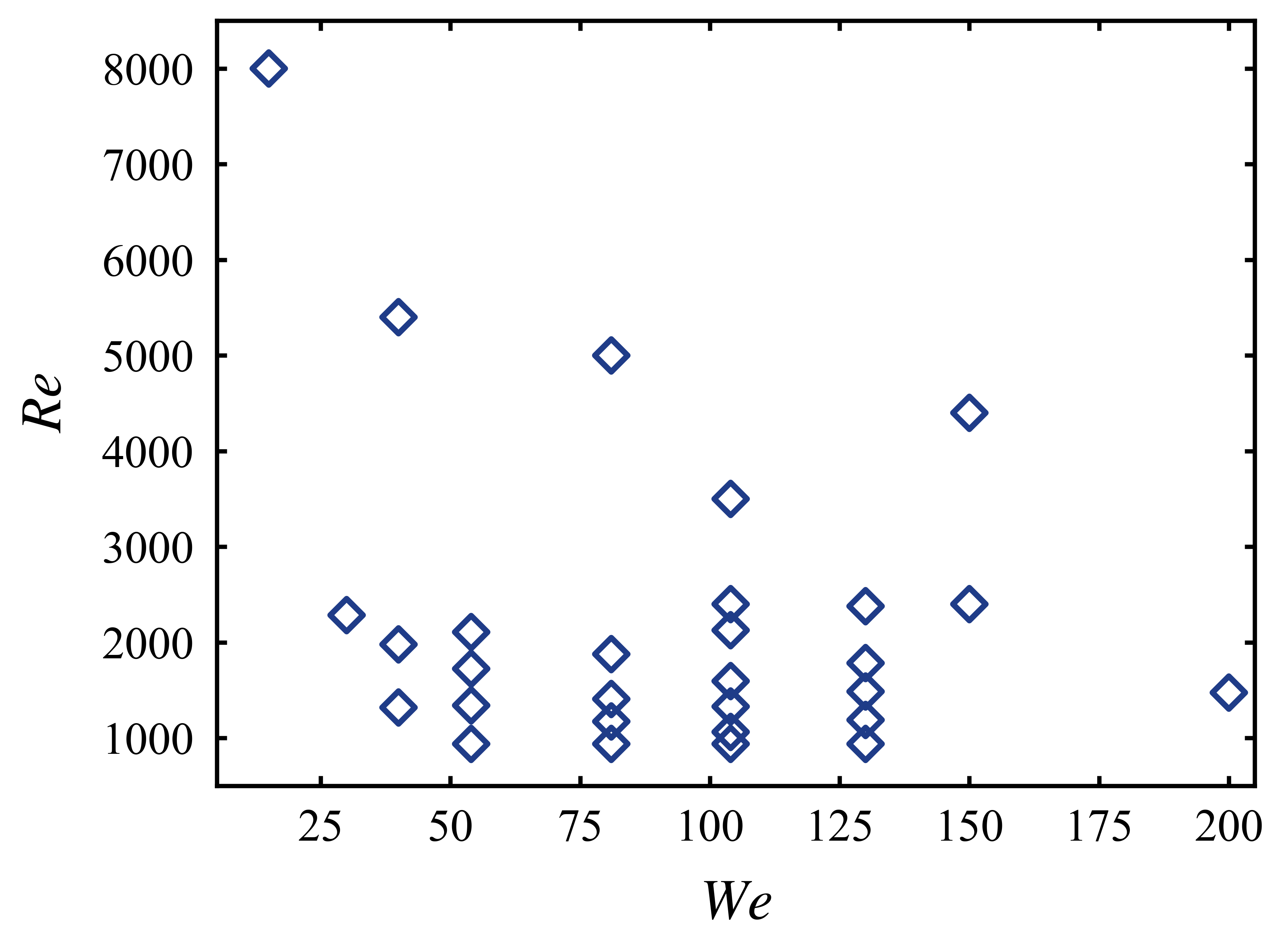}
\end{tabular}
\caption{Computational parameter space in terms of $We$ and $Re$ for simultaneous pair-drop impacts}
\label{fig:parameter_space}
\end{figure}

All of the validation and verification efforts indicate that our numerical platform captures well the physical process of simultaneous pair-drop impacts. The remainder of the study will describe the results of a comprehensive and cohesive analysis of the dynamics of two simultaneous impact drops across a large regime of parameter space. Since the splashing phenomenon and growth of the central sheet instability are not the principal focus of this paper, in the study we mainly examine impact parameters corresponding to drops of a $\phi=40\%$ glycerol-water solution. The parameter space as shown in Fig. \ref{fig:parameter_space} covers existing experimental parameters in \citet{GoswamiPhDthesis} and extends them. In this work, the cases are set with $L=12$ in pair impact configuration. For reference, example dimensional properties for $\phi=40\%$ are: $\rho_l=1114.5kg/m^3,~\mu_l=4.814mPa,~\sigma=69.6mN/m,~D=3.25mm$.

% and table \ref{tab:parameter_space_2} 

\section{Scaling comparison at different drop separation distance}\label{appC}
\begin{figure}
\centering
\begin{tabular}{cc}
  \includegraphics[width=0.475\columnwidth]{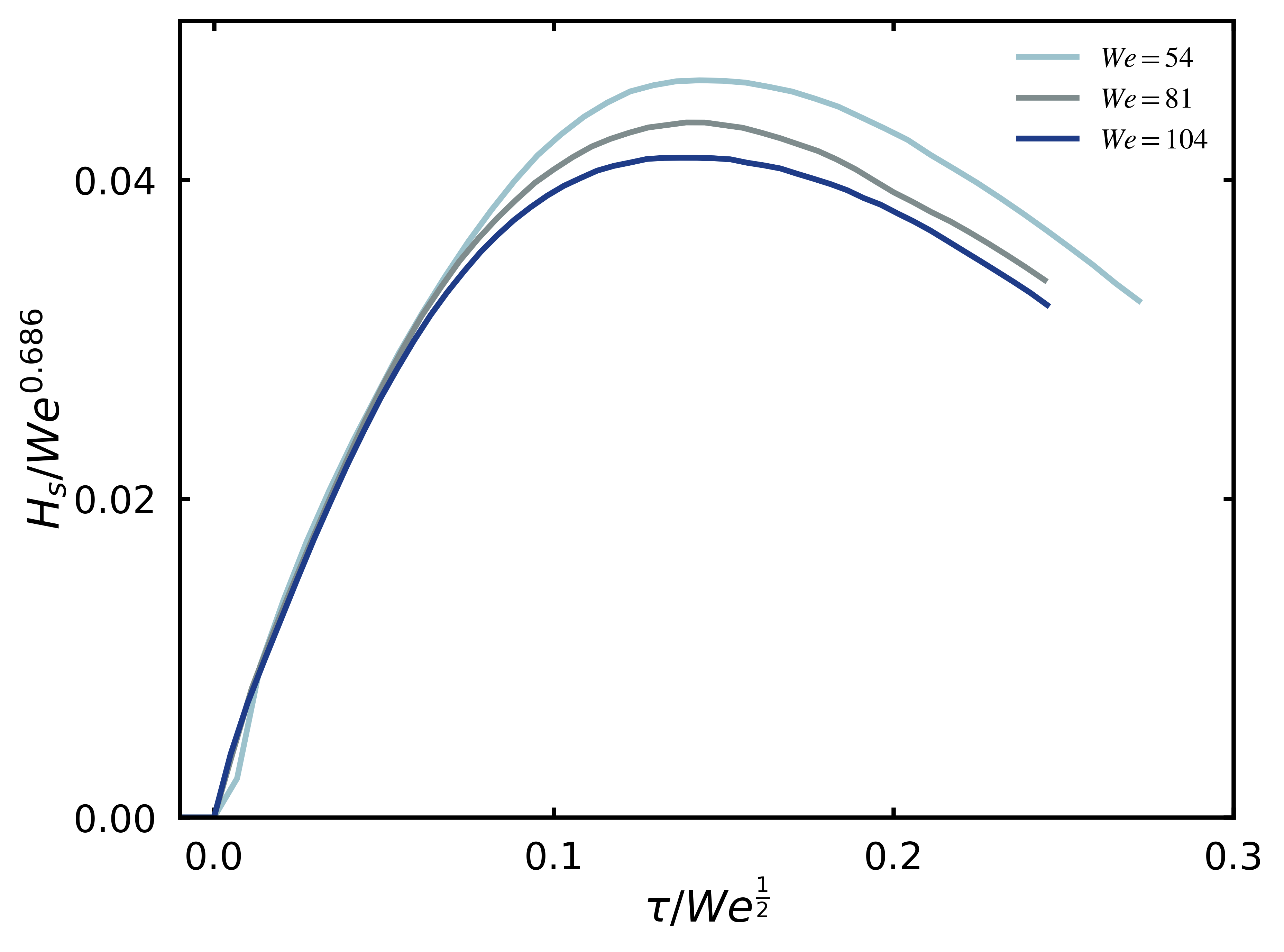} &
  \includegraphics[width=0.475\columnwidth]{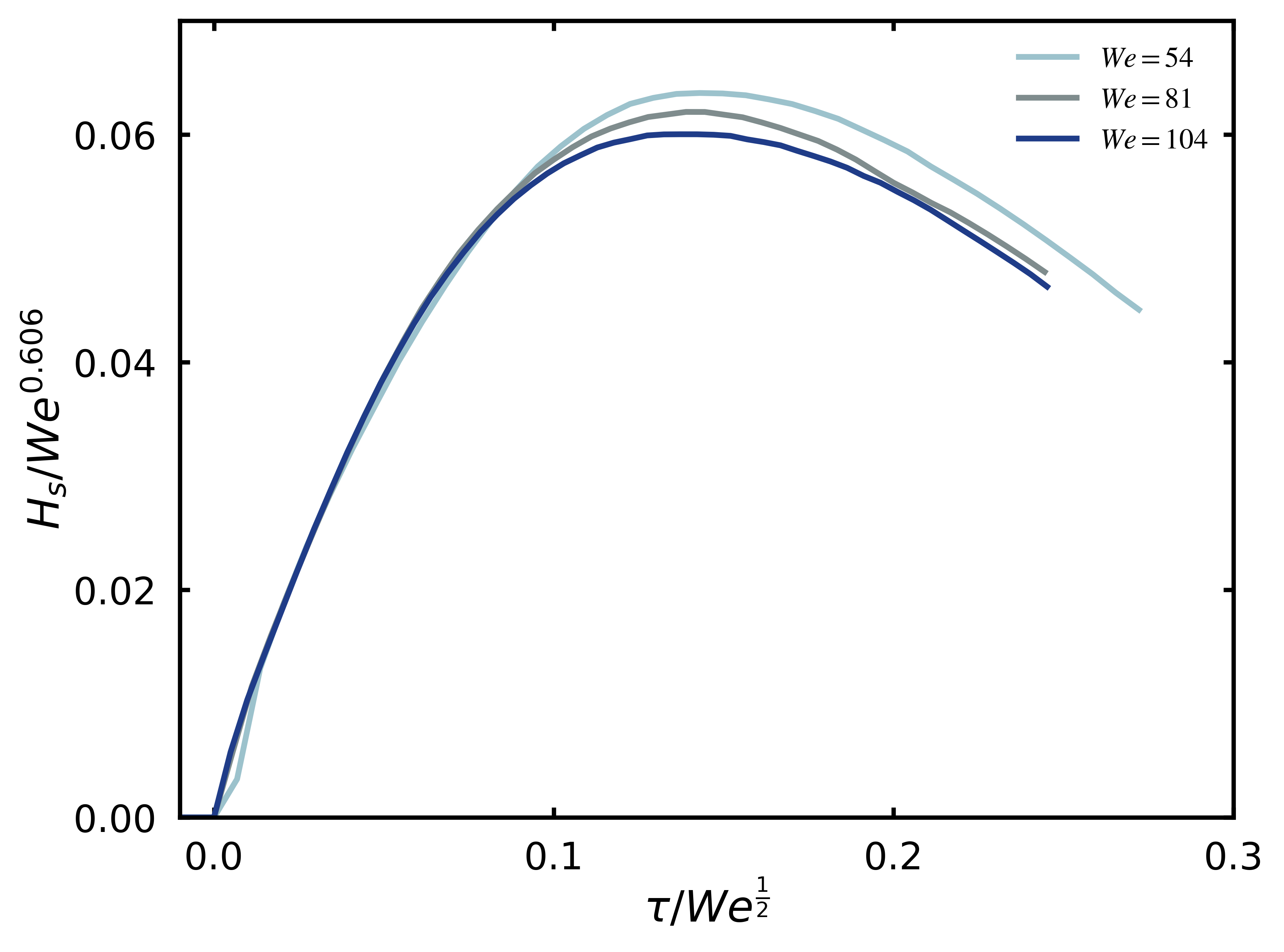} \\
  (a) & (b)\\
    \includegraphics[width=0.475\columnwidth]{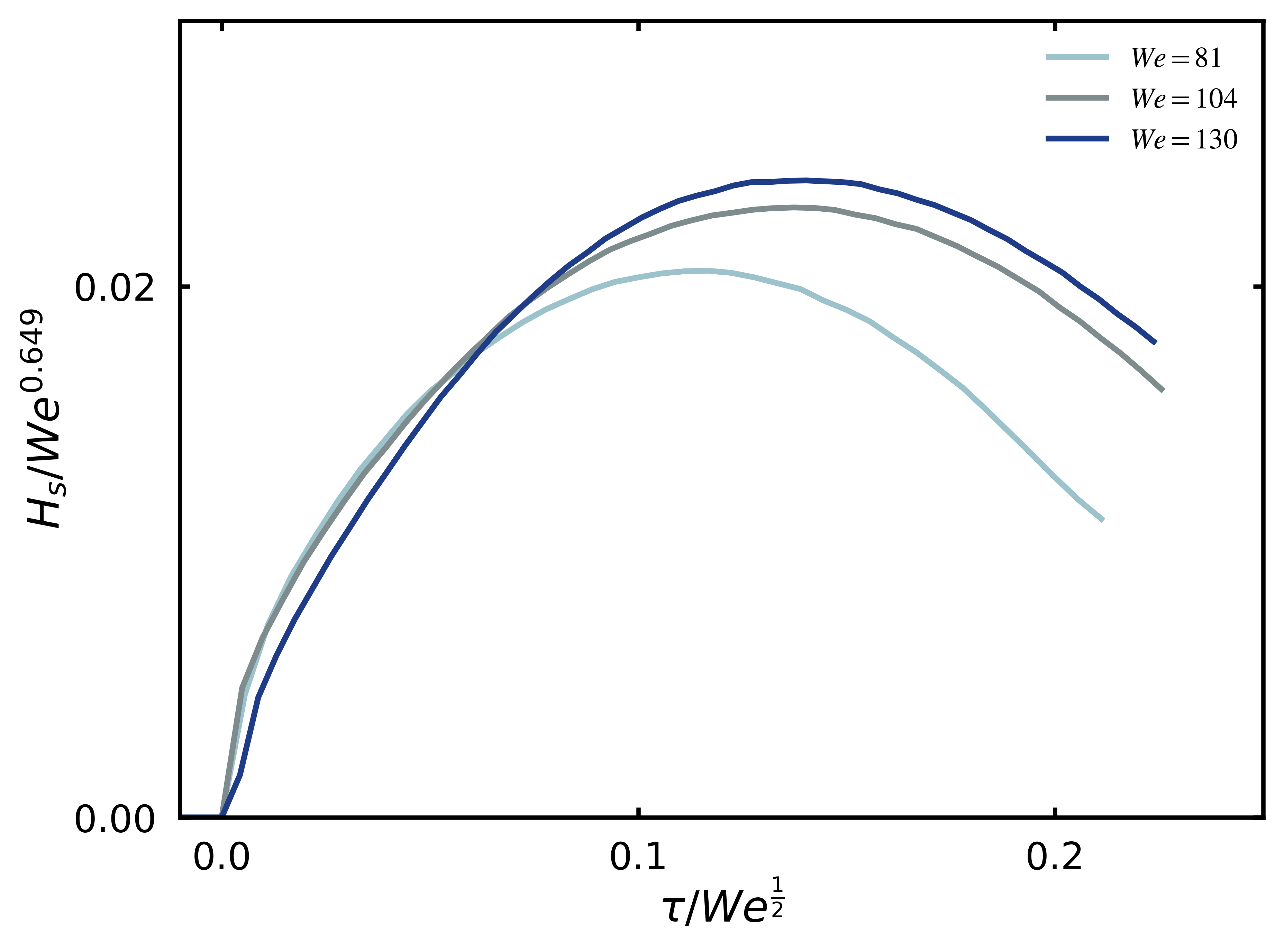} &
  \includegraphics[width=0.475\columnwidth]{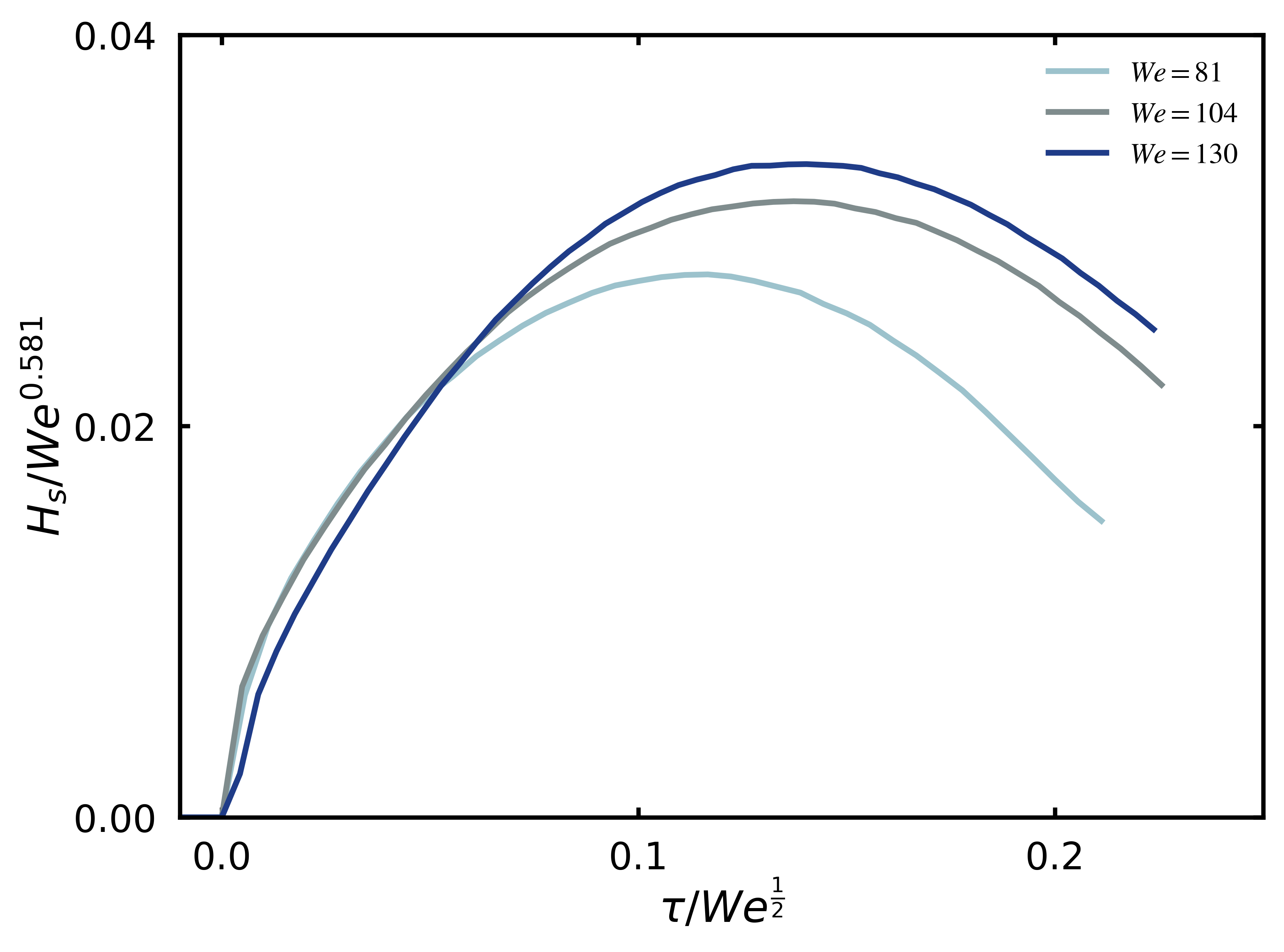} \\
  (c) & (d)
\end{tabular}
\caption{Validation of numerical simulation results collapsed using the scaling law for $H_{s,\max}$ in terms of $We$, as predicted by (a,c) the lollipop model and (b,d) the cylindrical disk model at $\Delta x = 1.5$ and $2.2$, respectively, with fixed $Re=938$}
\label{fig:scaling_law_validation}
\end{figure}

We obtain the scalings for $H_{s,max}$ at different $\Delta x$ through the energetic model in the large $Re$ regime. In this section, we consider the effectiveness of the proposed scalings. We plot the rescaled height evolution with fixed $Re=938$ for three different $We=54,81,104$ at $\Delta x=1.5$ in Figure \ref{fig:scaling_law_validation}(a) and (b), and $We=81,104,130$ at $\Delta x=2.2$ in Figure \ref{fig:scaling_law_validation}(c) and (d). $H_s$ is rescaled by the scaling obtained from lollipop model and cylindrical disk model, and $\tau$ is rescaled by capillary time $\tau_{cap}\equiv\sqrt{\hat\rho_l\hat D_0^3/\hat\sigma}=We^{\frac{1}{2}}$. The height profiles are aligned well with the cylindrical disk model $H_{s,max}\sim We^{0.606}$ and lollipop scaling $H_{s,max}\sim We^{0.686}$ at $\Delta x=1.5$, and the maximum relative error for $H_{s,max}$ at $t_{H_s}$ is 6.25\% and 12.8\% respectively. At $\Delta x=2.2$, the maximum relative error for $H_{s,max}$ at $t_{H_s}$ is 12.5\% and 15.1\% for lollipop model and cylindrical disk model respectively. It indicates that the energetic model works well even considering different $\Delta x$, and the cylindrical disk asymptotic scaling works better than lollipop model at smaller drop separation distance, and vise versa. In addition, $t_{H_s}$ is well aligned with $\tau_{cap}$.

\section{Morphology of the central sheet}

\subsection{Morphology and height of the central rising sheet}\label{subsubsec:Morph_central}
A central sheet rises after two drops interact on the substrate, as reported in \citet{Ersoy2020POF, Liang2020ActaM, Goswami2023JFM} as seen in Fig. \ref{fig:0_We54_f}. This typical semilunar central sheet plays a crucial role in pair-drop impact and even multiple drop impact, and may lead to splashing, where it would otherwise not be expected for single-drop impacts \citep{Ersoy2020POF,Goswami2023JFM}. In this section, a detailed investigation of its morphology will be carried out. 

\begin{figure}
\centering
 \includegraphics[width=0.7\columnwidth]{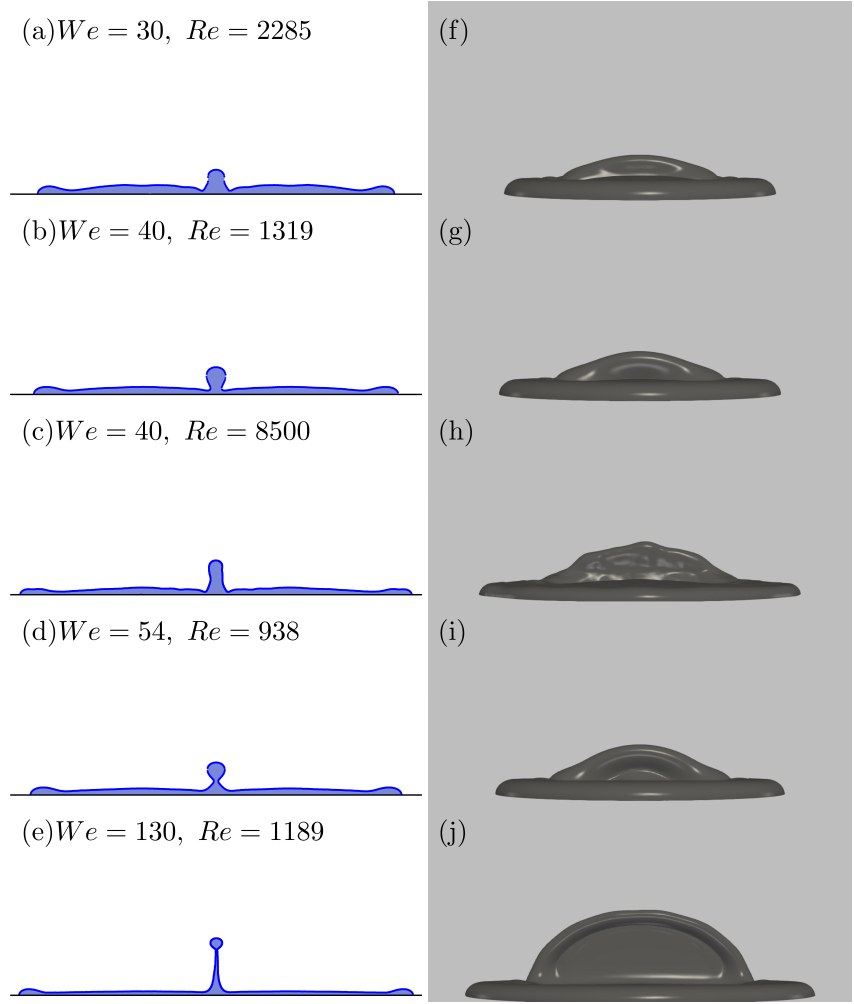}
\caption{Morphological changes of the central rising sheet with increasing $We$, shown by front views with two-dimensional slices (left column) and side views represented by the volume fraction $f$ (right column)}
\label{fig:Morphologies}
\end{figure}

We compare the morphology of the central sheet at the moment it reaches the maximum height. In Fig. \ref{fig:Morphologies}, different morphologies are shown for different $We$ and $Re$. The left column is a 2D slice across the centre of the configuration from the front view, and the right column is from the side view. From the front view, the central sheet is low and thick for low $We$. Even though the sheet becomes thinner with increasing $Re$ as presented by Fig. \ref{fig:Morphologies}(b) and (c), a more obvious rim on the central sheet appears for $We\geq54$. As $We$ increases to 130, the central sheet becomes thinner and reaches a higher maximum height. In other words, when $We<40$, central sheet is thick and low and close to a cylindrical disk. As $We\geq54$, it tends to resemble a "lollipop". This indicates that $We$ dominates the morphology of central sheet.

Moreover, from the side view, the typical semilunar structure is not present when $We$ is low. For higher $Re$ e.g. Fig. \ref{fig:Morphologies}(h), rim corrugations appear, owing to a combination of Rayleigh-Taylor and Rayleigh-Plateau instabilities, and dominated by the fastest growing mode Rayleigh-Plateau instability at large time \citep{Goswami2023JFM}. As $We$ increases, the semilunar structure appears as shown in Fig. \ref{fig:Morphologies}(j). These observations suggest that $We$ plays a more important role in the morphologies of central sheets than $Re$, and as $We$ increases the sheet at first features a low and thick cylindrical disk and later on a high and thin lollipop model.

%%The thickness turning regime of rising sheet decreases with the increase of $Re$, while it gets close to lamella thickness with the increase of vertical position, which suggests that the thickness of the central lamella is weakly dependent on $We$ and $Re$

\begin{figure}
\centering
\begin{tabular}{cc}     
 \includegraphics[height=0.35\columnwidth]{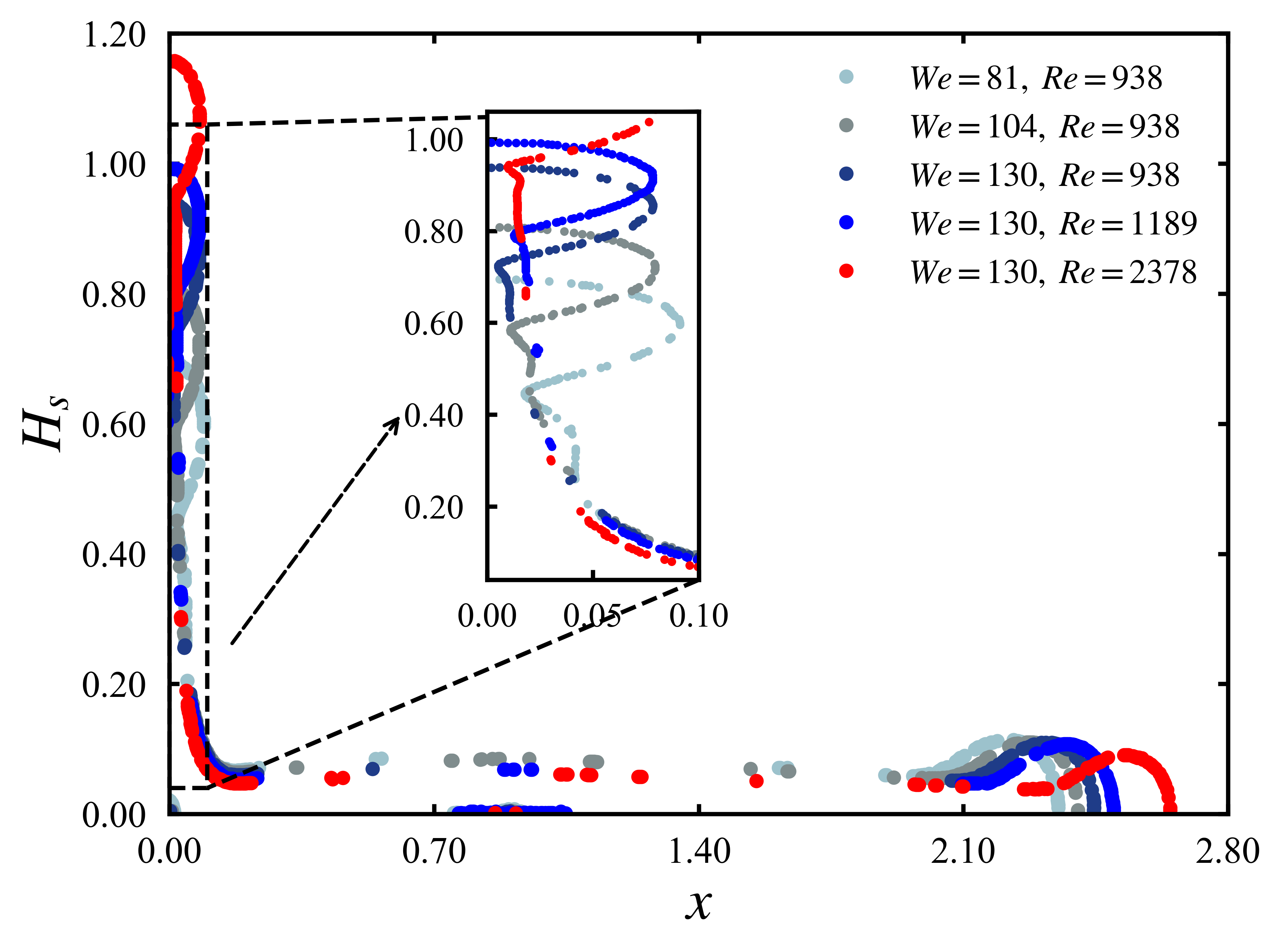} &
 \includegraphics[height=0.35\columnwidth]{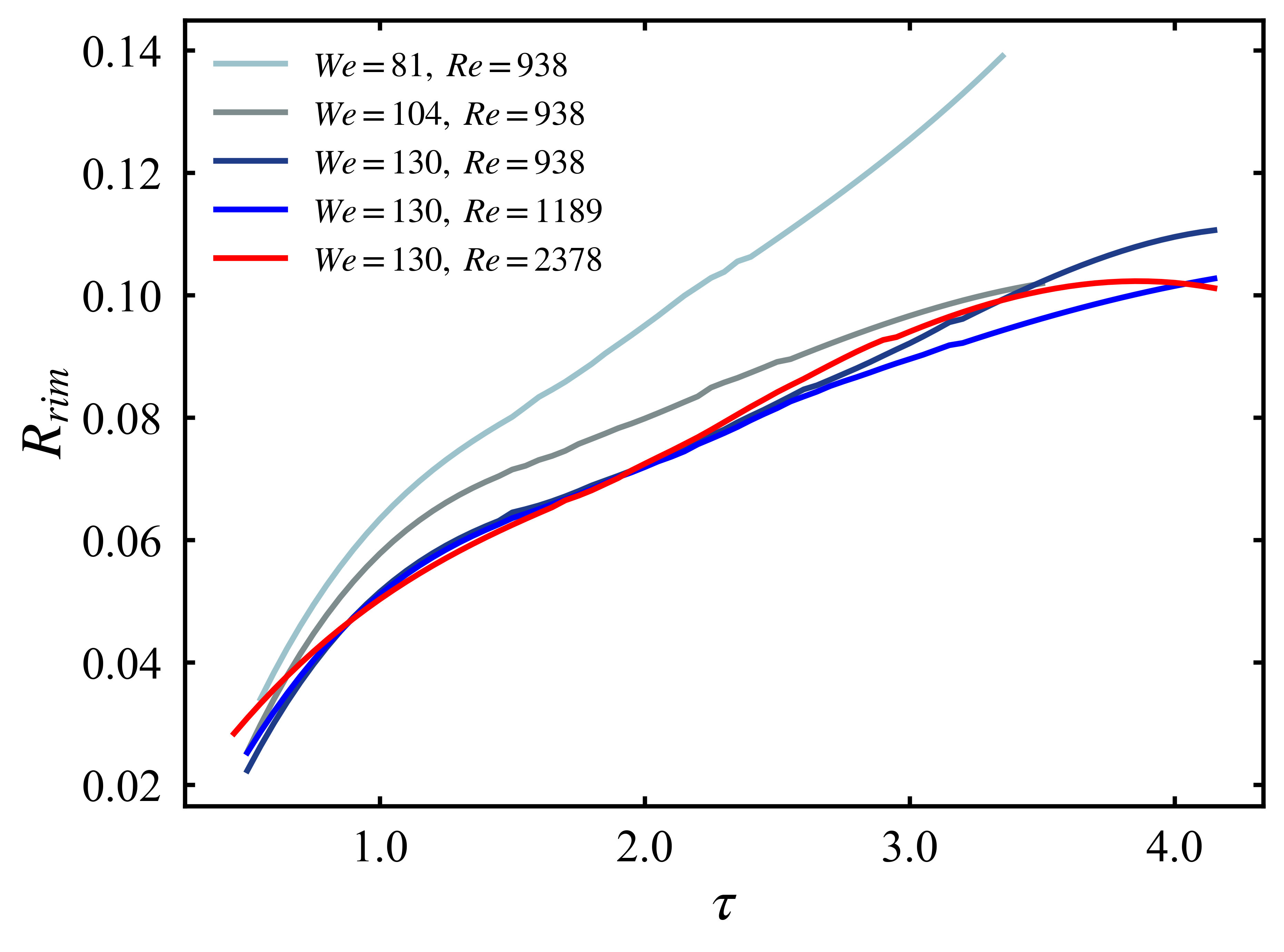} \\
 (a) & (b)
\end{tabular}
\caption{Dimensionless measurements of the central rising sheets: (a) half of the rising sheet including the bottom spreading region, and (b) evolution of the rim radius on the central sheet for different $We$ and $Re$}
\label{fig:central_hmtRrim}
\end{figure}

We further investigate the thickness of the central sheet at its maximum height and the radius of the central rim in Fig. \ref{fig:central_hmtRrim}. In Fig. \ref{fig:central_hmtRrim}(a), we plot the thickness of the central sheet at its maximum height. The central rising sheet consists of a lamella and a rim, similarly to the structure of the spreading drop on the substrate. It shows that the thickness of lamella part varies little with $We$ and $Re$. Given Fig. \ref{fig:central_hmtRrim}(b), the evolution of $R_{rim}$ on the central sheet is shown for the same values of $We$ and $Re$. Generally, $R_{rim}$ increases with time and the time evolution of rims is strongly dependent on $We$ and insensitive to $Re$. The dynamics of the lamella and the rim on the central sheet is similar to the behaviours of drop spreading on the substrate or that of rim collision \citep{Wang2017JFM, Gordillo2019JFM, Tang2024JFM}. In addition, the rim is only affected by surface tension in early time. Further details on how $R_{rim}$ is measured are given in section \S\ref{sec:data_process}.

\begin{figure}
\centering
\begin{tabular}{cc}     
 \includegraphics[height=0.35\columnwidth]{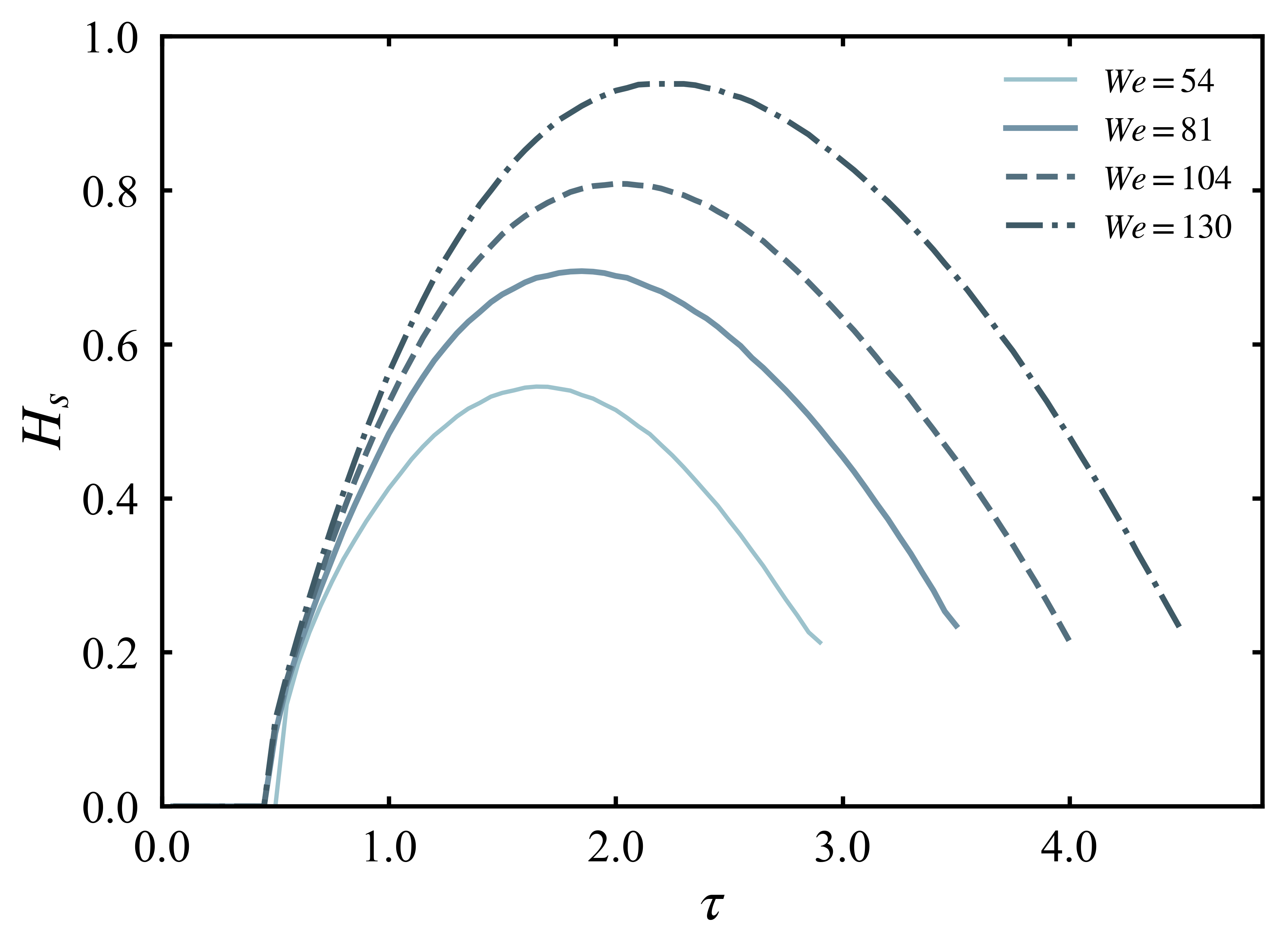} &
 \includegraphics[height=0.35\columnwidth]{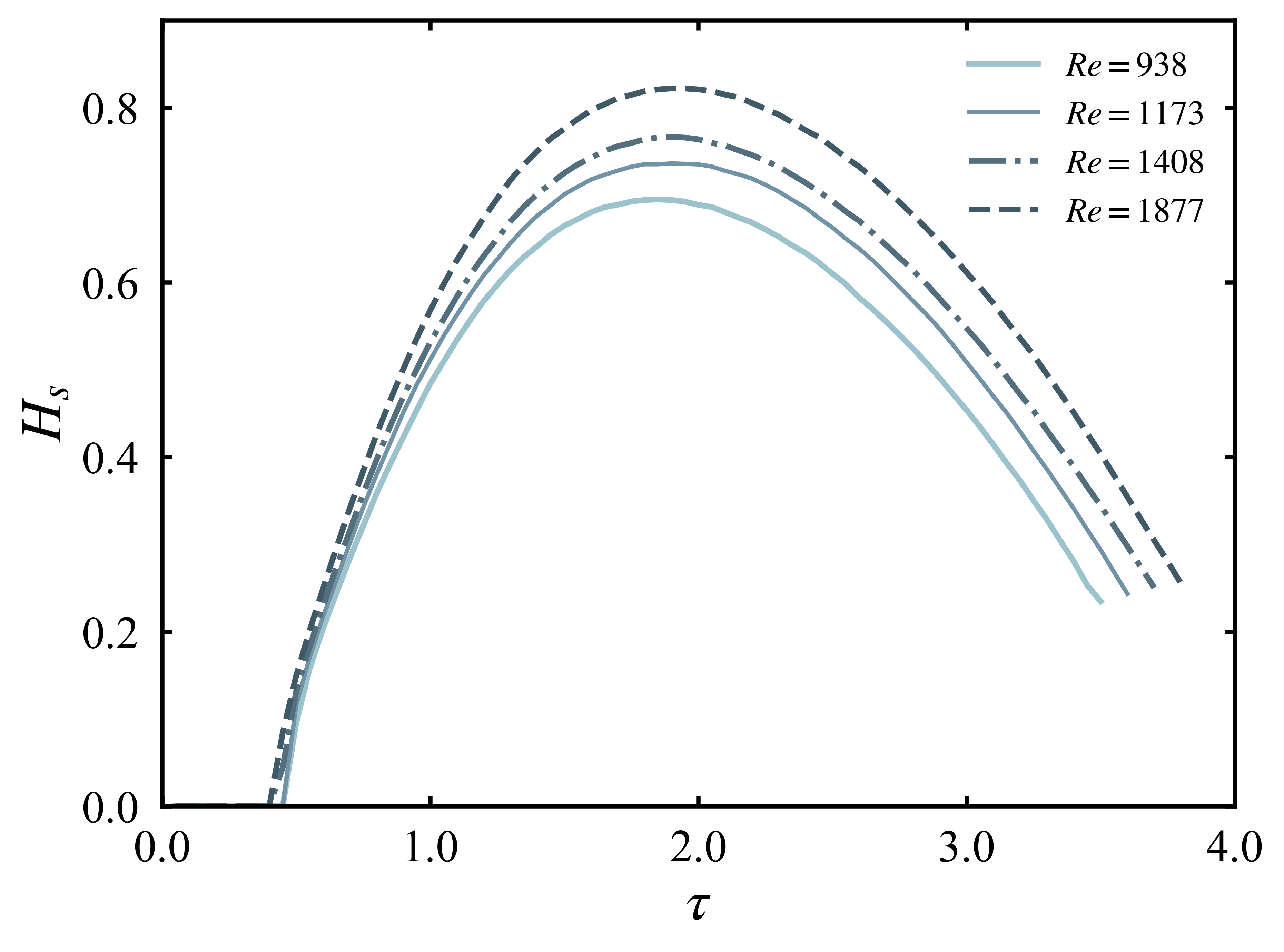} \\
 (a) & (b)
\end{tabular}
\caption{Variations of the central rising sheet height with changes in (a) $We$ and (b) $Re$. Note: the reference case in each comparison is $We=80$, $Re=938$, and only one dimensionless number is varied in each panel}
\label{fig:Height_variation}
\end{figure}

%\subsubsection{Height of the central rising sheet}\label{subsubsec:heights}
Finally, Fig. \ref{fig:Height_variation} shows the effects of surface tension and viscosity on the height of central rising sheet, through $We,~Re$. In each of Fig. \ref{fig:Height_variation}(a,b), the evolution of $H_s$ is shown in terms of one varying dimensionless number at a time, where the reference case is $We=80,~Re=938$. 

%even though delete it here, but i think this reveals something for the underlying mechanisms, or maybe not
% Given Fig. \ref{fig:Height_variation}(b), the interaction of drops starts at around the same time, and these lines collapse at the very beginning interaction stage. The inertia dominates at this stage. With the development of the lamella, the rim of central sheet starts to form as shown in Fig. \ref{fig:40_We130_f}, and surface tension starts to play an important role.

Both $We,~Re$ affect the dynamics of the central sheet, though to different extents, as suggested in Fig. \ref{fig:Height_variation}(a,b). With the increase of $We$, the maximum heights and the corresponding maximum times $t_{H_s}=t(H_S=H_{s,max})$ increase, because the central sheet rises over a longer time to its maximum value. Similar observations can be made from Fig. \ref{fig:Height_variation}(c) regarding $Re$, although here $t_{H_s}$ is approximately invariant; thus clearly in these problems surface tension controls $t_{H_s}$.

%because intertia part is deleted, the conclusion should also be changed: only inertia plays the most important role at the very beginning stage
%%%%%%%%maybe put this at the very beginning, and then we do not talk about the effects of Fr

\section{Kinematics of the central sheet}
\begin{figure}[!h]
\centering
\begin{tabular}{c}
  \includegraphics[width=0.7\columnwidth]{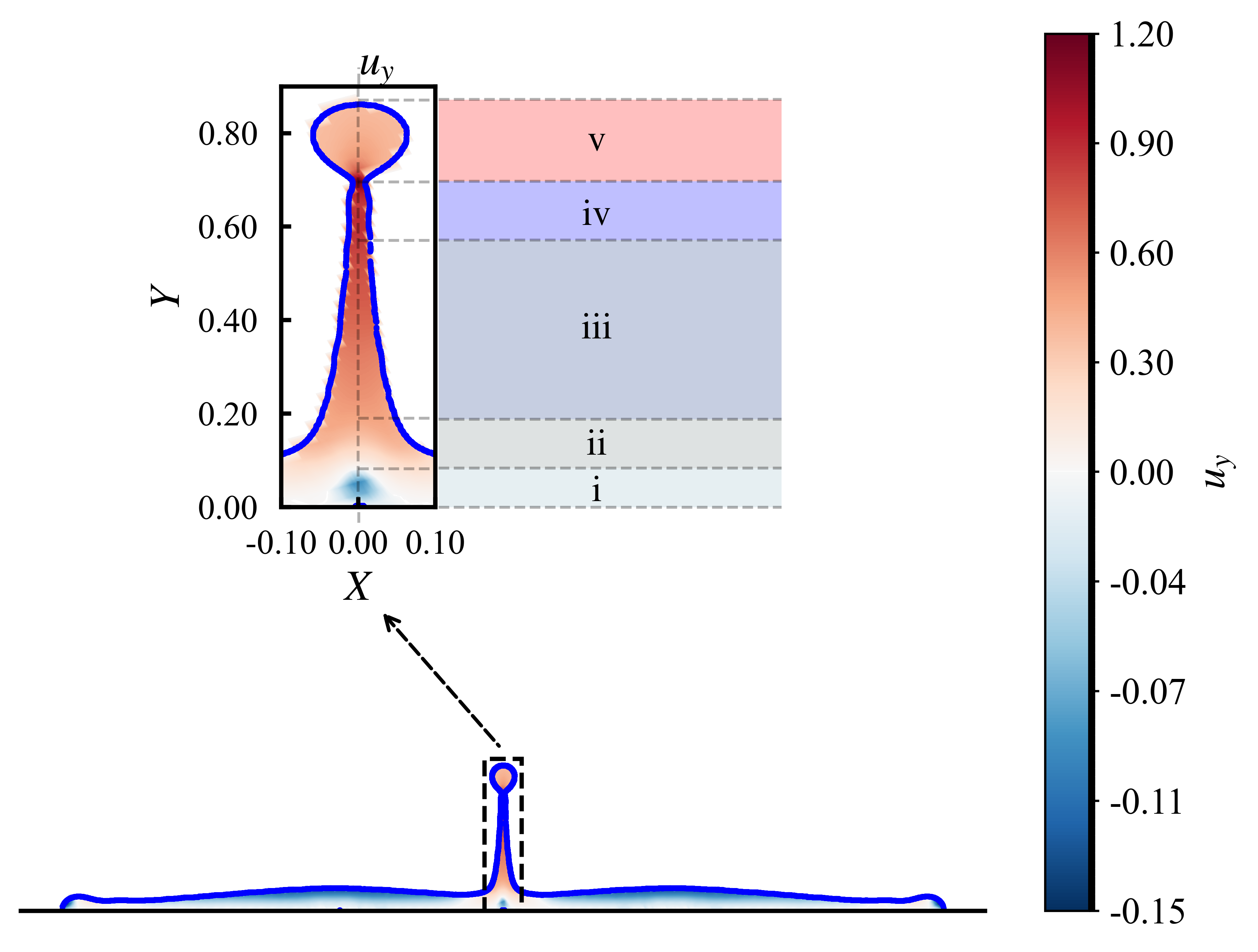}\\
  (a)
\end{tabular}
\begin{tabular}{c}
 \includegraphics[height=0.45\columnwidth]{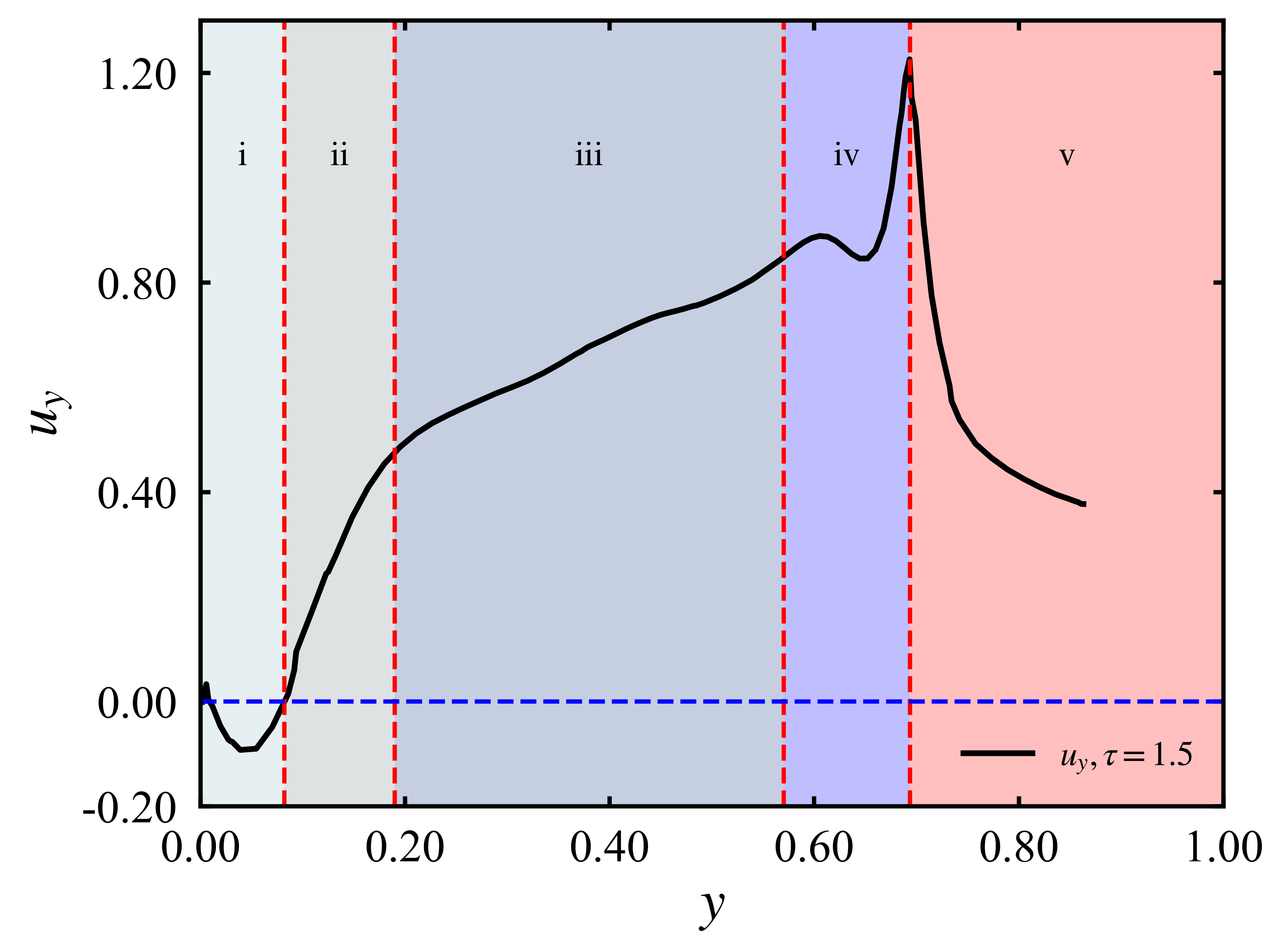}\\
 (b)
\end{tabular}
\caption{(a) Two-dimensional slice of the vertical velocity field of the central rising sheet: region $\mathrm{i}$ the vortex region; region $\mathrm{ii}$ the impact pulse region; region $\mathrm{iii}$ the main lamellar part of the rising sheet; region $\mathrm{iv}$ the transition from the lamella to the rim; and region $\mathrm{v}$ the rim. (b) Vertical velocity $u_y$ along the central line at $We=130$ and $Re=1189$}
\label{fig:velocity_region}
\end{figure}
\begin{figure}[t]
\centering
\begin{tabular}{cc}     
 \includegraphics[height=0.35\columnwidth]{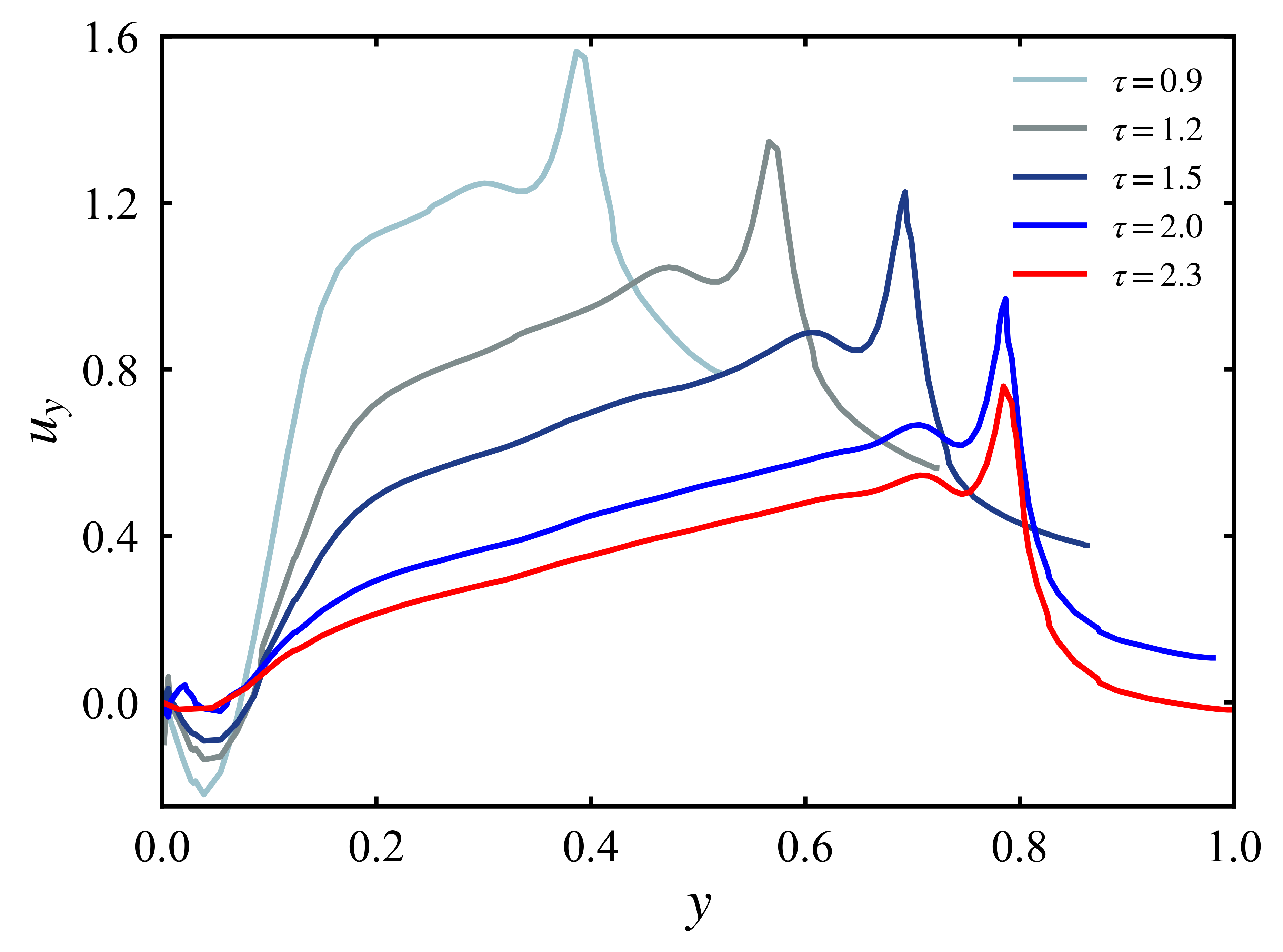} &
 \includegraphics[height=0.36\columnwidth]{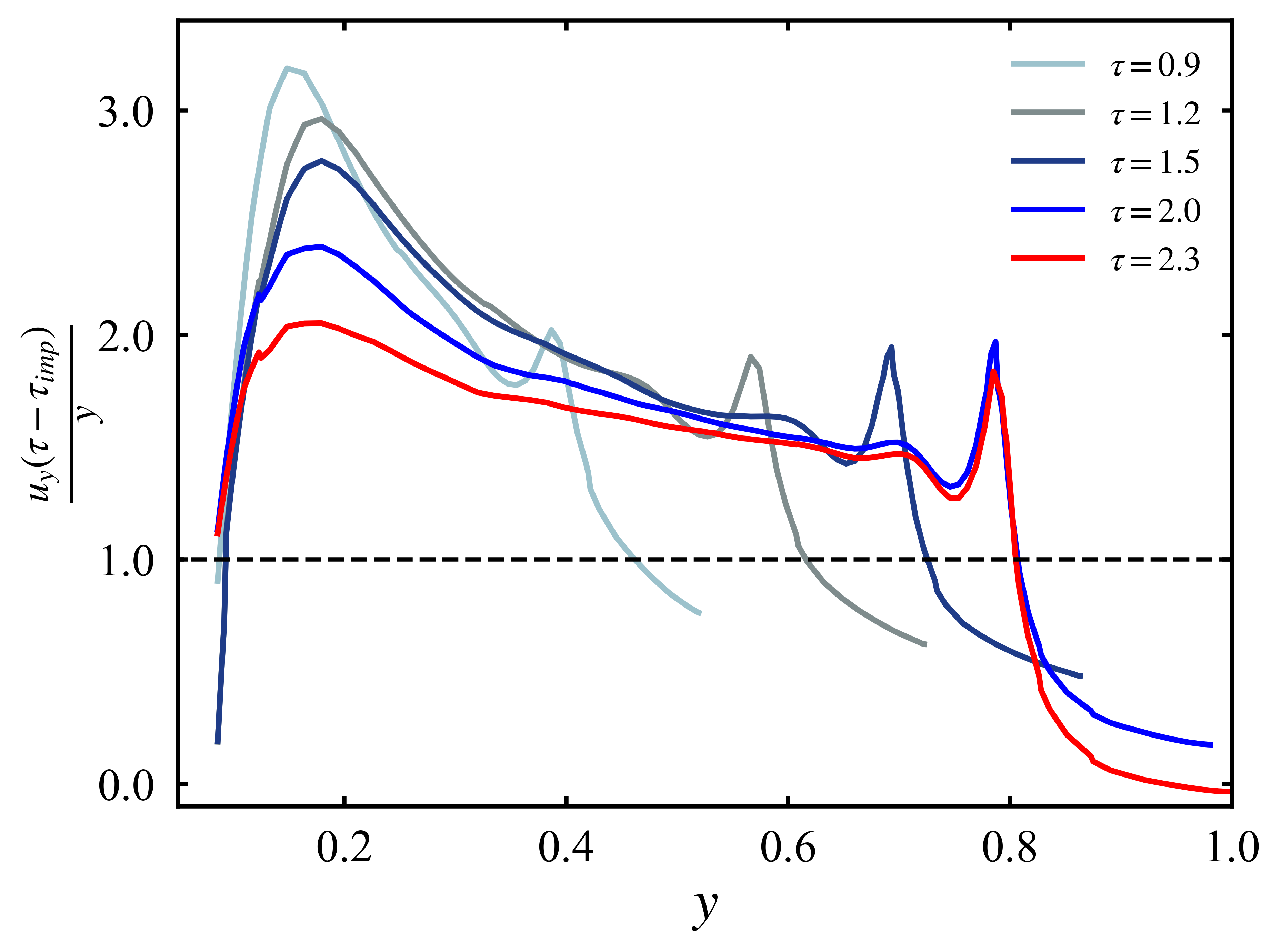} \\
 (a) & (b)
\end{tabular}
\caption{(a) Vertical velocity $u_y$ along the central line of the central sheet over time and (b) rescaled vertical velocity profiles at $We=130$ and $Re=1189$}
\label{fig:velocity_rescale}
\end{figure}

In this section, we turn to the kinematics in the central rising sheet. As given in Fig. \ref{fig:velocity_region}, we plot vertical velocity ($u_y$) of the central sheet at $t_{H_s}$ for $We=130,~Re=1189,~Fr=78$. $u_y$ is divided into five regions according to the characteristics: in region $\mathrm{i}$ (the vortex region ) at the base of the rising sheet, there is a vortex and flow recirculates near the substrate. The thickness of this region is around the same order as the thickness of bottom lamella. Region $\mathrm{ii}$, the impact pulse region, shows a rapid gain in velocity which decays over time as show in Fig. \ref{fig:velocity_rescale}(a). Region $\mathrm{iii}$ describes the main lamellar part of the rising sheet, and shows a gentler increase in velocity. Region $\mathrm{iv}$, describes the transition from the lamella part to the rim, and shows a rapid increase in velocity because of necking behaviour. Region $\mathrm{v}$ describes the rim. 

For the same case, we further investigate $u_y$ of the central sheet over time. In Fig. \ref{fig:velocity_rescale}(a), we plot $u_y$ at different times from early rising time to $t_{H_s}$. Generally, $u_y$ decreases with time at fixed $y$. Regions $\mathrm{i}$ and $\mathrm{ii}$ almost disappear at large time, such as $\tau=2.3$. It is well-known that, for single-drop spreading on a substrate, velocity in the lamellar part follows e.g. an $r/\tau$ scaling \citep{Yarin1995JFM,Wang2017JFM,Gordillo2019JFM,Tang2024JFM}, but Fig. \ref{fig:velocity_rescale}(b) shows that the corresponding scaling of $u_y\approx\frac{y}{\tau-t_{imp}}$ for the present case, where $t_{imp}$ is the rim collision time marking the formation of the central sheet, does not explain the data. This suggests that Regions $\mathrm{i}$ and $\mathrm{ii}$, which are missing from the cited cases, plays a strong role in setting the velocity profile in the central sheet. A clear dynamic model accounting for this velocity profile is left for future work.

\begin{figure}[!h]
\centering
\begin{tabular}{c}
  \includegraphics[width=0.7\columnwidth]{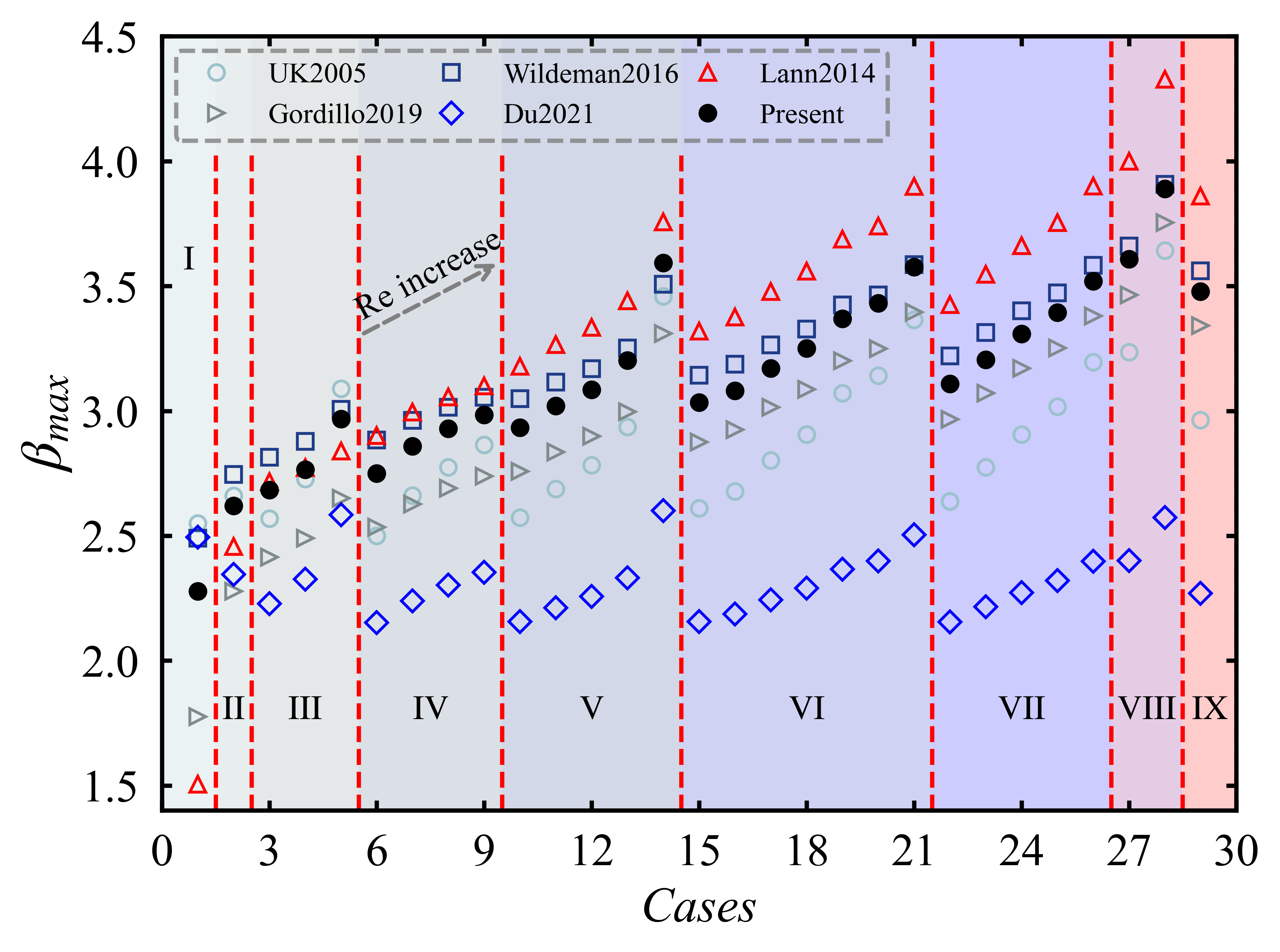}
\end{tabular}
\caption{Comparison of maximum spreading radius between simulation results and different models obtained from one drop impact \citep{UK2005Lang, Wildeman2016JFM, Laan2014PRA, Gordillo2019JFM, Du2021Lang}. Note: different regimes stand for different $We$ noted by Roman numerals, namely $We=15$ in ${\mathrm{I}}$; $We=30$ in ${\mathrm{II}}$; $We=40$ in ${\mathrm{III}}$; $We=54$ in ${\mathrm{IV}}$; $We=81$ in ${\mathrm{V}}$;  $We=104$ in ${\mathrm{VI}}$; $We=130$ in ${\mathrm{VII}}$; $We=150$ in ${\mathrm{VIII}}$; $We=200$ in ${\mathrm{IX}}$, and in each regime $Re$ increases as arrow shows}
\label{fig:Rmax_comparison}
\end{figure}

Furthermore, we compare the maximum spreading factor with previous theories. As shown in Fig. \ref{fig:Rmax_comparison}, all numerical cases are plotted as black circles, and the hollow points indicate various predictions for $\beta_{max}$ from the literature for single-drop impact at the given parameters. Different regimes are stand for different $We$ ranging from 15 to 200, and in each regime predicted and measured $\beta_{max}$ increases with $Re$. $\beta_{max}$ is not affected significantly by central interaction, which is also specified in \citet{Goswami2023JFM}. Our numerical data is in general agreement with most models in the literature with respect to both $We$ and $Re$. Moreover, our data is in closest agreement with  \citet{Wildeman2016JFM}, and this fact is used in the energetic model development. 

\section{Data processing details}\label{sec:data_process}
\begin{figure}[!h]
\centering
\begin{tabular}{c}     
 \includegraphics[height=0.4\columnwidth]{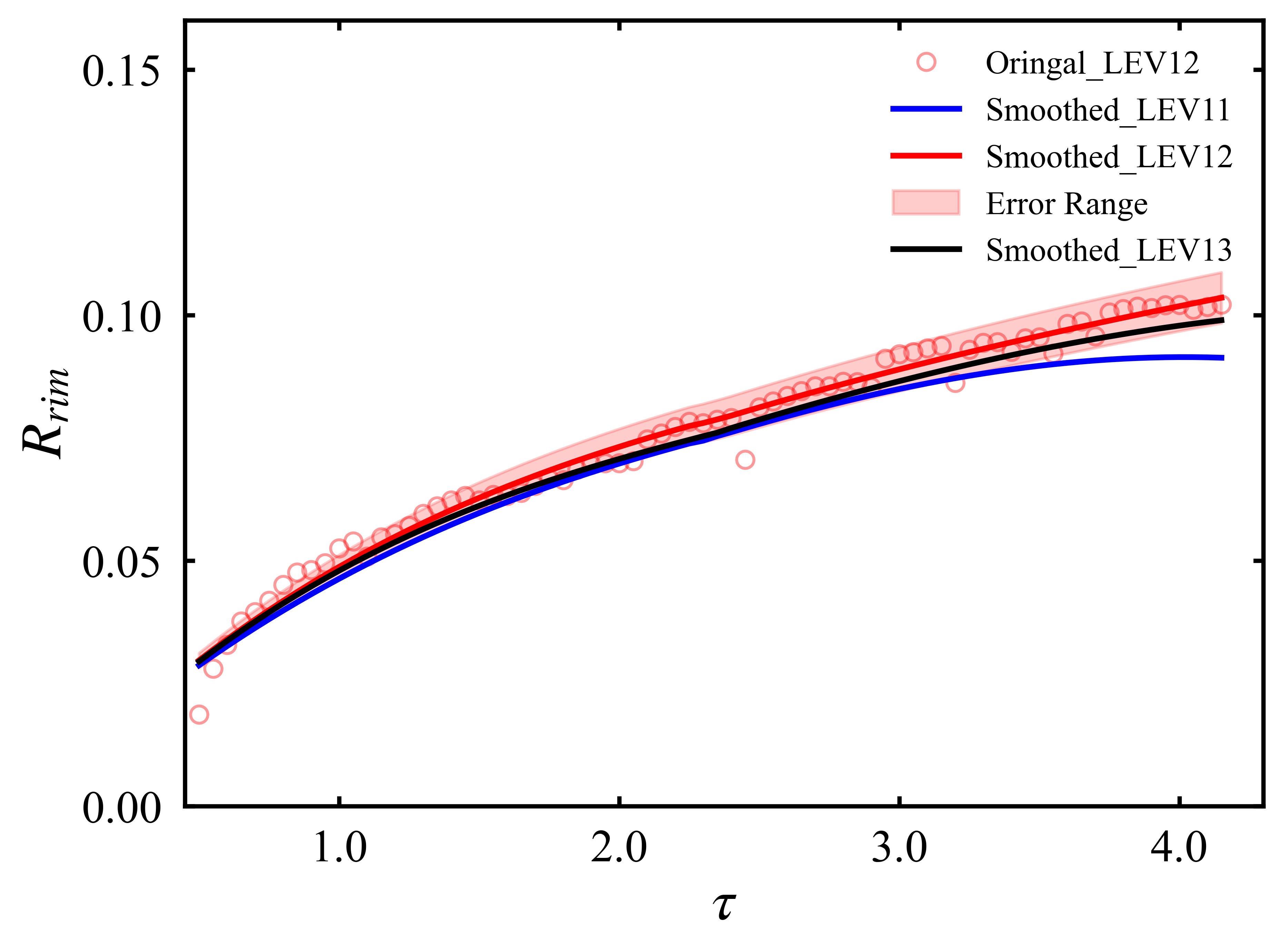}
\end{tabular}
\caption{Convergence test of the evolution of $R_{rim}$ on the central sheet for different grid resolution levels at $We=130$ and $Re=1189$, with original data shown by hollow symbols and post-processed results by solid lines}
\label{fig:centralrim_width}
\end{figure}

In this section, we provide a detailed explanation of the method used to plot $R_{rim}$, which is located at the maximum width of the rim on the central sheet. At early times the rim has not yet formed or its position is obscured by the central sheet, or by the drop bulk on the bottom, but we locate it by searching from the top of the central sheet. If it assumed that the rim grows monotonically (complicated somewhat by the adaptive nature of the grid) , we then obtain an approximate location at each time as the original data shown in Fig. \ref{fig:centralrim_width}. Only original data for level 12 are presented, while other results are shown with the smoothed lines for clarity. The error bar for the $R_{rim}$ is 5\%. The relative error of the smoothed $R_{rim}$ at $t_{H_s}$ are 5.63\% and 1.65\% respectively with different levels. We conclude that the data, processed using this method, are grid converged to obtain $R_{rim}$, and the evolution profiles of $R_{rim}$ shown in the main text are smoothed profiles within a 5\% error bar.

% \clearpage
% References using BibTeX
%\bibliographystyle{elsarticle-num-names}
%\bibliography{main}

\end{document}